\documentclass[11pt,letterpaper]{article}
\usepackage[T1]{fontenc}
\usepackage[margin=1in]{geometry}
\usepackage{amsmath,amsthm,mathtools}
\usepackage{newtxtext}
\usepackage[subscriptcorrection]{newtxmath}
\usepackage{bm,mathrsfs,graphicx,booktabs,array,enumitem}
\input{ursfs.fd}
\DeclareFontShape{U}{rsfs}{m}{n}{<-6.5> rsfs5 <6.5-8> rsfs7 <8-> rsfs10}{}
\usepackage{microtype}
\usepackage[round,authoryear]{natbib}
\usepackage[hidelinks,bookmarksnumbered]{hyperref}
\hypersetup{pdftitle={Single-fit inference for the realized prediction error of a logistic model with many predictors},
  pdfauthor={Seyoung Park and Eun Ryung Lee}}
\allowdisplaybreaks[2]

\newtheorem{theorem}{Theorem}
\newtheorem{lemma}{Lemma}
\newtheorem{corollary}{Corollary}
\newtheorem{proposition}{Proposition}
\theoremstyle{remark}
\newtheorem{remark}{Remark}
\newenvironment{tabnote}{\par\smallskip\noindent\begin{minipage}{\linewidth}\raggedright\footnotesize}{\end{minipage}\par}
\newcommand{\E}{\mathbb{E}}
\newcommand{\Prob}{\mathbb{P}}
\newcommand{\R}{\mathbb{R}}
\newcommand{\ind}{\mathbf{1}}
\newcommand{\pto}{\xrightarrow{P}}
\newcommand{\dto}{\Rightarrow}
\newcommand{\norm}[1]{\lVert#1\rVert}
\newcommand{\op}{{\mathrm{op}}}
\newcommand{\Var}{\operatorname{Var}}
\newcommand{\Cov}{\operatorname{Cov}}
\newcommand{\logit}{\operatorname{logit}}
\newcommand{\diag}{\operatorname{diag}}
\newcommand{\OptProofStep}[2]{\par\smallskip\noindent\textbf{Step #1: #2.}\ }
\DeclareMathOperator*{\argmin}{arg\,min}
\newcommand{\appendixserial}[1]{%
  \setcounter{#1}{0}%
  \expandafter\def\csname the#1\endcsname{S\arabic{#1}}%
  \expandafter\def\csname theH#1\endcsname{appendix.\arabic{#1}}}

\title{Single-fit inference for the realized prediction error of a logistic model with many predictors}
\author{Seyoung Park\thanks{Department of Applied Statistics, Yonsei University,
50 Yonsei-ro, Seodaemun-gu, Seoul 03722, South Korea.
\href{mailto:ishspsy@yonsei.ac.kr}{\nolinkurl{ishspsy@yonsei.ac.kr}}.}
\and Eun Ryung Lee\thanks{Department of Statistics, Sungkyunkwan University,
25-2 Sungkyunkwan-ro, Jongno-gu, Seoul 03063, South Korea.
\href{mailto:erlee@skku.edu}{\nolinkurl{erlee@skku.edu}}.}}
\date{}

\begin{document}
\maketitle

\begin{abstract}

When independent validation data are unavailable, evaluating a fitted logistic model on new patients requires estimating prediction error and uncertainty. Evaluating the fit on training data is often optimistic. Bootstrap validation is widely used to correct this bias, but with many predictors, it can retain bias and yield unreliable confidence intervals. In this article, we study these limitations and develop single-fit estimation and inference without resampling. When the number of Gaussian predictors grows proportionally with sample size, we analyze unpenalized logistic models. We prove that standard bootstrap correction can have nonzero limiting bias even as the resample count grows and location-shifted intervals can have zero limiting coverage. Instead, we adjust the fitted scores to consistently estimate realized Brier and log losses. We account for common fitting and adjustment error to construct valid marginal intervals and joint regions. We also establish these single-fit estimation and inference guarantees for certain non-Gaussian designs in proportional dimensions. Under moderate dimension growth and possible misspecification with general predictors, we extend single-fit inference to ridge regularization and construct asymptotically valid confidence intervals for realized log loss. Simulations and analyses of hospital and diabetes datasets support our theory and demonstrate the effectiveness of the proposed methods.

\end{abstract}

\noindent\textbf{Keywords.}\ Bootstrap; Calibration; Cross-validation; Logistic regression; Prediction error; Regularization.

\section{Introduction}\label{sec:intro}

When a fitted logistic model is used to predict clinical outcomes for new patients, it is essential to evaluate its realized performance and quantify estimation uncertainty \citep{collins2024tripodai}. Here, realized performance refers to the performance of this fit in independent new patients, conditional on the training sample. Without independent validation data, evaluating the same fixed fitted model on the training sample yields apparent performance \citep{austin2017events}. This apparent performance is typically optimistic relative to realized performance. To correct this optimism, bootstrap validation is widely used \citep{harrell1996multivariable,steyerberg2001internal,riley2020calculating,iba2021reevaluation}. In Harrell's correction \citep{harrell2015regression,iba2021reevaluation,harrell2026rms}, for example, we evaluate each bootstrap refit on its resample and the original sample, and subtract the average difference from apparent performance. With more bootstrap replicates, Monte Carlo error around the conditional bootstrap mean decreases.

However, with many predictors, performance evaluation on new patients remains difficult even at recommended sample sizes \citep{martin2021developing}. In Gaussian logistic regression without an intercept, unpenalized maximum likelihood exhibits upward asymptotic slope bias along the true direction in proportional dimensions with finite fits and produces overconfident predicted probabilities for small positive limiting dimension ratios \citep{sur2019modern,bai2021dont}. These distortions can increase optimism in apparent performance and make reliable optimism correction challenging. Indeed, with few events per predictor, simulation studies show that bootstrap correction reduces optimism in estimated discrimination, but positive bias can remain in some designs \citep{smith2014correcting,austin2017events,iba2021reevaluation}. In such small training samples, a recent study also found substantial variability in bootstrap shrinkage-factor estimates \citep{pate2026agreement}. Together, these results indicate the need for a theoretical characterization of the limiting bias and interval coverage of bootstrap validation for a logistic fit with many predictors.

In this article, we study standard bootstrap validation in large dimensions and develop a new single-fit inference procedure for the realized performance of a fitted logistic model. Under unpenalized Gaussian logistic models, we adjust fitted scores and account for shared estimation error to evaluate realized Brier and log losses. To handle general predictors, possible separation, and misspecification, we develop ridge-regularized inference for realized log loss with bounded features. In simulations, our estimates usually have smaller absolute bias than standard bootstrap estimates \citep{efron1983estimating,harrell2015regression} and root mean squared errors similar to those of approximate leave-one-out estimates \citep{rahnamarad2020scalable}. Without refitting, our intervals attain near-nominal coverage in most Gaussian simulations and in repeated clinical samples for ridge fits. Location-shifted bootstrap intervals \citep{noma2021confidence} or intervals without the covariance correction can undercover.

Prior studies analyzed prediction error without independent validation data in Gaussian linear models, where least-squares fitted values are linear in the responses and have explicit expected squared error \citep{rosset2020fixed}. In proportional dimensions, generalized and leave-one-out cross-validation are consistent for the realized squared error of least-squares and ridge fits \citep{patil2021uniform}. On the other hand, these results do not yield inference for the realized prediction error of the original logistic fit. Note that in logistic regression, no unpenalized fit exists when predictors separate the binary responses \citep{candes2020phase}. When a fit exists, its fitted scores satisfy nonlinear limiting equations that depend on the true signal even under Gaussian predictors \citep{sur2019modern}. In addition, discrimination, calibration, Brier loss, and log loss depend on the joint distribution of responses and fitted scores. Thus, bootstrap validation and single-fit inference require score and covariance expansions.

We establish three main theoretical and methodological results. First, for unpenalized Gaussian logistic fits, we show that Harrell's correction, out-of-bag validation, and the .632 estimator \citep{efron1983estimating} can have nonzero limiting bias even when the number of resamples is large. A typical multinomial resample repeats some patients and omits others. This resampling changes the patient weights and the ratio of predictors to distinct patients. When the dimension is fixed, both the original and resample ratios converge to zero. On the other hand, in proportional dimensions, they converge to distinct positive limits. In addition, such changes in weights and ratios can cause the limiting bootstrap optimism to differ from the true optimism of the original fit. For this fit, we establish limiting score distributions under the multinomial bootstrap and derive probability limits for all three estimators.

In particular, under weak signals, we show that Harrell's bootstrap estimator can overestimate model discrimination and underestimate realized Brier loss. When we subtract this estimated optimism from bootstrap quantiles of apparent performance \citep{noma2021confidence}, the coverage of the resulting location-shifted confidence interval can converge to zero. In high-dimensional logistic regression, prior studies analyzed existence of the maximum likelihood fit, slope estimation \citep{sur2019modern,candes2020phase,zhao2022asymptotic}, and prediction error limits for penalized fits \citep{sterzinger2026proportional}. Related resampling studies investigated coefficient uncertainty and the estimation errors of averaged coefficient estimators \citep{elkaroui2018can,clarte2024analysis,zhao2025adaptively,bellec2026asymptotics}. However, none of these studies establish high-dimensional bootstrap validation limits for the original unpenalized logistic fit or show that location-shifted performance intervals can have zero limiting coverage.

Second, for Gaussian predictors, we develop consistent estimation and joint inference without resampling for realized Brier and log losses of the original unpenalized fit with an intercept. Specifically, when the predictor dimension grows proportionally with sample size, we estimate a common scalar correction from fitted probabilities to adjust the fitted scores. Note that the adjusted training losses remain dependent across patients because they share this fit and scalar correction. If we omit this dependence, variance estimates are biased downward and intervals undercover. Hence, we account for both shared errors to construct valid marginal and joint confidence regions and confidence bounds for loss reduction over an intercept-only fit. In Theorem~\ref{thm:optboot-nongaussian}, we extend these single-fit consistency and joint inference results to certain designs with binary, categorical, or skewed predictors. We do not resample or estimate predictor covariance. Our Gaussian method applies in the interior of the original fit's existence region, including the part where bootstrap refits fail asymptotically.

In Gaussian single-index models, \citet{bellec2025observable} adjusted fitted scores and estimated the correlation between the fitted and true indices. Other work developed inference for logistic regression coefficients \citep{yadlowsky2021sloe} and estimated concordance for regularized Cox models \citep{massa2025observable}. \citet{rahnamarad2020scalable} approximated leave-one-out prediction error, and \citet{xu2021consistent} proved consistency of risk estimators in proportional linear regression. In Gaussian single-index models under strong convexity, \citet{bellec2026simultaneous} bounded the difference between approximate and exact leave-one-out estimates. Cross-validation studies analyzed risk estimation for ridge regression \citep{liu2020ridge,patil2021uniform,patil2022estimating} and also constructed confidence intervals for prediction error via nested cross-validation \citep{bates2024crossvalidation}. For an unpenalized logistic fit with an intercept and Gaussian predictors, we derive the joint limiting error distribution for estimating realized Brier and log losses inside the fit's existence region. We consistently estimate the full covariance matrix from the fit, and accounting for shared fitting and score-adjustment errors yields valid marginal intervals and joint confidence regions.

Third, to handle general predictors without guaranteed unpenalized fits, we study realized log loss under possible model misspecification. Under both classes, bounded features, a positive slope penalty, and an unpenalized intercept guarantee a single fit for one-step leave-one-out estimation of realized log loss. For $d=o(n^{2/3})$ under population moment, curvature, norm, and nondegenerate variance assumptions, we construct nominal confidence intervals. In a corollary for arbitrary $d$, we establish valid coverage under bounded full feature norm, uniformly positive slope penalty, and $\Pr(Y=1)$ away from $0$, $1/2$, and $1$. For the realized log loss of a ridge logistic fit, \citet{rahnamarad2020error} established an exact leave-one-out mean squared error bound with a population covariance constant and no sample-based standard error. For a full ridge fit with an unpenalized intercept, we derive an asymptotically normal one-step estimator of this loss and consistently estimate its variance.

The remainder of this article is organized as follows. In Section~\ref{sec:optboot-model}, we introduce the model and realized targets, and in Section~\ref{sec:optboot-scores}, we establish the limits of multinomial bootstrap validation. In Sections~\ref{sec:optboot-onefit} and~\ref{sec:general-inference}, we develop single-fit inference for the unpenalized model under Gaussian predictors and predictors with non-Gaussian components and for the ridge-regularized model, respectively. In Sections~\ref{sec:numerical} and~\ref{sec:support}, we present simulation studies and data analyses, and in Section~\ref{sec:discussion}, we conclude with a discussion. The proofs are in the appendices.

\section{Model}\label{sec:optboot-model}

\subsection{Gaussian logistic model and target performance}\label{subsec:optboot-logistic}

Suppose we observe independent training data $\mathcal D=\{(X_i,Y_i):1\le i\le n\}$ for $n$ patients, where $X_i\in\R^p$ is a vector of predictors and $Y_i\in\{0,1\}$ is a binary response. Let $s(x)=(1+e^{-x})^{-1}$ and $\ell_y(x)=\log(1+e^x)-yx$ denote the logistic function and negative log-likelihood for $y\in\{0,1\}$. We consider the Gaussian logistic model
\begin{equation}\label{eq:optboot-logistic}
 X_i\sim N_p(0,\Sigma_n),\qquad
 \Prob(Y_i=1\mid X_i)=s(\theta_n+X_i^\top\beta_n),\qquad \Sigma_n\succ0,
\end{equation}
where the intercept $\theta_n$, slope vector $\beta_n$, and predictor covariance $\Sigma_n$ are unknown. Define the signal variance $\gamma_n^2=\beta_n^\top\Sigma_n\beta_n$ and the ratio $\kappa_n=p/n$ of the number $p$ of predictors, excluding the intercept, to sample size. We assume $(\theta_n,\gamma_n,\kappa_n)\to(\theta,\gamma,\kappa)$ with $\gamma>0$ and $\kappa>0$.

On $\mathcal D$, we minimize $n^{-1}\sum_{i=1}^n\ell_{Y_i}(b+X_i^\top\beta)$ to obtain the unpenalized maximum likelihood estimator $(\widehat b,\widehat\beta)$ and risk scores $\widehat\eta_i=\widehat b+X_i^\top\widehat\beta$. Without penalization or variable selection, invertible linear transformations of $X_i$ reparameterize the slope and leave $\widehat\eta_i$ unchanged. In theory, we analyze standardized predictors without eigenvalue bounds on $\Sigma_n$ or covariance estimation. By \citet{candes2020phase} and Theorem~\ref{F-thm-exist}, the threshold $h(\theta,\gamma)\in(0,1)$ in equation~\eqref{S-boundary} determines whether $(\widehat b,\widehat\beta)$ exists. With probability tending to one, $(\widehat b,\widehat\beta)$ exists uniquely without linear separation when $0<\kappa<h(\theta,\gamma)$. When $\kappa>h(\theta,\gamma)$, separation occurs and no finite fit exists with probability tending to one.

We evaluate $(\widehat b,\widehat\beta)$ on the six measures $F\in\{A,B,R,a,c,d\}$ in equations~\eqref{S-risks}--\eqref{S-calibration} and subsection~\ref{S-reader-performance}. The area under the receiver operating characteristic curve (AUC) $A$ is the probability that the score for a random case exceeds that of an independent non-case, with half credit for ties. We define Brier loss $B$ as the mean squared difference between predicted probability and binary response, and log loss $R$ as the mean negative log-likelihood. The calibration intercept $a$ and slope $c$ minimize the mean logistic loss of the response on the risk score. Calibration-in-the-large $d$ minimizes this loss with the slope fixed at one.

We define apparent performance $F_{\rm app}$ by evaluating $F$ empirically on training pairs $\{(Y_i,\widehat\eta_i):1\le i\le n\}$ in $\mathcal D$. For an independent new patient $(X,Y)$ from \eqref{eq:optboot-logistic}, we define realized performance $F_{\rm ext}$ by evaluating the fixed fit $(\widehat b,\widehat\beta)$ on $(X,Y)$ conditional on $\mathcal D$. In particular, for log loss,
\[
 R_{\rm ext}=\E[\ell_Y(\widehat b+X^\top\widehat\beta)\mid\mathcal D].
\]
This expectation averages over a new patient with $(\widehat b,\widehat\beta)$ fixed and varies with the sample $\mathcal D$. Note that our target $F_{\rm ext}$ is distinct from its unconditional mean over the sample $\mathcal D$. It also differs from the performance of models fitted on smaller samples \citep{bates2024crossvalidation}.

Throughout, $\pto$ and $\dto$ denote convergence in probability and distribution, $\Phi$ the standard normal distribution function, $I_m$ the $m\times m$ identity matrix, $\operatorname{tr}$ the matrix trace, and $\|\cdot\|$ the Euclidean norm. We suppress subscript $n$ on data and performance measures. For positive deterministic $a_n$, $O_P(a_n)$ and $o_P(a_n)$ denote random quantities that are bounded in probability and converge in probability to zero after division by $a_n$.

\subsection{Bootstrap and cross-validation procedures}\label{subsec:optboot-procedures}
We first compare existing bootstrap validation methods for estimating realized performance. In each of $B_n$ bootstrap replicates, we draw $n$ patients with replacement from $\mathcal D$, independently conditional on $\mathcal D$, and refit the model by maximum likelihood. Let $c_B=1-e^{-1}$ denote the limiting proportion of distinct resampled patients. When $0<\kappa<c_Bh(\theta,\gamma)$, the refit exists with probability $1-o(1)$ (Theorem~\ref{S-state}).

To estimate optimism in $F$, we define bootstrap apparent performance by evaluating each refit on its resample. We average this performance minus the equal-weight evaluation of the refit on $\mathcal D$ over the $B_n$ replicates to obtain $\overline D_F$. With $\overline D_F$ and original apparent performance $F_{\rm app}$, we define Harrell's estimator $\widehat H_F=F_{\rm app}-\overline D_F$ \citep{harrell2015regression,harrell2026rms}, and also consider replacing the $B_n$-replicate average with conditional bootstrap expectation given $\mathcal D$.

For interval estimation, let $q^*_{u,F}$ denote the empirical $u$-quantile of bootstrap apparent performance across the $B_n$ replicates. For $0<\alpha<1$, we construct the location-shifted (LS) confidence interval $I_F^{\rm LS}=[q^*_{\alpha/2,F}-\overline D_F,\,q^*_{1-\alpha/2,F}-\overline D_F]$ \citep[Algorithm~1]{noma2021confidence}.

We also compare three resampling estimators (Sections~\ref{S-reader-bootstrap} and~\ref{S-alternatives}). Specifically, in out-of-bag (OOB) validation as in \citet{noma2021confidence}, we evaluate each refit with equal weights on omitted patients and average $F$ over $B_n$ replicates. In the .632 estimator \citep{efron1983estimating}, we weight original apparent performance and the OOB average by $1-c_B$ and $c_B$. For $K$-fold cross-validation with fixed $K\ge2$, we split $\mathcal D$ into $K$ nearly equal folds independently of the data. For each fold, we fit on the remaining folds, where the fraction tends to $1-1/K$. Then, we average $F$ over held-out folds or evaluate it on pooled held-out scores.

\section{Limits of bootstrap validation}\label{sec:optboot-scores}

\subsection{Limiting score distributions and bias of Harrell's correction}\label{sec:optboot-limits}
To study Harrell's estimator when $p/n\to\kappa>0$, we decompose its error relative to realized external performance as $\widehat H_F-F_{\rm ext}=(F_{\rm app}-F_{\rm ext})-\overline D_F$. In Theorem~\ref{S-empirical} and Proposition~\ref{S-functional}, we establish limiting joint distributions of response and score and limits of $F$. Here we obtain the probability limit of $F_{\rm app}-F_{\rm ext}$ as true limiting optimism, namely limiting apparent minus external performance of the original fit. On the other hand, for a refit, bootstrap limiting optimism is its limiting apparent performance minus limiting performance under equal weights on $\mathcal D$, and we prove $\overline D_F$ has this limit. Because these limits can differ, we define $\Delta_F$ as true limiting optimism minus this bootstrap limit.

When $\kappa>0$, two resampling distortions can make $\Delta_F$ nonzero. First, sampling with replacement retains a limiting fraction $c_B=1-e^{-1}$ of distinct patients and increases the refit's effective ratio of predictors to distinct patients from $\kappa$ to $\kappa/c_B$. Second, under limiting $\operatorname{Poisson}(1)$ resample counts, empirical evaluation on the resample weights patients by resample multiplicity and gives mean count $2$. However, equal-weight evaluation across $\mathcal D$ gives mean count $1$ and mixes omitted and resampled patients (Theorems~\ref{S-state}--\ref{S-empirical} and Proposition~\ref{S-functional}).

We express each performance measure as a functional $F$ of the joint distribution of the response and score. Because $F$ may be nonlinear, we first evaluate $F$ on each refit in $\widehat H_F$ before averaging optimism differences in $\overline D_F$. Under equal weights on $\mathcal D$, a refit's limiting joint distribution is a mixture of resampled and omitted patients with weights $c_B$ and $1-c_B=e^{-1}$ (Theorem~\ref{S-empirical}).
Because larger AUC and smaller Brier loss indicate better prediction, optimistic bias corresponds to $\Delta_A>0$ and $\Delta_B<0$. Evaluating $F$ at these limiting distributions, we show in the following theorem that the error of $\widehat H_F$ converges in probability to $\Delta_F$.

\begin{theorem}\label{thm:optboot-Harrell}
Assume \eqref{eq:optboot-logistic}, $\gamma>0$, and $0<\kappa<c_Bh(\theta,\gamma)$. For $F\in\{A,B,R,a,c,d\}$ and each fixed $B_n$, we have $\widehat H_F-F_{\rm ext}\pto\Delta_F$, and $\widehat H_F$ converges in probability to the limiting external performance plus $\Delta_F$. For $F=A,B$, both conclusions also hold for every deterministic $B_n\ge1$ and the conditional bootstrap expectation.
\end{theorem}

By Theorem~\ref{thm:optboot-Harrell} (Theorem~\ref{S-Harrell}), Harrell's estimator for $F_{\rm ext}$ is not consistent if $\Delta_F\ne0$. For AUC and Brier loss, this limiting bias does not vary with the number of resamples. To study weak signals at $\theta=0$, we define all limiting performance quantities and bootstrap biases at $\theta=\gamma=0$ by continuous extension as $\gamma\downarrow0$. Let $\pi_{\rm circ}=3.14159\ldots$ be the circle constant.

\begin{theorem}\label{thm:optboot-signs}
Under \eqref{eq:optboot-logistic}, when $\theta=0$ and $\gamma=0$, as $\kappa\downarrow0$,
\begin{equation}\label{eq:optboot-nullsigns}
 \Delta_A=(1-2^{-1/2})\sqrt{\kappa/\pi_{\rm circ}}+o(\sqrt\kappa),\qquad
 \Delta_B=-\frac12\kappa^2+O(\kappa^3).
\end{equation}
For every sufficiently small fixed $\kappa>0$, there exists $\epsilon(\kappa)>0$ such that for each fixed $\gamma\in(0,\epsilon(\kappa))$ and $\theta=0$, we have $\kappa<c_Bh(0,\gamma)$, $\Delta_A>0$, $\Delta_B<0$, and $\Delta_c>0$.
\end{theorem}

By Theorem~\ref{thm:optboot-signs}, in weak-signal models with $\theta=0$, small $\gamma>0$, and small $\kappa>0$, Harrell's estimator overestimates realized AUC ($\Delta_A>0$) and calibration slope ($\Delta_c>0$) and underestimates realized Brier loss ($\Delta_B<0$). From~\eqref{eq:optboot-nullsigns}, at $\theta=\gamma=0$, we see that $1-2^{-1/2}\approx29.3\%$ of apparent AUC optimism remains as $\kappa\downarrow0$ (Proposition~\ref{S-null}).

\begin{remark}[Numerical evaluation and slope shrinkage]
In 48 settings with prevalences $0.05$--$0.4$, AUCs of the true linear score $0.65$--$0.85$, and $\kappa\in[0.02,0.2]$, we obtain the existence threshold by numerical integration. The bootstrap fit exists asymptotically in 38 settings, where $\Delta_A>0$ with values up to $0.101$, $\Delta_B$ takes both signs, and $|\Delta_B|$ is below $0.05$ times the limiting external Brier loss. Because the unpenalized fit has apparent calibration slope $1$, $\widehat{H}_c$ converges in probability to the limiting slope of the bootstrap fit with equal weights on $\mathcal{D}$. At $\theta=\gamma=0$, this limit is $1/2+O(\kappa)$, but the limiting external slope is $0$. For $\theta=0$ and $0<\gamma<\epsilon(\kappa)$ in Theorem~\ref{thm:optboot-signs}, scaling fitted slopes by $\widehat{H}_c$ gives a limiting external calibration slope below $1$ (Section~\ref{S-shrinkage}).
\end{remark}

\subsection{Limits of cross-validation and bootstrap intervals}\label{sec:optboot-alternatives}
In the Gaussian model~\eqref{eq:optboot-logistic}, held-out resampling changes the effective dimension ratio (Theorem~\ref{S-altlimit}). For $F\in\{A,B,R,a,c,d\}$, fixed $K\ge2$ and $\kappa K/(K-1)<h(\theta,\gamma)$, the $K$-fold cross-validation estimator converges in probability to the limiting external performance at $\kappa K/(K-1)>\kappa$. When $0<\kappa<c_Bh(\theta,\gamma)$ and $B_n$ is fixed, the OOB and .632 estimators converge in probability to the bootstrap fit's limiting external performance and the $(1-c_B,c_B)$ weighted combination of the original fit's limiting apparent performance and this OOB limit. If $F\in\{A,B\}$, both limits also hold for all deterministic $B_n\ge1$ and conditional bootstrap expectations.

From Theorem~\ref{S-altlimit}, for $F\in\{A,B\}$ and each deterministic $B_n\ge1$, the OOB and .632 estimation errors have probability limits equal to the differences between the estimator limits and the limiting external performance of the original fit. At $\theta=\gamma=0$, by Proposition~\ref{S-null}, the limiting OOB and .632 Brier errors are $\kappa/4+O(\kappa^2)$ and $(1-3/e)\kappa/4+O(\kappa^2)$ as $\kappa\downarrow0$. By continuity, for sufficiently small fixed $\kappa>0$, these opposite signs hold for $\theta=0$ and small $\gamma>0$ in the bootstrap existence region. Here, OOB overestimates realized Brier loss and the .632 estimator underestimates it.

For interval estimation of optimism-corrected performance, \citet{harrell2025bootstraplimits} studied asymmetric bootstrap limits. On the other hand, for the location-shifted interval $I_F^{\rm LS}$ \citep[Algorithm~1]{noma2021confidence}, we subtract the estimated optimism from the bootstrap quantiles of apparent performance, and we derive its limiting center and coverage for AUC and Brier loss in the next proposition.

\begin{proposition}\label{prop:optboot-ls}
Assume \eqref{eq:optboot-logistic}, $\gamma>0$, $0<\kappa<c_Bh(\theta,\gamma)$, $F\in\{A,B\}$, and $0<\alpha<1$. For every deterministic $B_n\ge1$, both lower and upper limits $\inf I_F^{\rm LS}$ and $\sup I_F^{\rm LS}$ converge in probability to the limiting performance of the bootstrap fit evaluated with equal weights on the original sample. If this limit differs from the limiting external performance of the original fit, then $\Prob\{F_{\rm ext}\in I_F^{\rm LS}\}\to0$.
\end{proposition}

In Proposition~\ref{prop:optboot-ls}, because bootstrap apparent performance concentrates across resamples, $q^*_{\alpha/2,F}$ and $q^*_{1-\alpha/2,F}$ converge in probability to its limit, which cancels when we subtract $\overline D_F$. Thus, in probability, both bounds of $I_F^{\rm LS}$ converge to the bootstrap fit's limiting performance under equal weights on $\mathcal D$ and its width converges to zero. If this center differs from the original fit's limiting external performance, coverage converges to zero. This center's bias minus the bias $\Delta_F$ of $\widehat H_F$ equals the bootstrap fit's limiting apparent performance minus that of the original fit. For Brier loss at $\theta=\gamma=0$ as $\kappa\downarrow0$, this center bias is $-\kappa/4+O(\kappa^2)$ and exceeds in magnitude $\Delta_B=-\kappa^2/2+O(\kappa^3)$ in~\eqref{eq:optboot-nullsigns} (Section~\ref{S-ls-intervals} and Proposition~\ref{S-null}).

\section{Joint inference for realized Brier and log losses from one fit}\label{sec:optboot-onefit}

\subsection{Estimation under the Gaussian model}\label{subsec:method-scalar}

To avoid bootstrap dimension and weight distortions, we evaluate the original unpenalized fit of~\eqref{eq:optboot-logistic} at $\kappa_n=p/n$ with unit weights and no resampling. When $\gamma>0$ and $0<\kappa<h(\theta,\gamma)$, linear invariance implies that its score distributions depend on $\Sigma_n$ only through $\gamma_n$. Conditional on $\mathcal{D}$, a new patient's pair $(Y,\widehat{b}+X^\top\widehat\beta)$ converges in distribution in probability to $(Y,V_0)$. On $\mathcal{D}$, the empirical distribution of $\{(Y_i,\widehat\eta_i):1\le i\le n\}$ converges in probability to the distribution of $(Y,T_0)$ (Theorems~\ref{S-state} and~\ref{S-empirical}). Here $(Y,T_0,V_0)$ satisfies $T_0+\lambda_0\{s(T_0)-Y\}=V_0$ (equation~\eqref{S-prox}), where the unique deterministic scalar $\lambda_0>0$ solves $\E[\lambda_0s'(T_0)/\{1+\lambda_0s'(T_0)\}]=\kappa$ (equation~\eqref{S-state-eq}). Because $T_0=V_0-\lambda_0\{s(T_0)-Y\}$, we have 
\[
T_0>V_0 \ \mbox{when}\ Y=1 \quad \mbox{and}\quad
 T_0<V_0 \  \mbox{when}\  Y=0,
 \]
which reduces apparent Brier and log losses. To correct this optimism, we form 
\[
\mbox{residuals}\ q_i=s(\widehat\eta_i)-Y_i,\quad \mbox{curvatures}\ d_i=s'(\widehat\eta_i)=s(\widehat\eta_i)\{1-s(\widehat\eta_i)\}.
\]
 For fixed $\lambda>0$, $n^{-1}\sum_{i=1}^n\lambda d_i/(1+\lambda d_i)$ converges in probability to $\E[\lambda s'(T_0)/\{1+\lambda s'(T_0)\}]$ (Theorem~\ref{S-empirical}). Because $d_i>0$ on any finite fit and $0<\kappa_n<1$, this average increases strictly from zero to one on $(0,\infty)$ and equals $\kappa_n$ at a unique root $\widehat\lambda>0$, where we obtain $\widehat\lambda$ without estimating $\Sigma_n$ or inverting a matrix. In Algorithm~1, we compute $\widetilde{V}_i=\widehat\eta_i+\widehat{\lambda}q_i$ to recover the limiting distribution of $(Y,V_0)$ in Theorem~\ref{S-augconsistent}. From $(Y_i,\widetilde{V}_i)$, we form the estimators $\widehat{F}$, and on new patients we use the original fit $(\widehat{b},\widehat\beta)$ for prediction.

\medskip
\noindent\begin{minipage}{\linewidth}

\hrule height 0.8pt
\vspace{3pt}
\noindent\textbf{Algorithm 1:} Performance estimation under the Gaussian model\par
\vspace{2pt}
\hrule height 0.4pt
\vspace{4pt}
\noindent\textbf{Step 1 (Unpenalized fit):} Fit the unpenalized logistic model with an intercept to $\mathcal D$ by maximum likelihood, and obtain $\widehat\eta_i=\widehat b+X_i^\top\widehat\beta$, $q_i=s(\widehat\eta_i)-Y_i$, and $d_i=s'(\widehat\eta_i)=s(\widehat\eta_i)\{1-s(\widehat\eta_i)\}$.\\[3pt]
\textbf{Step 2 (Score adjustment and point estimation):} For a finite fit with $0<\kappa_n=p/n<1$, solve for the unique scalar $\widehat\lambda>0$, compute the adjusted scores $\widetilde V_i$, and evaluate each performance measure $F\in\{A,B,R,a,c,d\}$ on the adjusted pairs $(Y_i,\widetilde V_i)$ to obtain $\widehat F$:

\[
 \frac1n\sum_{i=1}^n\frac{\widehat\lambda d_i}{1+\widehat\lambda d_i}=\kappa_n,\qquad
 \widetilde V_i=\widehat\eta_i+\widehat\lambda q_i.
\]
\vspace{2pt}
\hrule height 0.8pt
\end{minipage}\par
\medskip

To construct confidence intervals for $(B_{\rm ext},R_{\rm ext})$, we estimate the asymptotic covariance matrix of $\sqrt{n}(\widehat{B}-B_{\rm ext},\widehat{R}-R_{\rm ext})^\top$. The adjusted losses on $\mathcal{D}$ depend on the shared fit and $\widehat\lambda$, and thus the sample variance of these losses underestimates the asymptotic variance. Let $\ell_B(y,v)=(s(v)-y)^2$, $\ell_R(y,v)=\ell_y(v)$, and $\ell'_F(y,v)=\partial_v\ell_F(y,v)$ for $F\in\{B,R\}$. At $(\kappa,\theta,\gamma)$, we linearize the equation for $\widehat\lambda$ around $\lambda_0$ with derivative $J=\E[s'(T_0)/\{1+\lambda_0s'(T_0)\}^2]>0$ and define $a_F=\E[\{s(T_0)-Y\}\ell'_F(Y,V_0)]$ (Lemma~\ref{S-xa-multiplier}). By Proposition~\ref{S-xa-covariance-identity}, from errors in $\widehat\lambda$ and the squared norm of orthogonal slopes and their covariances with empirical losses, for $F,G\in\{B,R\}$, we obtain $\operatorname{Cov}\{\ell_F(Y,V_0),\ell_G(Y,V_0)\}+\lambda_0a_Fa_G/J$ (equation~\eqref{S-xa-integrals}). Here, the first term is the independent-patient loss covariance, and the rank-one term accounts for shared fitting and adjustment error. To estimate both terms, let $\operatorname{Cov}_n$ denote the empirical covariance with divisor $n$. Replacing population expectations and $\lambda_0$ with sample averages on $(Y_i,\widetilde{V}_i)$ and $\widehat\lambda$ from Algorithm~1 (equation~\eqref{S-xa-variance-estimator}), we define $\widehat\Sigma=(\widehat\Sigma_{FG})_{F,G\in\{B,R\}}$ by
\[
 \widehat\Sigma_{FG}=\operatorname{Cov}_n\{\ell_F(Y_i,\widetilde V_i),\ell_G(Y_i,\widetilde V_i)\}+\frac{\widehat\lambda\{n^{-1}\sum_{i=1}^nq_i\ell_F'(Y_i,\widetilde V_i)\}\{n^{-1}\sum_{i=1}^nq_i\ell_G'(Y_i,\widetilde V_i)\}}{n^{-1}\sum_{i=1}^nd_i/(1+\widehat\lambda d_i)^2}.
\]
Let $z_a=\Phi^{-1}(a)$ denote the standard normal $a$-quantile. For $F\in\{B,R\}$ and $0<\alpha<1$, we form the $(1-\alpha)$ confidence interval
\[
 \left[\widehat F\pm z_{1-\alpha/2}(\widehat\Sigma_{FF}/n)^{1/2}\right],
\]
clipped to $[0,1]$ for $F=B$ or $[0,\infty)$ for $F=R$.

\subsection{Consistency and joint inference}\label{subsec:onefit-guarantees}

In the following theorem, we establish consistency of the corrected estimates for the realized performance of the original fit throughout $0<\kappa<h(\theta,\gamma)$, including $c_Bh(\theta,\gamma)<\kappa<h(\theta,\gamma)$, where a bootstrap refit fails to exist with probability $1-o(1)$.

\begin{theorem}\label{thm:optboot-onefit}
Assume \eqref{eq:optboot-logistic}, $\gamma>0$, and $0<\kappa<h(\theta,\gamma)$. Let $\lambda_0>0$ denote the deterministic positive correction factor in the limiting equations for the original unpenalized fit (with unit weight for every observation) in Theorem~\ref{S-state}. Then $\widehat\lambda\pto\lambda_0$, and for $F\in\{A,B,R,a,c,d\}$, we have $\widehat F-F_{\rm ext}\pto0$.
\end{theorem}

We prove Theorem~\ref{thm:optboot-onefit} in Theorem~\ref{S-augconsistent}. Prior studies compare approximate and exact leave-one-out errors under strongly convex regularization in Gaussian single-index models \citep{bellec2026simultaneous}. For $F\in\{B,R\}$, both $\widehat F$ and $F_{\rm ext}$ vary with $\mathcal D$ at order $n^{-1/2}$. When we decompose $\widehat\beta$ into signal-direction and covariance-orthogonal slopes, first-order intercept and signal-slope errors cancel in $\widehat F-F_{\rm ext}$ (Proposition~\ref{S-xa-head-cancellation}). The remaining estimation error comes from $\widehat\lambda$, the squared norm of the orthogonal slopes, and the empirical losses. For $\widehat\Sigma$ in Section~\ref{subsec:method-scalar}, let $\Sigma=(\Sigma_{FG})_{F,G\in\{B,R\}}$ denote the population matrix at $(\kappa_n,\theta_n,\gamma_n)$ in equation~\eqref{S-xa-integrals}. For $n$-dependent $\Sigma$, we define convergence in distribution to $N_2(0,\Sigma)$ as Gaussian approximation in the bounded Lipschitz metric. In the next theorem, we establish the joint normal approximation and the consistency of the covariance estimator for $\Sigma$.

\begin{theorem}\label{thm:optboot-onefitci}
Let $\mathcal K$ be a compact subset of $\{(\kappa,\theta,\gamma):\gamma>0,\ 0<\kappa<h(\theta,\gamma)\}$. Assume \eqref{eq:optboot-logistic} and $(\kappa_n,\theta_n,\gamma_n)\in\mathcal K$ for every $n$. For $0<\alpha<1$ and $F\in\{B,R\}$, uniformly on $\mathcal K$, we have $\sqrt n(\widehat B-B_{\rm ext},\widehat R-R_{\rm ext})^\top\dto N_2(0,\Sigma)$, $\widehat\Sigma-\Sigma\pto0$ with $\Sigma\succ0$, and $\Prob\{F_{\rm ext}\in[\widehat F\pm z_{1-\alpha/2}(\widehat\Sigma_{FF}/n)^{1/2}]\}\to1-\alpha$.
\end{theorem}

We prove Theorem~\ref{thm:optboot-onefitci} in Theorem~\ref{S-xa-clt}, Proposition~\ref{S-xa-variance-consistency}, and Corollary~\ref{S-xb-clt}. With the covariance correction, we obtain asymptotically nominal marginal coverage across training samples for the realized losses of the respective fitted models.

\begin{remark}[Undercoverage without the variance correction]\label{rem:undercoverage}
Under the conditions of Theorem~\ref{thm:optboot-onefitci}, the rank-one covariance term is strictly positive on both the Brier and log-loss diagonal entries. When we construct a nominal $(1-\alpha)$ marginal normal interval from only the empirical variance of adjusted losses, the upper limit of coverage for $F_{\rm ext}$ is strictly below $1-\alpha$ uniformly over the compact parameter set (equation~\eqref{eq:independent-coverage}).
\end{remark}

\begin{remark}[Joint confidence regions]
When the fit is finite and $\widehat\Sigma$ is nonsingular, we construct the $(1-\alpha)$ Wald confidence ellipse centered at $(\widehat B,\widehat R)^\top$ for Brier and log losses. Here, we use covariance $\widehat\Sigma/n$ and the $(1-\alpha)$ quantile of the chi-squared distribution with two degrees of freedom. We may intersect this ellipse with the natural loss range $[0,1]\times[0,\infty)$. Under the conditions of Theorem~\ref{thm:optboot-onefitci}, where $\Sigma\succ0$ and $\widehat\Sigma-\Sigma\pto0$, we obtain asymptotic $(1-\alpha)$ coverage for $(B_{\rm ext},R_{\rm ext})^\top$ across training samples (Section~\ref{S-joint-evaluation}). Alternatively, we can form a rectangular joint confidence region from two Bonferroni-adjusted marginal intervals.
\end{remark}

\begin{remark}[Comparison with an intercept-only fit]\label{subsec:loss-reduction}
When both response classes occur in $\mathcal D$, we also compare the full fit with an intercept-only fit based on the sample prevalence. Specifically, we subtract $\widehat F$ in Algorithm~1 from the empirical mean reference loss for $F\in\{B,R\}$. In Section~\ref{S-reference-gain} (Corollary~\ref{cor:reference-gain}), we establish uniform normal approximation and covariance consistency on $\mathcal K$. We then construct two-sided normal intervals and a one-sided $(1-\alpha)$ normal lower confidence bound for the realized loss reduction. When this lower bound is positive, we declare improvement. For each fixed loss $F\in\{B,R\}$, the asymptotic joint probability of declaring improvement and a nonpositive realized loss reduction is at most $\alpha$, uniformly on $\mathcal K$.
\end{remark}

\subsection[Predictors with non-Gaussian components]{Predictors with non-Gaussian components}\label{subsec:optboot-nongaussian}
We extend Theorems~\ref{thm:optboot-onefit} and~\ref{thm:optboot-onefitci} from Gaussian predictors as follows. The pairs $(X_i,Y_i)$ and pair $(X,Y)$ for a new patient are independent and identically distributed with $\Prob(Y_i=1\mid X_i)=s(\theta_n+X_i^\top\beta_n)$ and $X_i=(X_{i1}^\top,X_{i2}^\top)^\top$. For fixed $m_0$, $X_{i1}\in\R^{m_0}$ has a fixed distribution with nonsingular covariance and sub-Gaussian tails, which allows binary, categorical, or skewed predictors. Let $X_{i2}=\mu_n+\Psi_nX_{i1}+L_n\xi_i$ with deterministic unknown $\mu_n$, $\Psi_n$, and invertible matrix $L_n$, and $\xi_i\in\R^{p-m_0}$ independent of $X_{i1}$. In case (a), $\xi_i\sim N_{p-m_0}(0,I_{p-m_0})$ and $\beta_n$ is not restricted. In case (b), $\xi_i$ has independent coordinates with mean zero, unit variance, and uniformly bounded sub-Gaussian norms, and the coefficients of $X_{i2}$ in $\beta_n$ equal zero. For case (b), we do not require symmetry, zero third moments, or fourth moments equal to those under Gaussianity. We define standardized parameters $(\vartheta_n,\varpi_n)$ and threshold $h_Z(\vartheta,\varpi)$ by equations~\eqref{S-NG-eq-affineparameters} and~\eqref{S-NG-eq-headboundary} with $\Omega_n=L_nL_n^\top$, where $\vartheta_n$ is the standardized intercept ($\theta_n$ in Appendix~\ref{S-model}). Under \eqref{eq:optboot-logistic}, $h_Z(\vartheta_n,\varpi_n)=h(\theta_n,\gamma_n)$.

\begin{theorem}\label{thm:optboot-nongaussian}
Suppose case (a) or case (b) holds and $(\kappa_n,\vartheta_n,\varpi_n)$ lies in a compact subset $\mathcal K$ of $\{(\kappa,\vartheta,\varpi):0<\kappa<h_Z(\vartheta,\varpi)\}$ for all $n$. Let $\lambda_{0,n}$ denote the correction factor of the state in equation~\eqref{S-NG-eq-newstate} and $\Sigma$ denote the matrix in equation~\eqref{S-NG-eq-cov}, both evaluated at $(\kappa_n,\vartheta_n,\varpi_n)$. With probability tending to one, $(\widehat b,\widehat\beta)$ exists and is unique. For Algorithm~1 without modification, $\widehat\lambda-\lambda_{0,n}\pto0$, $\widehat F-F_{\rm ext}\pto0$ for $F\in\{A,B,R,a,c,d\}$, and all conclusions of Theorem~\ref{thm:optboot-onefitci} hold with this $\Sigma$. These conclusions are uniform on $\mathcal K$.
\end{theorem}

We prove cases (a) and (b) in Theorems~\ref{S-NG-thm-head} and~\ref{S-NG-U-main}. There exist scale mixtures of Gaussian vectors for which $\widehat B$ and $\widehat R$ have nonzero limiting biases (Proposition~\ref{S-NG-prop-mixture}).

\section{Inference for realized log loss of the penalized fit}\label{sec:general-inference}

\subsection{Estimation under general covariates}

We consider independent and identically distributed pairs $(X_i,Y_i)$ with general covariates $X_i\in\R^p$. Here, we estimate realized log loss under possible logistic model misspecification. Let $\phi:\R^p\to\R^{d-1}$ be a fixed feature map and define $z_i=(1,\phi(X_i)^\top)^\top\in\R^d$. Note that for an independent new pair $(X,Y)$ from the same population with $z=(1,\phi(X)^\top)^\top$ and deterministic $u\in\R^d$, the population log loss is $R(u)=\E\ell_Y(z^\top u)$ over $(X,Y)$. With a deterministic penalty $\Lambda_n\succeq0$, we compute the penalized fit
\begin{equation}\label{eq:practical-fit}
 \widehat u=\arg\min_u\left\{\frac1n\sum_{i=1}^n\ell_{Y_i}(z_i^\top u)+\frac12u^\top\Lambda_nu\right\},
\end{equation}
and target its realized log loss $R(\widehat u)$ conditional on $\mathcal D$.

For general covariates, the sample can separate and a finite unpenalized logistic fit may not exist. In~\eqref{eq:practical-fit}, we use bounded features $\phi(X_i)$ and a positive ridge slope penalty $\Lambda_n$. When both response classes are present, we obtain a unique finite fit and bounded observation weights. We then apply approximate leave-one-out (ALO) estimation \citep{rahnamarad2020scalable,rahnamarad2020error} to remove optimism without $n$ refits. Because an unnormalized leave-one-out sum multiplies the per-observation penalty by $n/(n-1)$, we define $F_{-i}(u)=(n-1)^{-1}\sum_{j\ne i}\ell_{Y_j}(z_j^\top u)+u^\top\Lambda_nu/2$ at the same penalty $\Lambda_n$ (equation~\eqref{S-reader-loo}). Taking one Newton step on $F_{-i}$ from $\widehat u$ yields the adjusted score
\[
 \widetilde\eta_i=z_i^\top[\widehat u-\{\nabla^2F_{-i}(\widehat u)\}^{-1}\nabla F_{-i}(\widehat u)].
\]
At $\widehat u$, we define 
\[
q_i=s(z_i^\top\widehat u)-Y_i,\ w_i=s'(z_i^\top\widehat u), \ r=n^{-1}\sum_{j=1}^nq_jz_j+\Lambda_n\widehat u,\ M=n^{-1}\sum_{j=1}^nw_jz_jz_j^\top+(1-1/n)\Lambda_n.
\]
 Subtracting $\ell_{Y_i}(z_i^\top u)+u^\top\Lambda_nu/2$ from $n$ times the objective in~\eqref{eq:practical-fit} yields $(n-1)\nabla F_{-i}(\widehat u)=nr-(q_iz_i+\Lambda_n\widehat u)$ and $(n-1)\nabla^2F_{-i}(\widehat u)=n(M-n^{-1}w_iz_iz_i^\top)$. Applying the Sherman--Morrison identity to $M-n^{-1}w_iz_iz_i^\top$ reduces the Newton step above identically to $\widetilde\eta_i$ in Step~2 of Algorithm~2. We average $\ell_{Y_i}(\widetilde\eta_i)$ to define the estimator $\widehat C_{\rm ALO}$ of $R(\widehat u)$.

\smallskip
\noindent\begin{minipage}{\linewidth}
\setlength{\abovedisplayskip}{4pt}
\setlength{\belowdisplayskip}{4pt}
\setlength{\abovedisplayshortskip}{4pt}
\setlength{\belowdisplayshortskip}{4pt}

\hrule height 0.8pt
\vspace{3pt}
\noindent\textbf{Algorithm 2:} Penalized estimation of log loss\par
\vspace{2pt}
\hrule height 0.4pt
\vspace{4pt}
\noindent\textbf{Step 1 (Ridge-penalized fit):} Compute the penalized fit $\widehat u$ from \eqref{eq:practical-fit}, and obtain $\eta_i=z_i^\top\widehat u$, $q_i=s(\eta_i)-Y_i$, and $w_i=s'(\eta_i)$ for $i=1,\ldots,n$.\\[3pt]
\textbf{Step 2 (Score adjustment and point estimation):} With $P_n(a_i)=n^{-1}\sum_{i=1}^n a_i$, form $M=P_n(w_iz_iz_i^\top)+(1-1/n)\Lambda_n$ and $r=P_n(q_iz_i)+\Lambda_n\widehat u$. When $M$ is invertible, compute $h_i=n^{-1}z_i^\top M^{-1}z_i$ and $c_i=n^{-1}z_i^\top M^{-1}\Lambda_n\widehat u$ for $i=1,\ldots,n$. If $1-w_ih_i\ne0$ for every $i$, compute the adjusted scores $\widetilde\eta_i$ and the log-loss estimator $\widehat C_{\rm ALO}$:
\[
 \widetilde\eta_i=\eta_i+\frac{q_ih_i+c_i-z_i^\top M^{-1}r}{1-w_ih_i},\qquad
 \widehat C_{\rm ALO}=\frac1n\sum_{i=1}^n\ell_{Y_i}(\widetilde\eta_i).
\]
\vspace{2pt}
\hrule height 0.8pt
\end{minipage}\par
\medskip

In $\widetilde\eta_i$, the term $q_ih_i$ comes from the removed $i$th loss gradient, and we obtain $M$ and $c_i$ when we divide by $n-1$ at fixed $\Lambda_n$. From the rank-one Hessian update, we also obtain the denominator $1-w_ih_i$. When $\Lambda_n=0$ and $r=0$, $\widetilde\eta_i$ reduces to $\eta_i+\{h_i/(1-w_ih_i)\}q_i$, where $h_i/(1-w_ih_i)$ replaces the common scalar $\widehat\lambda$ in Algorithm~1, and we evaluate each performance functional on $(Y_i,\widetilde\eta_i)$ for unpenalized ALO estimation. When $\widehat u$ is an exact fit of~\eqref{eq:practical-fit}, we have $r=0$, and for an inexact fit we retain $-z_i^\top M^{-1}r$ to remove the full-sample residual in the Newton step. We show in Appendix~\ref{S-moderate} that Theorem~\ref{thm:moderate} still holds if $\|r\|=o_P(n^{-1/2})$. We compute all $n$ scores $\widetilde\eta_i$ from one factorization of $M$ without refitting and use $\widehat u$ for prediction.

We next estimate the variance of $\sqrt n\{\widehat C_{\rm ALO}-R(\widehat u)\}$. Note that in equation~\eqref{S-moderate-risk}, expanding $P_n\ell_{Y_i}(\eta_i)-R(\widehat u)$ around the penalized population minimizer gives a quadratic term in the centered empirical score whose mean is the leading bias removed in $\widehat C_{\rm ALO}$. Distinct pairs of observations contribute nonnegative variance to this term through the population Hessian and score covariance in~\eqref{S-moderate-risk}--\eqref{S-moderate-expansion}. Specifically, here $M+\Lambda_n/n$ is the Hessian in~\eqref{eq:practical-fit} at $\widehat u$, and $\widehat J_c$ denotes the centered empirical covariance matrix of $q_iz_i$ with divisor $n$. If we multiply this distinct-pair term by $n$, replace $1-1/n$ by $1$, and substitute $M+\Lambda_n/n$ and $\widehat J_c$, we obtain a nonnegative trace correction that we add to the empirical variance of the adjusted losses, and we define $\widehat V_{\rm ALO}$ by
\[
 \widehat V_{\rm ALO}=P_n\{\ell_{Y_i}(\widetilde\eta_i)-\widehat C_{\rm ALO}\}^2+\frac2n\operatorname{tr}\!\left[\{(M+\Lambda_n/n)^{-1}\widehat J_c\}^2\right].
\]
For $0<\alpha<1$, we form the $(1-\alpha)$ confidence interval
\[
 I_{\rm ALO}=\left[\widehat C_{\rm ALO}\pm z_{1-\alpha/2}\sqrt{\widehat V_{\rm ALO}/n}\right]\cap[0,\infty).
\]
Note that given $\widehat u$, computing $\widehat C_{\rm ALO}$ and $I_{\rm ALO}$ requires $O(nd^2+d^3)$ operations, and in practice we use $\Lambda_n=\operatorname{diag}(0,\rho I_{d-1})$ with fixed $\rho>0$ and an unpenalized intercept.
\subsection{Inference for the full fit}\label{subsec:moderate}

\citet{commenges2015universal} derived asymptotic distributions of approximate cross-validation criteria in fixed dimensions. Beyond fixed dimensions, for general covariates without assuming a correctly specified logistic model, let $u_n^*$ minimize $R(u)+u^\top\Lambda_nu/2$. We also compare $\widehat C_{\rm ALO}$ with the penalized form of Takeuchi's information criterion (TIC), $\widehat C_{\rm TIC}=P_n\ell_{Y_i}(\eta_i)+n^{-1}\operatorname{tr}\{(M+\Lambda_n/n)^{-1}\widehat J_c\}$ \citep{commenges2015universal}. The following theorem establishes consistency of $\widehat V_{\rm ALO}$ and asymptotic normality of the studentized estimation error as $d$ grows with $n$.
\begin{theorem}\label{thm:moderate}
Suppose the observations are independent and identically distributed and there exist fixed positive constants $K,D,c,\sigma_0$ such that
\[
 \sup_{\|a\|=1}\E\exp\{(a^\top z)^2/K^2\}\le2,\qquad
 \|u_n^*\|\le D,\qquad
 \E\{s'(z^\top u_n^*)zz^\top\}+\Lambda_n\succeq cI,
\]
and $\sigma_n^2:=\operatorname{Var}\{\ell_Y(z^\top u_n^*)\}\ge\sigma_0^2$. If $d=o(n^{2/3})$, the minimizer in \eqref{eq:practical-fit} is finite and unique with probability tending to one, $\widehat V_{\rm ALO}/\sigma_n^2\pto1$, and
\begin{equation}\label{eq:moderate-clt}
 \frac{\sqrt n\{\widehat C_{\rm ALO}-R(\widehat u)\}}{\sqrt{\widehat V_{\rm ALO}}}\dto N(0,1).
\end{equation}
The same theorem holds for the TIC point estimator $\widehat C_{\rm TIC}$ studentized with $\widehat V_{\rm TIC}$, defined just after equation~\eqref{S-reader-variance}.
\end{theorem}

In Theorem~\ref{thm:moderate}, we obtain asymptotically nominal coverage of $I_{\rm ALO}$ across training samples for the realized log loss of each full fitted model. There, we bound the loss moments from the exponential moment and minimizer norm conditions, bound $\|\widehat u-u_n^*\|$ from positive curvature, and ensure positive loss variance from $\sigma_n^2\ge\sigma_0^2$.

To examine the sufficient condition $d=o(n^{2/3})$ in~\eqref{eq:moderate-clt}, we expand $P_n\ell_{Y_i}(\eta_i)-R(\widehat u)$ at $u_n^*$ in Appendix~\ref{S-moderate}. Because $\widehat C_{\rm ALO}-P_n\ell_{Y_i}(\eta_i)=n^{-1}\operatorname{tr}(M^{-1}\widehat J_c)+O_P(d^2/n^2)$, we offset the mean of the leading quadratic error in $\widehat C_{\rm ALO}$, and from equation~\eqref{S-moderate-expansion} we obtain
\[
 \widehat C_{\rm ALO}-R(\widehat u)=P_n\ell_{Y_i}(z_i^\top u_n^*)-R(u_n^*)+O_P\!\left\{\frac{\sqrt d}{n}+\left(\frac{d}{n}\right)^{3/2}\right\},
\]
where $P_n\ell_{Y_i}(z_i^\top u_n^*)-R(u_n^*)$ is the centered empirical loss at $u_n^*$. Here, $\sqrt d/n$ and $(d/n)^{3/2}$ come from the centered quadratic error and the third-order remainder in the parameter expansion. When we multiply this remainder by $\sqrt n$ in~\eqref{eq:moderate-clt}, we obtain $O_P(\sqrt{d/n}+d^{3/2}/n)=o_P(1)$ under $d=o(n^{2/3})$.

Recall from Theorem~\ref{thm:optboot-onefitci} that under Gaussian predictors with $p/n$ converging to a positive constant, shared estimation error contributes a nonzero first-order rank-one covariance term with strictly positive diagonal entries for both losses. When $d=o(n^{2/3})$ in Theorem~\ref{thm:moderate} for the ridge fit, we instead have $d/n\to0$ and the trace term in $\widehat{V}_{\rm ALO}$ has order $O_P(d/n)=o_P(1)$, a nonnegative finite-sample correction of smaller than first order under these assumptions.

When $\|\phi(X)\|$ is uniformly bounded and the slope penalty is bounded below by a positive constant, the unpenalized empirical intercept is finite whenever both classes occur, and for $\Prob(Y=1)$ in a compact subset of $(0,1)$, the quantities $h_i$, $c_i$, and $\|\widehat{u}-u_n^*\|^2$ are $O_P(n^{-1})$ with constants independent of $d$. Note that at prevalence $1/2$ without predictor effects, we have $u_n^*=0$, $\ell_Y(z^\top u_n^*)=\log 2$, and $\sigma_n^2=0$. On the other hand, $\sigma_n^2$ is uniformly bounded away from $0$ outside a fixed open interval around $1/2$. At $1/2$, we apply Theorem~\ref{thm:moderate} if $\sigma_n^2\ge\sigma_0^2$ and its other conditions hold. The following corollary establishes uniform asymptotic coverage of $I_{\rm ALO}$ for arbitrary feature dimension $d$.

\begin{corollary}\label{cor:bounded-ridge}
For independent pairs $(X_i,Y_i)$ from a common distribution, let $d=d_n$ be any sequence of positive integers, and fix $L<\infty$, $\rho_0>0$, and $0<\pi_0<1/2$, $\pi_1>0$. For every $n$, suppose $\|\phi(X)\|\le L$, $\Lambda_n=\operatorname{diag}(0,\Lambda_s)$ with $\Lambda_s\succeq\rho_0I$, and $\pi:=\Prob(Y=1)\in[\pi_0,1-\pi_0]$ with $|\pi-1/2|\ge\pi_1$. On $0<\sum_{i=1}^nY_i<n$, the full fit is finite and unique, and for each fixed $0<\alpha<1$, uniformly over these joint distributions of $(X_i,Y_i)$, $\Prob\{R(\widehat u)\in I_{\rm ALO}\}\to1-\alpha$.
\end{corollary}

Under Corollary~\ref{cor:bounded-ridge}, by algorithmic stability, we have an $O_P(n^{-1})$ difference between the average conditional risk of leave-one-out fits and the realized risk of the full fit at the same penalty. Because the normalized one-step approximation has $o_P(n^{-1/2})$ error, $I_{\rm ALO}$ from the single full fit is first-order equivalent to leave-one-out intervals \citep{bayle2020confidence}.

\begin{remark}[Penalty selection]\label{rem:penalty-selection}
When selecting among a fixed number $m$ of candidate penalties, we minimize $\widehat C_{\rm ALO}$ over the $m$ fits and replace $z_{1-\alpha/2}$ with the Bonferroni quantile $z_{1-\alpha/(2m)}$. Under Theorem~\ref{thm:moderate} or Corollary~\ref{cor:bounded-ridge} for each candidate, by a union bound, we obtain asymptotic coverage at least $1-\alpha$ for the realized risk of the selected fit (Section~\ref{S-bounded-ridge}).
\end{remark}

\section{Simulations}\label{sec:numerical}

\subsection{Bias of validation estimates}\label{subsec:sim-gaussian}

In this section, we study point estimation bias under model~\eqref{eq:optboot-logistic} with $X_i\sim N_p(0,I_p)$ and $\beta_n=\gamma e_1$, where $e_1$ is the first standard basis vector. Specifically, when $n=1000$, we consider $\theta=0$, $\gamma\in\{1,0.3\}$, and $\kappa\in\{0.05,0.1,0.2\}$, and $\theta=-2$, $\gamma=1$, and $\kappa\in\{0.05,0.1\}$. On the other hand, when $n=2000$, we consider $\theta=0$, $\gamma\in\{1,0.3\}$, and $\kappa=0.1$. We compare $\widehat F$ in Algorithm~1 (One-fit) with Harrell's estimator (100 bootstrap resamples), the .632 estimator, and pooled ten-fold cross-validation (CV). In Table~\ref{tab:sim-logistic}, we define each AUC error as the estimate minus $A_{\rm ext}$ over 200 replications, and in Table~\ref{S-tab-sim-brier} we report the corresponding results for Brier loss.

\begin{figure}[htbp]
\centering
\includegraphics[width=\linewidth]{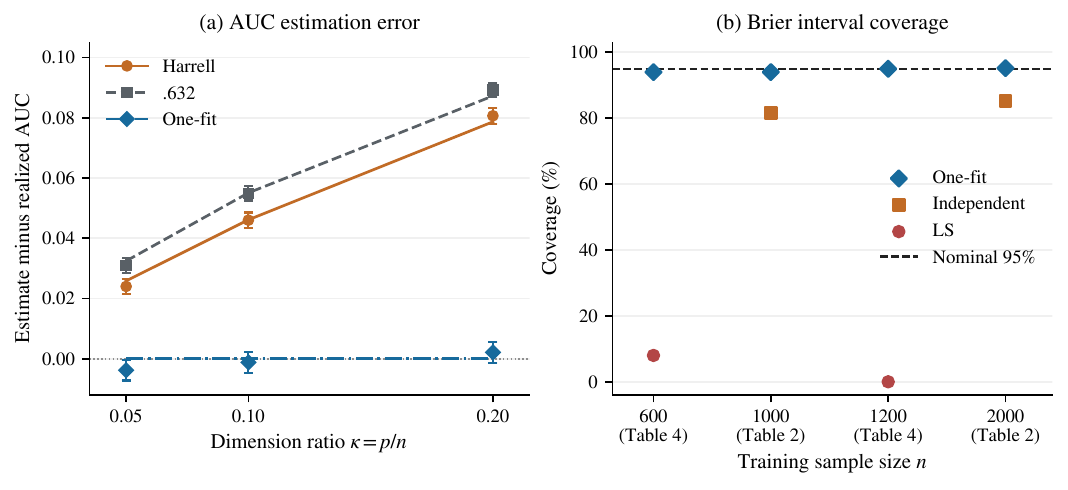}
\caption{Results at $\theta=0$, $\gamma=0.3$. (a) $200$-replication mean AUC error $\pm2$ Monte Carlo standard errors ($n=1000$, bars) and limits for Harrell (solid, circles), .632 (dashed, squares), and One-fit (dash-dot, diamonds), with zero dotted. (b) Brier coverage (Tables~\ref{tab:cov-onefit}, \ref{tab:noma-intervals}) for One-fit (diamonds), Independent (squares), and location-shifted bootstrap, LS (circles), with $95\%$ dashed.}
\label{fig:gaussian-validation}
\end{figure}

When we increase $n$ from $1000$ to $2000$ at fixed $\kappa$ in Table~\ref{tab:sim-logistic}, Harrell's mean AUC error remains nearly unchanged around the asymptotic fixed-ratio limit of Theorem~\ref{thm:optboot-Harrell}. At $(\theta,\gamma,\kappa)=(0,0.3,0.1)$, Harrell's empirical mean signed AUC error rounds to $+0.046$ for both $n$ and equals the rounded limit $+0.046$, and $\widehat A$ has mean error $-0.001$ at both sample sizes. In Fig.~\ref{fig:gaussian-validation}(a), at $(0,0.3,0.2)$ and $n=1000$, Harrell's mean error is $+0.081$ with theoretical limit $+0.079$, and $\widehat A$ has mean error $+0.002$. Across all ten settings in Table~\ref{tab:sim-logistic}, $\widehat A$ has mean signed AUC error between $-0.004$ and $+0.002$.

\begin{table}[htbp]
{\centering
\setlength{\arrayrulewidth}{0.3pt}
\setlength{\tabcolsep}{3pt}
\caption{Mean realized AUC and mean AUC errors, with theoretical limiting errors in brackets.}
\label{tab:sim-logistic}
\begin{tabular}{@{}rrrrrrrrr@{}}
\toprule
$\theta$ & $\gamma$ & $\kappa$ & $n$ & External & Harrell & .632 & 10-fold CV & One-fit \\
\midrule
$0$ & $1$ & $0.05$ & $1000$ & 0.713 & $+$0.006 [$+$0.006] & $+$0.008 [$+$0.008] & $-$0.006 [$-$0.003] & $-$0.001 \\
$0$ & $1$ & $0.1$ & $1000$ & 0.691 & $+$0.019 [$+$0.019] & $+$0.021 [$+$0.021] & $-$0.005 [$-$0.004] & $-$0.001 \\
$0$ & $1$ & $0.2$ & $1000$ & 0.657 & $+$0.047 [$+$0.049] & $+$0.045 [$+$0.047] & $-$0.011 [$-$0.006] & $-$0.003 \\
$0$ & $0.3$ & $0.05$ & $1000$ & 0.546 & $+$0.024 [$+$0.026] & $+$0.031 [$+$0.032] & $-$0.006 [$-$0.002] & $-$0.004 \\
$0$ & $0.3$ & $0.1$ & $1000$ & 0.534 & $+$0.046 [$+$0.046] & $+$0.055 [$+$0.055] & $-$0.004 [$-$0.002] & $-$0.001 \\
$0$ & $0.3$ & $0.2$ & $1000$ & 0.524 & $+$0.081 [$+$0.079] & $+$0.089 [$+$0.087] & 0.000 [$-$0.001] & $+$0.002 \\
$-2$ & $1$ & $0.05$ & $1000$ & 0.705 & $+$0.016 [$+$0.016] & $+$0.016 [$+$0.016] & $-$0.009 [$-$0.004] & $-$0.002 \\
$-2$ & $1$ & $0.1$ & $1000$ & 0.674 & $+$0.042 [$+$0.043] & $+$0.036 [$+$0.038] & $-$0.010 [$-$0.006] & $-$0.003 \\
$0$ & $1$ & $0.1$ & $2000$ & 0.691 & $+$0.019 [$+$0.019] & $+$0.021 [$+$0.021] & $-$0.005 [$-$0.004] & 0.000 \\
$0$ & $0.3$ & $0.1$ & $2000$ & 0.533 & $+$0.046 [$+$0.046] & $+$0.055 [$+$0.055] & $-$0.003 [$-$0.002] & $-$0.001 \\
\bottomrule
\end{tabular}\par}
\begin{tabnote}
CV, cross-validation. We use separate sets of 200 replications for $n=1000$ and $n=2000$. For fixed $\kappa$, the theoretical limiting errors do not depend on $n$.
\end{tabnote}
\end{table}

\subsection{Marginal and joint interval coverage}\label{subsec:interval-comparisons}

In Table~\ref{tab:cov-onefit}, we evaluate confidence interval coverage for $(B_{\rm ext},R_{\rm ext})$ over $500$ training samples per setting. We compare intervals based on $\widehat\Sigma_{FF}$ (One-fit) with naive intervals that omit its rank-one term (Independent). Both intervals are centered at $\widehat B$ and $\widehat R$, and One-fit has $94.0$--$96.6\%$ Brier coverage and $93.0$--$96.2\%$ log-loss coverage. At $(n,\theta,\gamma,\kappa)=(1000,0,1,0.351)$, log-loss coverage is $96.0\%$ for One-fit versus $75.0\%$ for Independent (equation~\eqref{eq:independent-coverage}). More generally, coverage for Independent decreases to $75.0$--$84.0\%$ at $\kappa=0.351$ and $81.6$--$85.6\%$ at $\gamma=0.3$ when the rank-one term is omitted.

\begin{table}[htbp]
{\centering
\setlength{\arrayrulewidth}{0.3pt}
\caption{Coverage (\%) of nominal $95\%$ intervals for Brier and log losses with and without the variance correction.}
\label{tab:cov-onefit}
\setlength{\tabcolsep}{4pt}
\begin{tabular}{@{}rrrrrrrr@{}}
\toprule
 & & & & \multicolumn{2}{c}{Brier} & \multicolumn{2}{c}{Log loss}\\
\cmidrule(lr){5-6}\cmidrule(lr){7-8}
$n$ & $\theta$ & $\gamma$ & $\kappa$ & One-fit & Independent & One-fit & Independent \\
\midrule
1000 & $0$ & $1$ & $0.05$ & 96.2 & 92.2 & 96.2 & 92.0 \\
1000 & $0$ & $1$ & $0.1$ & 94.8 & 88.4 & 95.4 & 87.6 \\
1000 & $-2$ & $1$ & $0.1$ & 94.8 & 91.8 & 93.0 & 89.0 \\
1000 & $0$ & $0.3$ & $0.1$ & 94.0 & 81.6 & 93.6 & 81.8 \\
1000 & $0$ & $1$ & $0.351$ & 96.6 & 83.6 & 96.0 & 75.0 \\
2000 & $0$ & $1$ & $0.05$ & 95.8 & 93.0 & 95.6 & 92.4 \\
2000 & $0$ & $1$ & $0.1$ & 94.2 & 88.4 & 94.6 & 87.6 \\
2000 & $-2$ & $1$ & $0.1$ & 94.0 & 91.6 & 94.0 & 89.6 \\
2000 & $0$ & $0.3$ & $0.1$ & 95.2 & 85.2 & 95.6 & 85.6 \\
2000 & $0$ & $1$ & $0.351$ & 94.2 & 84.0 & 94.6 & 77.0 \\
\bottomrule
\end{tabular}\par}
\begin{tabnote}
At $(0,1,0.1)$ for $n=1000$ and $2000$, we compute the Independent coverages from separate sets of 500 samples.
\end{tabnote}
\end{table}

When Brier and log losses are evaluated on the same patients, the scaled errors $\sqrt n(\widehat B-B_{\rm ext})$ and $\sqrt n(\widehat R-R_{\rm ext})$ have strong positive covariance. Table~\ref{tab:joint-covariance} reports results for the $500$ training samples per setting from four selected $\theta=0$ rows of Table~\ref{tab:cov-onefit}, where we compute $(B_{\rm ext},R_{\rm ext})$ by Gaussian quadrature. In all four settings, the mean of $\widehat\Sigma_{BR}$ is close to the empirical covariance. At $(n,\gamma,\kappa)=(1000,1,0.351)$, this mean is $1.6718$ and the empirical covariance is $1.6928$. In this setting, $95\%$ ellipse coverage is $94.2\%$ for One-fit and $77.4\%$ for Independent. Joint coverage is $94.2$--$95.6\%$ for One-fit and $76.6$--$88.2\%$ for Independent across all four settings.

\begin{table}[htbp]
\centering
\setlength{\arrayrulewidth}{0.3pt}
\caption{Joint inference for Brier and log loss.}\label{tab:joint-covariance}
\begin{tabular}{@{}rrrrrrr@{}}
\toprule
 & & & \multicolumn{2}{c}{Covariance} & \multicolumn{2}{c}{Joint coverage (\%)}\\
\cmidrule(lr){4-5}\cmidrule(lr){6-7}
$n$ & $\gamma$ & $\kappa$ & Empirical & Estimated & One-fit & Independent\\
\midrule
1000 & 0.3 & 0.1 & 0.1584 & 0.1478 & 94.4 & 84.4 \\
1000 & 1 & 0.351 & 1.6928 & 1.6718 & 94.2 & 77.4 \\
2000 & 0.3 & 0.1 & 0.1408 & 0.1476 & 95.6 & 88.2 \\
2000 & 1 & 0.351 & 1.7549 & 1.6554 & 95.4 & 76.6 \\
\bottomrule
\end{tabular}
\end{table}

In Table~\ref{tab:noma-intervals}, we run $100$ replications per design with $1{,}000$ resamples each at $\theta=0$, $\kappa=0.1$, $\gamma\in\{0.3,1.0\}$, and $n\in\{600,1200\}$. Fig.~\ref{fig:gaussian-validation}(b) shows the $\gamma=0.3$ rows. For all four settings, the midpoint of $I_B^{\rm LS}$ has more negative mean signed Brier error than $\widehat H_B$. At $\gamma=0.3$ and $n=1200$, these errors are $-0.029$ and $-0.005$, while $\widehat B$ has error $-0.002$ and One-fit has $95.0\%$ Brier coverage. As $n$ increases at fixed $\kappa$, empirical coverage of $I_F^{\rm LS}$ decreases or remains zero for AUC and decreases for Brier loss from $8.0\%$ to $0.0\%$ at $\gamma=0.3$ and $15.0\%$ to $3.0\%$ at $\gamma=1.0$. By Proposition~\ref{prop:optboot-ls}, the width of $I_F^{\rm LS}$ tends to zero and both shifted limits converge in probability to a common limit. If this limit differs from the limiting external performance of the original fit, coverage tends to zero.

\begin{table}[htbp]
{\centering
\setlength{\arrayrulewidth}{0.3pt}
\caption{Coverage (\%) and mean Brier errors of bootstrap and one-fit procedures.}\label{tab:noma-intervals}
\begin{tabular}{@{}rrrrrrrr@{}}
\toprule
$\gamma$ & $n$ & LS AUC & LS Brier & One-fit Brier & Harrell error & LS error & One-fit error\\
\midrule
0.3 & 600 & 0.0 & 8.0 & 94.0 & $-$0.003 & $-$0.028 & 0.000\\
0.3 & 1200 & 0.0 & 0.0 & 95.0 & $-$0.005 & $-$0.029 & $-$0.002\\
1.0 & 600 & 5.0 & 15.0 & 94.0 & $-$0.003 & $-$0.026 & $-$0.001\\
1.0 & 1200 & 1.0 & 3.0 & 94.0 & $-$0.003 & $-$0.026 & $-$0.001\\
\bottomrule
\end{tabular}\par}
\begin{tabnote}
LS, location-shifted bootstrap.
\end{tabnote}
\end{table}

As in Remark~\ref{subsec:loss-reduction}, we compare the full fit with the intercept-only fit based on sample prevalence on Brier and log losses. We run $500$ replications per setting for the four settings in Table~\ref{tab:noma-intervals} and $(\theta,\gamma,n)=(-2,1,1000)$ at $\kappa=0.1$. Across all five settings and both losses, we obtain $94.6$--$95.6\%$ empirical coverage for the realized loss reduction. We also test improvement over sample prevalence. For each loss and setting, at most $2.8\%$ of the $500$ training samples have both a positive one-sided lower confidence bound and a nonpositive true realized loss reduction.

In Appendix~\ref{S-bio-additional}, we examine the finite-sample sensitivity of Algorithm~1 under correlated Gaussian, binary, mixed, and heavy-tailed $t_5$ designs with $n=1000$ and 500 replications per setting. We also run 1,000 replications per setting at $n\in\{1000,2000\}$ for four non-Gaussian designs that satisfy the conditions of Theorem~\ref{thm:optboot-nongaussian}. Across all settings, nominal $95\%$ Brier and log-loss coverage is $93.3$--$97.0\%$, with minimum coverage $83.7\%$ without covariance correction.

\section{Real data analysis}\label{sec:support}

\subsection{Datasets and spline comparison}\label{subsec:clinical-design}

In this section, we apply our methods to the Study to Understand Prognoses and Preferences for Outcomes and Risks of Treatments (SUPPORT) \citep{knaus1995support,support2data} and the Pima diabetes dataset in \texttt{MASS} \citep{smith1988adap,pimadata}. In SUPPORT, there are 9,105 hospitalized adults with 2,360 in-hospital deaths and measurements mostly from day~3, and we use all 9,105 in spline comparisons. For the repeated-sampling experiments in Section~\ref{subsec:practical-comparators}, our discrete SUPPORT target population has the 8,758 patients and 2,222 deaths with hospitalization and follow-up beyond day~3, where we use 15 measurements and three disease indicators without both duration variables. On the other hand, for the Pima dataset of 532 complete records with 177 cases, we exclude glucose because it is in the diagnostic definition and use the other six predictors. In Algorithm~2, we keep binary indicators at zero or one and impute missing values with data-independent centers $c_j$. For each numerical predictor $x_j$, we use a data-independent scale $s_j>0$ and define $\phi_j(x)=3\tanh[\operatorname{asinh}\{(x_j-c_j)/s_j\}/3]$ as in Section~\ref{subsec:ridge-numerics}. In addition, in Section~\ref{subsec:practical-comparators}, we use $\rho=0.05$ for the fixed penalty and also consider data-dependent grid selection.

We first evaluate unpenalized logistic fits on 200 random splits of SUPPORT at $n\in\{1700,680,340,170\}$ with $p=34$ ($\kappa=p/n\in\{0.02,0.05,0.1,0.2\}$). The 34 predictors include restricted cubic splines with four knots for eight measurements, seven linear terms, and three disease indicators. On each split, we estimate training knots and preprocessing and fix them for the $9105-n$ held-out patients and 200 bootstrap resamples. We define AUC error as estimated minus held-out AUC. At $n=170$ ($\kappa=0.2$), 18/200 original fits separate and 17,673/36,400 ($49\%$) bootstrap refits fail across 182 finite original fits. All 200 refits fail for one finite original fit, and we compute Harrell's estimator from successful refits within each split on 181 paired splits. In Table~\ref{tab:support-paired}, Harrell's mean AUC error rises from $0.001$ at $\kappa=0.02$ ($n=1700$) to $0.076$ at $\kappa=0.2$. On these 181 splits, mean AUC error at $\kappa=0.2$ is $0.012$ for One-fit ($\widehat A$) and $-0.006$ for ALO. For the three larger dimension ratios, both estimates have smaller absolute mean errors than Harrell's correction.

\begin{table}[htbp]
{\centering
\setlength{\arrayrulewidth}{0.3pt}
\caption{Mean AUC errors in SUPPORT, with Monte Carlo standard errors in brackets.}
\label{tab:support-paired}
\begin{tabular}{@{}rrrrr@{}}
\toprule
$\kappa$ & Splits & Harrell & ALO & One-fit\\
\midrule
0.02 & 200 & $+$0.001 [0.001] & $-$0.002 [0.001] & $-$0.002 [0.001] \\
0.05 & 200 & $+$0.007 [0.002] & $-$0.002 [0.002] & $-$0.001 [0.002] \\
0.1 & 200 & $+$0.017 [0.003] & $-$0.009 [0.003] & $-$0.005 [0.003] \\
0.2 & 181 & $+$0.076 [0.005] & $-$0.006 [0.005] & $+$0.012 [0.005] \\
\bottomrule
\end{tabular}\par}
\begin{tabnote}
ALO, approximate leave-one-out.
\end{tabnote}
\end{table}

\begin{figure}[htbp]
\centering
\includegraphics[width=\linewidth]{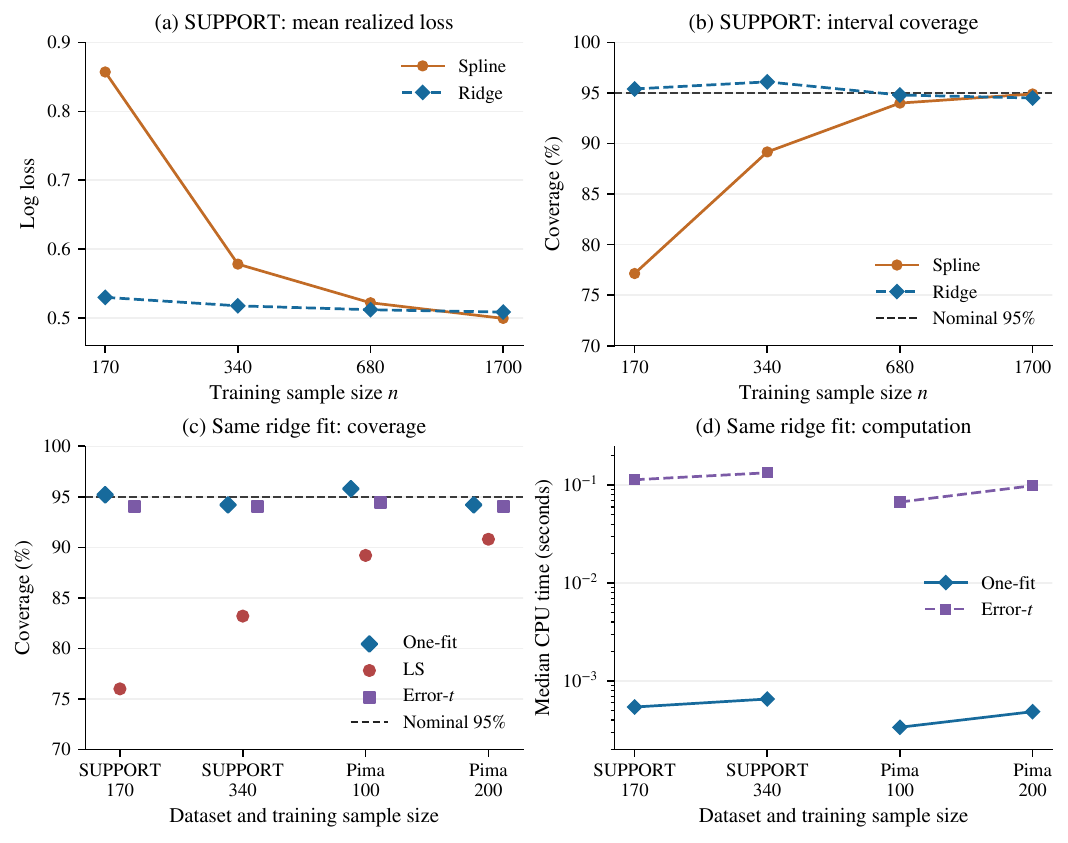}
\caption{Panels (a)--(b) show SUPPORT average realized log loss and coverage for finite spline fits and all ridge fits. Panel (c) shows One-fit, LS, and Error-$t$ coverage with horizontal dashed $95\%$ lines in (b)--(c), and (d) compares only One-fit and Error-$t$ median CPU time.}
\label{fig:clinical-validation}
\end{figure}

We next generate binary responses from the 34-predictor spline model fitted to SUPPORT once we scale its linear predictor to unit variance at fixed prevalence. On each of 1,000 samples per size, we estimate preprocessing and fit this unpenalized model to evaluate its log-loss interval (Section~\ref{subsec:method-scalar}) on finite fits. Under misspecification, we fit ridge logistic regression with 18 bounded features and form $I_{\rm ALO}$ (Section~\ref{sec:general-inference}). At $n=170$, only 958 spline fits are finite (mean risk $0.857$), and coverage is $77.1\%$ with median width $0.445$. All 1,000 ridge fits are finite (mean risk $0.530$), and $I_{\rm ALO}$ has $95.4\%$ coverage and median width $0.182$. For $n=340$, finite-fit coverage is $89.2\%$ for spline versus $96.1\%$ for ridge. When $n=1700$, mean risk is $0.500$ for spline versus $0.509$ for ridge. On non-Gaussian covariates, bounded-feature ridge coverage is near nominal at $n=170$ and $340$, and the mean risks at $n=170$ are for finite spline fits and all ridge fits (Fig.~\ref{fig:clinical-validation}(a)--(b)).

\subsection{Ridge inference}\label{subsec:practical-comparators}

We evaluate $I_{\rm ALO}$ at $n\in\{170,340,680,1700\}$ on SUPPORT and $n\in\{100,200,400\}$ on Pima by treating each full eligible dataset as a discrete target population. Specifically, for each $n$, we draw 1,000 independent samples with replacement and compute the exact realized target $R(\widehat u)$ on this population. We also fit an intercept-only model on the training sample and define the log-loss gain as its realized mean log loss minus $R(\widehat u)$. Note that all 7,000 intervals $I_{\rm ALO}$ are finite, and the fitted models have positive mean log-loss gains. In Table~\ref{tab:practical-calibration}, we obtain $94.8$--$96.1\%$ coverage for $I_{\rm ALO}$ across the seven settings, and coverage decreases to $91.7$--$95.5\%$ without the trace correction. On the other hand, if we replace $\widehat C_{\rm ALO}$ by the training mean log loss of the same fit at fixed standard error and realized target, coverage decreases to $76.1$--$93.4\%$. As in Section~\ref{subsec:moderate}, we center $I_{\rm ALO}$ at $\widehat C_{\rm ALO}$ to remove the leading optimism bias, and the $O_P(d/n)$ trace term mainly improves coverage at $n=170$ and $n=100$.

\begin{table}[htbp]
\centering
\setlength{\arrayrulewidth}{0.3pt}
\caption{Log-loss intervals for ridge fits with bounded features.}\label{tab:practical-calibration}
\begin{tabular}{@{}lrrrrr@{}}
\toprule
Dataset & $n$ & Coverage (\%) & Median width & Mean risk & Mean gain\\
\midrule
SUPPORT & 170 & 94.9 & 0.182 & 0.509 & 0.060 \\
SUPPORT & 340 & 94.8 & 0.115 & 0.497 & 0.071 \\
SUPPORT & 680 & 96.1 & 0.077 & 0.491 & 0.076 \\
SUPPORT & 1700 & 95.6 & 0.046 & 0.488 & 0.078 \\
Pima & 100 & 95.4 & 0.190 & 0.540 & 0.101 \\
Pima & 200 & 95.2 & 0.123 & 0.533 & 0.106 \\
Pima & 400 & 95.3 & 0.082 & 0.529 & 0.108 \\
\bottomrule
\end{tabular}
\end{table}

Regarding bootstrap procedures, we compare $I_{\rm ALO}$ (One-fit) with LS, basic, and studentized (Error-$t$) intervals for this Algorithm~2 fit at fixed $\phi$ and $\rho=0.05$. Specifically, with 200 resamples across 500 samples at $n\in\{170,340\}$ on SUPPORT and $n\in\{100,200\}$ on Pima, we form LS via Section~\ref{subsec:optboot-procedures}. Note that on each resample, the refit error is its ALO estimate minus its mean log loss on the original data. In Error-$t$, we divide it by its ALO standard error and scale by the original one. We subtract upper and lower raw or rescaled quantiles from $\widehat C_{\rm ALO}$ and clip both lower limits at zero. In Table~\ref{tab:practical-comparators} and Fig.~\ref{fig:clinical-validation}(c)--(d), $I_{\rm ALO}$ and Error-$t$ have similar coverage and median width. On the other hand, coverage is lower for LS and $90.4$--$93.2\%$ for basic. We compute the ridge fit and $I_{\rm ALO}$ in a median of $0.00034$--$0.00066$ seconds, about 200-fold faster than $0.067$--$0.134$ seconds for 200 refits and ALO calculations. In Section~\ref{S-practical-cv}, we compare $I_{\rm ALO}$ with naive ten-fold cross-validation for the realized log loss of this full fit over 500 Pima samples per size. At $n=100$ and $200$, naive ten-fold cross-validation has coverage $93.0\%$ and $93.8\%$ versus $95.8\%$ and $94.2\%$ for $I_{\rm ALO}$.

\begin{table}[htbp]
{\centering
\setlength{\arrayrulewidth}{0.3pt}
\caption{All three log-loss intervals from the same fitted ridge model per sample across the same 500 samples per row.}\label{tab:practical-comparators}
\begin{tabular}{@{}lrrrrrrr@{}}
\toprule
 & & \multicolumn{3}{c}{Coverage (\%)} & \multicolumn{3}{c}{Median width}\\
\cmidrule(lr){3-5}\cmidrule(lr){6-8}
Dataset & $n$ & One-fit & LS & Error-$t$ & One-fit & LS & Error-$t$\\
\midrule
SUPPORT & 170 & 95.2 & 76.0 & 94.0 & 0.181 & 0.145 & 0.175 \\
SUPPORT & 340 & 94.2 & 83.2 & 94.0 & 0.116 & 0.104 & 0.111 \\
Pima & 100 & 95.8 & 89.2 & 94.4 & 0.191 & 0.172 & 0.184 \\
Pima & 200 & 94.2 & 90.8 & 94.0 & 0.122 & 0.120 & 0.117 \\
\bottomrule
\end{tabular}\par}
\begin{tabnote}
LS, location-shifted bootstrap.
\end{tabnote}
\end{table}

In full-cohort ridge fits at $\rho=0.05$ for future eligible patients from the same source populations, the estimated log losses and $95\%$ intervals are $0.4872$ $[0.4773,0.4972]$ for SUPPORT and $0.5334$ $[0.4984,0.5683]$ for Pima. For $\rho\in\{0.01,0.05,0.2\}$, we minimize $\widehat C_{\rm ALO}$ on both full eligible datasets to select $\rho=0.01$ and apply the Bonferroni adjustment in Remark~\ref{rem:penalty-selection}. This rule has $97.6$--$99.2\%$ resampling coverage across 1,000 SUPPORT and 500 Pima samples per size (see Section~\ref{app:penalty-selection}).

\section{Discussion}\label{sec:discussion}

In this article, we studied validation of a fitted logistic model without independent test data, where bootstrap weights change the effective dimension. We use Algorithm~1 without tuning or refits to correct optimism and first-order covariance in the existence region for the Gaussian and non-Gaussian-component models of Theorems~\ref{thm:optboot-onefit}--\ref{thm:optboot-nongaussian}. When separation, general covariates, or misspecification arises, we apply Algorithm~2 with bounded features and a positive ridge slope penalty for a unique finite fit if both classes occur. By Theorem~\ref{thm:moderate} ($d=o(n^{2/3})$) or Corollary~\ref{cor:bounded-ridge}, the nonnegative $o_P(1)$ trace in $\widehat{V}_{\rm ALO}$ is asymptotically negligible in the variance, and we form valid log-loss intervals with a fixed penalty or finite-grid Bonferroni selection.

A natural extension of Theorem~\ref{thm:optboot-nongaussian} is first-order inference at the $n^{-1/2}$ scale for a growing number of independent non-Gaussian residual coordinates with nonzero regression coefficients. It is also of interest to develop single-fit inference for regularized Cox regression \citep{massa2025observable} under censoring and shared risk sets.
\section*{Data availability}

The \texttt{support2} dataset is distributed by Vanderbilt University \citep{support2data}. The complete-record splits \texttt{Pima.tr} and \texttt{Pima.te} are available in the \texttt{MASS} package \citep{pimadata}. We provide replication code and saved numerical results on \href{https://doi.org/10.5281/zenodo.24356958}{Zenodo} \citep{parklee2026replication}.

\section*{Declaration of the use of generative AI and AI-assisted technologies}

During this work, the authors used ChatGPT (GPT-6 pro)  for English editing and
to cross-check the mathematical arguments. The authors take full responsibility for the content of the publication.

\section*{Acknowledgement}

Seyoung Park's work was supported by the National Research Foundation of Korea (NRF), funded by the Ministry of Science and ICT (MSIT) (RS-2025-00517793). Eun Ryung Lee's work was supported by the NRF, funded by the MSIT (2022R1A2C1012798 and RS-2025-02216235).

Seyoung Park and Eun Ryung Lee are co-corresponding authors. The authors have no conflicts of interest.

\clearpage
\appendix
\renewcommand{\thesection}{S\arabic{section}}
\renewcommand{\theHsection}{appendix.\arabic{section}}
\appendixserial{equation}
\appendixserial{theorem}
\appendixserial{lemma}
\appendixserial{corollary}
\appendixserial{proposition}
\appendixserial{remark}
\appendixserial{table}
\appendixserial{figure}
\phantomsection
\section*{Appendices}
\addcontentsline{toc}{part}{Appendices}

Sections~\ref{S-model} and~\ref{S-onefit} prove the bootstrap limits and bias in Theorems~\ref{thm:optboot-Harrell}--\ref{thm:optboot-signs} and Proposition~\ref{prop:optboot-ls}, and the Gaussian single-fit results in Theorems~\ref{thm:optboot-onefit}--\ref{thm:optboot-onefitci}, respectively. Section~\ref{S-NG-section} proves the two cases of Theorem~\ref{thm:optboot-nongaussian} and gives a scale mixture for which score adjustment is biased. Section~\ref{S-moderate} proves Theorem~\ref{thm:moderate} and Corollary~\ref{cor:bounded-ridge} for penalized fits. Section~\ref{S-bio-additional} gives additional simulation and clinical results.

\section{Limits and bias of bootstrap validation}\label{S-model}
\subsection{Gaussian model and separation}

For scalars, vectors, and matrices, $\|A\|$ denotes the absolute value, Euclidean norm, and operator norm, respectively, where we also use $\|A\|_{\rm op}$ for the operator norm and $\|A\|_F$ for the Frobenius norm. Here, we define $\|A\|_{\max}=\max_{jk}|A_{jk}|$, $\|v\|_\infty=\max_j|v_j|$, and $\|f\|_\infty=\sup_x|f(x)|$, and let $\ind\{E\}$ denote the indicator of $E$ and $Qf=\int f\,dQ$ for integrable $f$. Also, for $q\ge1$, we define $\|U\|_{L^q}=(\E\|U\|^q)^{1/q}$, where for deterministic $a_n>0$, $U_n=O_{L^q}(a_n)$ means that $\|U_n\|_{L^q}\le C_q a_n$ for all large $n$ with $C_q$ independent of $n$. For $O_P(a_n)$ and $o_P(a_n)$, the ratio $\|U_n\|/a_n$ is bounded in probability or tends to zero in probability, respectively, as in Section~\ref{subsec:optboot-logistic}.

Next, let $s(x)=(1+e^{-x})^{-1}$, $\operatorname{logit}(u)=\log\{u/(1-u)\}$ for $0<u<1$, and $\ell_y(x)=\log(1+e^x)-yx$, where $\phi$ and $\Phi$ denote the standard normal density and distribution function. Here, for each $n$, the independent rows satisfy
\begin{equation}\label{S-experiment}
 X_i\sim N_p(0,\Sigma_n),\qquad
 \Prob(Y_i=1\mid X_i)=s(\theta_n+X_i^\top\beta_n),\qquad
 \Sigma_n\succ0,
\end{equation}
where the parameters are deterministic. Here, we define $\gamma_n^2=\beta_n^\top\Sigma_n\beta_n$ and $\kappa_n=p/n$, and our basic asymptotic regime is
\begin{equation}\label{S-regime}
 (\theta_n,\gamma_n,\kappa_n)\longrightarrow(\theta,\gamma,\kappa),
 \qquad \gamma>0,\quad 0<\kappa<c_Bh(\theta,\gamma),
 \qquad c_B=1-e^{-1}.
\end{equation}
Here, the intercept is excluded from $p$, and we require $0<\kappa<h(\theta,\gamma)$ for the ordinary-fit results and use the continuous extension at $\gamma=0$ for the zero-signal calculations.

For all sufficiently large $n$, let $Z_i=X_i^\top\beta_n/\gamma_n$ denote the standardized signal coordinate, which satisfies $Z_i\sim N(0,1)$.

Given independent standard normal variables $Z,G$, we define $Y\mid Z\sim\operatorname{Bernoulli}\{s(\theta+\gamma Z)\}$, $D=2Y-1$, and
\begin{equation}\label{S-boundary}
 h(\theta,\gamma)=\inf_{a,c\in\R}\E(aD+cDZ-G)_+^2,
 \qquad \pi=\E s(\theta+\gamma Z).
\end{equation}
Next, let $x_+=\max(x,0)$ and $D_i=2Y_i-1$, and for independent $G_i\sim N(0,1)$ that are also independent of the data, we define
\[
 Q_n=\inf_{a,c}P_n\{D_i(a+cZ_i)-G_i\}_+^2,\qquad
 m_0^2=\inf_{a^2+c^2=1}\E\{D(a+cZ)\}_+^2>0 ,
\]
where $P_n$ denotes the empirical average, and because both labels have positive conditional probability, we obtain the last inequality by compactness. For $(\theta_n,\gamma_n)\to(\theta,\gamma)$, we represent each label by thresholding an independent uniform variable, and by finite meshes, Gaussian moments, and dominated convergence, we get $\inf_{a^2+c^2=1}P_n\{D_i(a+cZ_i)\}_+^2\pto m_0^2$. By the Lipschitz property of the positive part, the empirical $L^2$ norm of $\{D_i(a+cZ_i)-G_i\}_+$ is at least that of $\{D_i(a+cZ_i)\}_+$ minus $(P_nG_i^2)^{1/2}$. From homogeneity and comparison at $(0,0)$, we can restrict all minimizers to a fixed ball with probability tending to one, and we apply the same mesh argument on this ball to obtain $Q_n\pto h(\theta,\gamma)$. Because $0\le Q_n\le P_n(-G_i)_+^2$ has bounded second moment, we have $\E|Q_n-h|\to0$ and $\E_G Q_n\pto h$, where $\E_G$ integrates out the $G_i$.

For $\mathcal C_n=\operatorname{span}\{(D_i)_i,(D_iZ_i)_i\}+\R_+^n$, by Lemma~3 in the arXiv v1 preprint of \cite{candes2020phase}, we have the statistical dimension $\delta(\mathcal C_n)/n=1-\E_GQ_n$. By considering both signs, we use the empirical lower bound to exclude any nonzero nonnegative vector in the two-dimensional span. Then, from Equation~(16) of the same preprint, we obtain separation with probability tending to one for $h<\kappa<1$. For $\kappa\ge1$, we retain the signal coordinate and enough nuisance coordinates to get a ratio in $(h,1)$ and then extend the separating vector by zeros.

Given the distinct bootstrap indices, their observations are independent of one another, and for every deterministic count $m_n/n\to c_B$, their dimension ratio tends to $\kappa/c_B$. Thus, the separation conclusion is uniform over $|m_n/n-c_B|\le\epsilon_n$ for any $\epsilon_n\downarrow0$. From Step~1 of the proof of Theorem~\ref{S-state}, the random count $M_n$ satisfies $M_n/n\pto c_B$. Above $c_Bh$, we average to get separation and apply Markov's inequality to obtain convergence in probability of the conditional bootstrap separation probability to one. Note that the prevalence satisfies $0<\pi<1$, and boundary equality is excluded. Below the thresholds, we obtain existence from Theorems~\ref{F-thm-exist} and~\ref{S-state}.

We apply a whitening transformation and an orthogonal rotation to obtain the analytic coordinates $(Z_i,A_i)$, where $A_i\sim N_{p-1}(0,I)$ is independent of $(Z_i,Y_i)$. In empirical design-matrix expressions, we let $Z=(Z_1,\ldots,Z_n)^\top\in\R^n$ denote the stacked signal vector and $A\in\R^{n\times(p-1)}$ denote the nuisance design matrix with $i$th row $A_i^\top$. Then, a fitted linear predictor has the form
\begin{equation}\label{S-coordinates}
 b+tZ_i+A_i^\top w,\qquad r=\norm w.
\end{equation}
Let $(\widehat b,\widehat t,\widehat w)$ denote the fitted analytic coefficients and define $\widehat r=\norm{\widehat w}$. Given a finite fit, an independent test score is $\widehat b+\widehat tZ+\widehat rG$, where $Z\sim N(0,1)$ and $G\sim N(0,1)$ is independent of $(Z,Y)$.

Let $\mathcal D$ denote the training data, let $\Prob_{\mathcal D}$ denote probability under the sampling distribution of $\mathcal D$, and let $\Prob_*,\E_*$ denote the conditional bootstrap probability and expectation given $\mathcal D$. Conditional convergence in probability of $H_n^*$ to $h_*$ means
\begin{equation}\label{S-conditional}
 \Prob_{\mathcal D}\{\Prob_*(|H_n^*-h_*|>\epsilon\mid\mathcal D)>\delta\}
 \longrightarrow0\qquad(\epsilon,\delta>0).
\end{equation}
In ordinary and bootstrap fits, we assign zero coefficients to undefined fits and fixed finite values to undefined statistics, with values in $[0,1]$ for the area under the receiver operating characteristic curve (AUC) and Brier loss and in $[0,\infty)$ for log loss. When an original fit or interval calculation fails, we assign the full target range, and in intercept-only comparisons we define $\widehat\pi=1/2$ when a response class is absent. Also, we record all failures and make no improvement declaration after fitting or evaluation failure or class absence. Within each existence region, these conventions apply with probability tending to zero for fixed $B_n$, and as $B_n$ grows, the failure proportion tends to zero and hence the AUC and Brier limits continue to hold.

Whenever any refit fails, the ridge bootstrap intervals are unavailable and we report the full target range for the two-stage bootstrap intervals.

\subsection{Existence and curvature of the unpenalized fit}\label{S-ordinary}
We analyze the maximum likelihood estimator (MLE) in the ordinary Gaussian logistic experiment
with $(p/n,\theta_n,\gamma_n)$ in a fixed compact subset of
$\{(\kappa,\theta,\gamma):\gamma>0,\ 0<\kappa<h(\theta,\gamma)\}$.
Let $W$ have rows $W_i^\top=A_i^\top$, let $\ind$ denote the vector of $n$ ones, let $B=[\ind,Z,W]$ and $D_i=2Y_i-1$, and define
$L_n(b,t,w)=n^{-1}\sum_i\ell_{Y_i}(b+tZ_i+W_i^\top w)$.
Let $\E_n$ denote the empirical average and $\|\cdot\|_2$ the
Euclidean norm or the $L^2$ norm of its argument.
Here, we use only parts (i) and (ii) of Theorem 3 of \cite{CGMT}, with Definition 2 and equations (10a)--(10b). In its notation, for compact sets $S_w,S_q$ and continuous $\psi$, we consider
\[
 \Phi=\min_{w\in S_w}\max_{q\in S_q}\{q^\top Ww+\psi(w,q)\},\qquad
 \phi=\min_{w\in S_w}\max_{q\in S_q}
 \{\norm w\,g^\top q+\norm q\,h^\top w+\psi(w,q)\},
\]
where all Gaussian entries are independent standard normals. Then, from that theorem,
\begin{equation}\label{F-eq-cgmt}
 \Prob(\Phi<c)\le2\Prob(\phi\le c).
\end{equation}
If the sets are convex and $\psi$ is convex in $w$ and concave in $q$, then, for every real $\mu$ and $a>0$,
\begin{equation}\label{F-eq-cgmt2}
 \Prob(|\Phi-\mu|>a)\le2\Prob(|\phi-\mu|>a).
\end{equation}
Here, we let $\odot$ denote the entrywise product, and note that $-\diag(D_i)W$ is standard Gaussian conditional on the fixed coordinates and labels. We then apply \eqref{F-eq-cgmt} on compact radius sets at those coordinates and \eqref{F-eq-cgmt2} on convex nuisance balls with a convex dual box. Also, for any $n\times q$ standard Gaussian matrix $W$, we use two $1/4$-nets of the unit spheres and the normal tail bound to obtain
\begin{equation}\label{F-eq-opbound}
 \Prob(\norm W_{\op}>2v\sqrt n)\le2\exp\{(n+q)\log9-nv^2/2\}.
\end{equation}
Indeed, each sphere has a $1/4$-net of size at most nine to the power of its dimension. On these nets, $\max_{x,y}|x^\top Wy|\ge\|W\|_{\op}/2$ by approximation at both arguments, and $x^\top Wy\sim N(0,1)$ for each pair.
\begin{theorem}\label{F-thm-exist}
There exist deterministic $R,c,C>0$ such that, for all sufficiently large $n$,
\[
 \Prob\!\left\{\text{MLE exists uniquely},\quad
 \widehat b_n^2+\widehat\beta_n^\top\Sigma_n\widehat\beta_n\le R^2\right\}
 \ge1-Ce^{-cn}.
\]
The same localization holds for restricted fits and fits with one row deleted, uniformly over their coordinates of fixed dimension in a fixed compact set.
\end{theorem}
\begin{proof}
For $Z_i=X_i^\top\beta_n/\gamma_n$, by whitening, we get $h_Z=h(\theta,\gamma)$ and
\[
 \widehat b_n^2+\widehat\beta_n^\top\Sigma_n\widehat\beta_n
 =\widehat b_n^2+\widehat t^2+\|\widehat w\|^2\le(\log2/c)^2
\]
by \eqref{S-NG-eq-fo-localization} and comparison with $L_n(0)=\log2$. Note that Lemma~\ref{S-NG-lem-localization} also includes restricted and deleted fits, and nonnegative quadratic penalties on $w$ preserve the lower bound and the comparison at $w=0$. For $p/n\le\bar\kappa<1$, we apply Lemma~\ref{S-NG-NG-mixed-geometry} to $[\ind,Z,W]$ to obtain \eqref{S-NG-eq-geo-curvature} with deletions. This completes the proof.
\end{proof}

Let $u\sim N_d(0,I)$ and let symmetric $B\ne0$ be independent of $u$, with eigenvalues $b_j$. For $|t|\|B\|\le1/4$, diagonalization and the Gaussian exponential moment give
\[
\log\E\{e^{t(u^\top Bu-\operatorname{tr}B)}\mid B\}
 =\sum_j\{-tb_j-\tfrac12\log(1-2tb_j)\}
 \le2t^2\operatorname{tr}(B^2).
\]
From the logarithmic power series, we have the inequality. By Chernoff's inequality for both signs, optimized over $t$, we get
\[
\Prob(|u^\top Bu-\operatorname{tr}B|>x\mid B)\le2\exp[-c\min\{x^2/\operatorname{tr}(B^2),x/\|B\|\}].
\]
For $d\le Cn$ and $\|B\|\le C$, from tail integration we obtain $\|(u^\top Bu-\operatorname{tr}B)/n\|_{L^q}\le C_qn^{-1/2}$ for every fixed $q\ge1$, with exponentially small probabilities for fixed positive deviations. Note that the form equals zero for $B=0$, and the bounds hold conditional on the remaining data when we define $B=0$ outside its boundedness event.

\subsection{State equations for bootstrap fits}\label{S-weighted}

The actual bootstrap counts and criterion are
\begin{equation}\label{S-multinomial}
\begin{gathered}
 (N_1,\ldots,N_n)\sim\operatorname{Multinomial}(n;1/n,\ldots,1/n),\\
 L_n^N(b,t,w)=\frac1n\sum_iN_i\ell_{Y_i}(b+tZ_i+A_i^\top w).
\end{gathered}
\end{equation}
Note that the bootstrap counts are independent of $\mathcal D$, and in the population system we use either $W=1$, indexed by $0$, or $W\sim\operatorname{Poisson}(1)$, indexed by $B$, where $W$ is independent of $(Z,G,Y)$ in both cases. For $r,\lambda>0$, we define
\begin{equation}\label{S-prox}
 V=b+tZ+rG,\qquad T+\lambda W\{s(T)-Y\}=V,\qquad
 U=W\{s(T)-Y\},\qquad d_T=s'(T).
\end{equation}
Here, this scalar equation has a unique root because the derivative of its left side in $T$ is at least one, and at $W=0$ we obtain $T=V$ and $U=0$.

\begin{theorem}\label{S-state}
Under \eqref{S-regime}, the Poisson system
\begin{equation}\label{S-state-eq}
 \E U=\E ZU=0,\qquad
 \E\frac{\lambda Wd_T}{1+\lambda Wd_T}=\kappa,\qquad
 \lambda^2\E U^2=\kappa r^2
\end{equation}
has a unique solution $(b_B,t_B,r_B,\lambda_B)$ with $(t_B,r_B,\lambda_B)\in(0,\infty)^3$. The state is locally $C^1$ in the strict interior and extends continuously to zero signal with $t_B=0$. The multinomial MLE exists uniquely with probability $1-o(1)$, and
\begin{equation}\label{S-fit-convergence}
 (\widehat b^*,\widehat t^*,\norm{\widehat w^*})\pto(b_B,t_B,r_B).
\end{equation}
Convergence is conditional as in \eqref{S-conditional} and joint with the original fit and any fixed number of bootstrap state summaries. For $W=1$ and $\gamma>0$, $0<\kappa<h(\theta,\gamma)$, the analogous state properties and MLE conclusions hold for the ordinary fit.
\end{theorem}
\begin{proof}
\OptProofStep{1}{Law of large numbers for the counts}
Note that for distinct indices and fixed nonnegative integers $j,k$, we have
\[
 \begin{gathered}
 \E z^{N_1}=\{1+(z-1)/n\}^n,\qquad
 \E z^{N_1}u^{N_2}=\{1+(z-1)/n+(u-1)/n\}^n,\\
 \E(N_1)_j(N_2)_k=(n)_{j+k}/n^{j+k}\le1.
 \end{gathered}
\]
Here, $(x)_j=x(x-1)\cdots(x-j+1)$ and $(x)_0=1$. These generating functions have independent Poisson limits, and we bound fixed moments by factorial moments. For bounded $f_i=f(N_i,Z_i,G_i,Y_i)$, we have
\[
\Var\Big(\frac1n\sum_i f_i\Big)
 =\frac1n\Var(f_1)+(1-1/n)\Cov(f_1,f_2)\longrightarrow0.
\]
From the joint Poisson limit, we obtain mean convergence, and we use moment truncation and finite meshes to extend this to functions and local Lipschitz constants of polynomial growth in $N+|Z|+|G|$, uniformly on compact parameter sets. We also extend the limit to moving $(\theta_n,\gamma_n)$ by thresholding independent uniforms. In particular, $M_n=\sum_i\ind\{N_i>0\}$ satisfies $M_n/n\pto c_B$.

\OptProofStep{2}{Existence and localization of the actual fit}
Note that conditional on the selected indices, the rows follow \eqref{S-experiment}. With $N_i\ge1$ on selected rows, we apply Lemma~\ref{S-NG-lem-localization} uniformly at dimension ratio $p/M_n\pto\kappa/c_B<h$ and have, with probability tending to one,
\begin{equation}
\label{S-localization}
 L_n^N(b,t,w)\ge\frac1n\sum_{N_i>0}\ell_{Y_i}(b+tZ_i+A_i^\top w)
 \ge c\sqrt{b^2+t^2+\norm w^2}.
\end{equation}
We observe from $L_n^N(0)=\log2$ that a minimizer exists on a fixed ball. Specifically, when $\kappa/c_B<h\le1/2$, we have $M_n>p+1$ with probability tending to one, and hence we get uniqueness from full rank of the selected Gaussian design and positive logistic weights.

\OptProofStep{3}{The population optimization problem}
We let $H(p)=p\log p+(1-p)\log(1-p)$ on $(0,1)$ and extend it continuously to $[0,1]$, and we define
\begin{equation}\label{S-popobjective}
\begin{aligned}
 \mathcal F_W(b,t,r)=\sup_{-WY\le U\le W(1-Y)}
 \big[\E\{U(b+tZ+rG)-WH\big(Y+\tfrac{U}{W\vee1}\big)\}-r\sqrt\kappa\norm U_{L^2}\big],\quad r\ge0.
\end{aligned}
\end{equation}
The supremum is over measurable $U$ with $U=0$ and zero entropy on $W=0$. From $|U|\le W$ and $0\le-WH\le W\log2$, we see that the objective is finite, convex, and locally Lipschitz. For Poisson $W$, we choose $U=\ind\{W>0\}q$ with $-Y\le q\le1-Y$ independent of $W$, and using $-WH\ge0$, we apply \eqref{S-NG-eq-fo-witness} at ratio $\kappa/c_B$ to obtain
\[
 \mathcal F_W(b,t,r)
 \ge c_B\sup_q\{\E(qV)-r\sqrt{\kappa/c_B}\|q\|_{L^2}\}
 \ge c\sqrt{b^2+t^2+r^2}.
\]

When $r>0$, the root $U_\lambda$ of \eqref{S-prox} has $\partial_\lambda|\lambda U_\lambda|=|U_\lambda|/(1+\lambda Wd_T)>0$ on $W>0$. For binary $Y$, we cannot have bounded $|\lambda U_\lambda|$ as $\lambda\to\infty$ because $T=V-\lambda U_\lambda$ would stay bounded with $s(T)-Y\to0$. Thus, $|\lambda U_\lambda|$ increases from zero to infinity, and
\begin{equation}\label{S-normroot}
 \lambda^2\E U_\lambda^2=\kappa r^2
\end{equation}
has a unique positive root $\lambda=r\sqrt\kappa/\norm U_{L^2}$. Specifically, by $H''(p)\ge4$ and the supporting inequality for the $L^2$ norm, this $U$ uniquely maximizes \eqref{S-popobjective} with an objective gap against $U'$ of at least
\begin{equation}
\label{S-entropygap}
 2\E\left[\frac{(U'-U)^2}{W\vee1}\right].
\end{equation}
Note that for $M>0$, we have
\[
\E(U'-U)^2\le M\E\left[\frac{(U'-U)^2}{W\vee1}\right]
       +4\E[W^2\ind\{W>M\}].
\]
We send the gap to zero and let $M\to\infty$ to obtain unweighted $L^2$ stability.

We next differentiate at the unique dual optimizer and get
\begin{equation}
\label{S-popgradient}
 \nabla\mathcal F_W=(\E U,\E ZU,\E GU-\sqrt\kappa\norm U_{L^2}).
\end{equation}
Because the right radial derivative at $r=0$ equals $-\sqrt\kappa\norm{W\{s(b+tZ)-Y\}}_{L^2}<0$, every minimizer has $r>0$. We have $\E GU=r\E[Wd_T/(1+\lambda Wd_T)]$ from Gaussian integration by parts and combine \eqref{S-normroot} with \eqref{S-popgradient} into \eqref{S-state-eq}, and its solutions minimize the convex objective.

From the affinity of the maximand along a minimizing segment and the unique dual optimizer, we obtain a common $U$ at both ends, and if we define $c=\sqrt\kappa/\norm U_{L^2}$, then
$b_j+t_jZ+r_jG=\logit(Y+U/W)+cr_jU\qquad(W>0).$
If $r_1\ne r_2$, by subtraction we see that $G-cU$ and hence $\logit(Y+U/W)$ are affine in $Z$. Then $U$ is measurable in $(W,Z,Y)$, and this contradicts independence of $G$ on $W=1$. Thus, we obtain $r_1=r_2$, $b_1=b_2$, and $t_1=t_2$, and uniqueness of $\lambda$ follows from \eqref{S-normroot}.

\OptProofStep{4}{Alignment and smoothness}
Let $p_0=s(\theta+\gamma Z)$ and let $U_y(V,W)$ denote the residual root at label $y$. Next, we apply integration by parts in $Z$ and differentiate the conditional label probabilities to get
\[
\E ZU=t\E\frac{Wd_T}{1+\lambda Wd_T}
  +\gamma\E[p_0(1-p_0)\{U_1(V,W)-U_0(V,W)\}].
\]
Because the $t$ coefficient is positive and $U_1<0<U_0$ on $W>0$, stationarity gives $t>0$ when $\gamma>0$ and $t=0$ when $\gamma=0$.

By compact localization, dominated convergence, and uniqueness, we get continuity and define
\[
a_T=\frac{Wd_T}{1+\lambda Wd_T},\quad
 D_0=\E\frac{U^2}{1+\lambda Wd_T}>0,\quad
 L_h=h_b+h_tZ+h_rG-\frac{\lambda h_r}{r}U.
\]
Differentiating \eqref{S-normroot} gives
\[
\dot\lambda=\lambda h_r/r-(\lambda/D_0)\E[a_TUL_h],\qquad
 \dot U=a_T\{L_h+(\lambda/D_0)U\E[a_TUL_h]\}.
\]
Substituting into the derivative of \eqref{S-popgradient} gives
\begin{equation}
\label{S-popHessian}
 h^\top\mathcal F_W''h=\E[a_TL_h^2]
            +\frac{\lambda}{D_0}\{\E[a_TUL_h]\}^2>0\quad(h\ne0).
\end{equation}
On $W=1$, $L_h$ is a nonzero affine function of $Z$ if $h_r=0$, and $\partial_G L_h=h_r/(1+\lambda Wd_T)\ne0$ if $h_r\ne0$. Hence, \eqref{S-popHessian} holds, and we differentiate under expectation because the norm derivative satisfies $2\lambda D_0>0$ and polynomial bounds hold in $W,|Z|,|G|$. Then, by the implicit function theorem, we get smoothness to every fixed order.

\OptProofStep{5}{Transfer of auxiliary values to the actual optimizer}
Conditional on $(N,Z,Y)$, $A$ is independent standard Gaussian, and Fenchel conjugacy on the localization ball and dual box $[-N_iY_i,N_i(1-Y_i)]$ gives the fixed-radius auxiliary value
\begin{equation}
\label{S-auxiliary}
 \mathcal F_n^{\rm aux}(b,t,r)=\max_U
 \left[\frac1n\sum_{i:N_i>0}\{U_i(b+tZ_i+rG_i)-N_iH(Y_i+U_i/N_i)\}
      -r d_n\Big(\frac1n\sum_iU_i^2\Big)^{1/2}\right],
\end{equation}
where $d_n=\norm{g_{p-1}}/\sqrt n\pto\sqrt\kappa$ for an independent standard Gaussian vector $g_{p-1}$. Here, we apply \eqref{F-eq-cgmt} on nonconvex restricted sets and \eqref{F-eq-cgmt2} only on convex compact sets with a convex--concave offset.

Next, we define $\lambda=rl$ by the root equation $l^2\E U(b+tZ+rG;rl)^2=\kappa$. As $r\to0$, we get $rl\to0$ from \eqref{S-normroot} because a subsequence bounded away from zero contradicts this equation on $W=1$ with bounded Gaussian variables. At $r=0$, we have the continuous value $l=\sqrt\kappa\,\norm{W\{s(b+tZ)-Y\}}_{L^2}^{-1}$, and because $2l\E[U^2/(1+rlWd_T)]>0$, we bound the root above and away from zero down to $r=0$ by compactness.

We also define $e_{W,Y}(v;\lambda)=\min_x\{W\ell_Y(x)+(x-v)^2/(2\lambda)\}$ for $\lambda>0$, extended continuously to $\lambda=0$, and by substitution we obtain
$\mathcal F_W(b,t,r)=\E e_{W,Y}(b+tZ+rG;rl)-\frac{r\kappa}{2l}.$
For the empirical identity, we use the empirical root and $d_n^2$. On compact parameter sets, the root integrands, Moreau values, and local Lipschitz constants have polynomial count and Gaussian bounds. From Step~1, we thus obtain uniform convergence of the empirical root and auxiliary objective down to $r=0$, namely
\begin{equation}
\label{S-uniformvalues}
 \sup_{(b,t,r)\in K}|\mathcal F_n^{\rm aux}(b,t,r)-\mathcal F_W(b,t,r)|\pto0.
\end{equation}
By uniqueness, we have a positive population gap outside each minimizer neighborhood. At the population $(b,t)$, we use two-sided comparison on the nuisance ball to bound the primary minimum above. From Equation~\eqref{F-eq-cgmt} on excluded radius sets and \eqref{S-uniformvalues}, we obtain the lower bound. Next, we obtain both bounds on a finite $(b,t)$ mesh because $P_n\{N_i(1+|Z_i|)\}=O_P(1)$. Then, we prove \eqref{S-fit-convergence} from \eqref{S-localization} by refining the mesh.

\OptProofStep{6}{Conditional and joint conclusions}
By Markov's inequality, for any event $E_n$ and $\delta>0$, we obtain
\[
\Prob_{\mathcal D}\{\Prob_*(E_n\mid\mathcal D)>\delta\}
 \le\Prob(E_n)/\delta
\]
and thus conditional nonexistence and state bounds. We get joint convergence for fixed numbers of refits and the ordinary fit by a union bound, and for $W=1$ with unit counts, this argument holds under $\kappa<h(\theta,\gamma)$. This completes the proof.\end{proof}

\subsection{Limiting score distributions}\label{S-scores}\label{S-reader-scores}

We define probability measures for the ordinary and bootstrap states of Theorem~\ref{S-state} by
\begin{equation}\label{S-five}
 \begin{array}{ll}
 Q_Af=\E f(Y,T_0),\quad Q_Ef=\E f(Y,V_0),\\
 Q_A^*f=\E Wf(Y,T_B),\quad Q_U^*f=\E f(Y,T_B),\quad Q_E^*f=\E f(Y,V_B).
 \end{array}
\end{equation}
Here, $A,E,U$ denote apparent, external, and unweighted evaluation on the original sample. Note that with $\E W=1$, the weighted measure has unit mass, each has prevalence $\pi$, and by \eqref{S-prox} at zero count we get
\begin{equation}\label{S-mixture}
 Q_U^*=e^{-1}Q_E^*+c_BQ_+^*,\qquad
 Q_+^*f=\E\{f(Y,T_B)\mid W>0\}.
\end{equation}

Here, $\widehat\eta_i^*=\widehat b^*+X_i^\top\widehat\beta^*$ denotes the fitted score for patient $i$ under the bootstrap fit $(\widehat b^*,\widehat\beta^*)$.

\begin{theorem}\label{S-empirical}
Under \eqref{S-regime}, for every bounded continuous $f$ on $\{0,1\}\times\R$,
\begin{align}
 \frac1n\sum_iN_if(Y_i,\widehat\eta_i^*)&\pto Q_A^*f,\label{S-empiricalBA}\\
 \frac1n\sum_if(Y_i,\widehat\eta_i^*)&\pto Q_U^*f,\label{S-empiricalBU}\\
 \frac1{n-M_n}\sum_{N_i=0}f(Y_i,\widehat\eta_i^*)&\pto Q_E^*f.\label{S-empiricalBE}
\end{align}
The conditional conclusions \eqref{S-conditional} hold. The ordinary empirical and conditional external distributions converge to $Q_A$ and $Q_E$. The ordinary conclusions also hold when $(\theta_n,\gamma_n,\kappa_n)\to(\theta,\gamma,\kappa)$ with $\gamma>0$ and $0<\kappa<h(\theta,\gamma)$.
\end{theorem}

\begin{proof}
\OptProofStep{1}{Perturbation of the dual coordinates for positive counts}
Specifically, for bounded $\psi(W,Z,Y,p)$ supported on $1\le W\le M$ and compactly in $Z$, twice continuously differentiable in $p\in[0,1]$, we perturb the dual entropy to
\begin{equation}
\label{S-perturb}
 WH(p)+\epsilon\psi(W,Z,Y,p).
\end{equation}
For $W>0$, let $g_\epsilon(u)$ denote the perturbed entropy at $u=W(p-Y)$. Note that when $|\epsilon|\|\psi_{pp}\|_\infty<2$, the conjugate changes by at most $|\epsilon|\|\psi\|_\infty$, and hence by localization we have
\[
 g_\epsilon''(u)=\frac{H''(p)}W+\frac{\epsilon\psi_{pp}}{W^2}\ge\frac2W,
 \qquad p=Y+u/W.
\]
Note that at zero count we have $U=0$. For $W>0$, from the divergence of $H'$ near $0$ and $1$ and boundedness of $\psi_p$, we get a unique interior solution with sign determined by $Y$. At fixed $V$, it satisfies
\[
 g_\epsilon'(U)+\lambda U=V,\qquad
 \partial_\lambda|\lambda U|
   =\frac{|U|g_\epsilon''(U)}{g_\epsilon''(U)+\lambda}>0.
\]
If $|\lambda U|$ stayed bounded as $\lambda\to\infty$, then $U\to0$ and $|g_\epsilon'(U)|\to\infty$, and this contradicts the first equality. Hence, \eqref{S-normroot} has a unique positive root for $r>0$.

At fixed $(b,t,r)$ with $r>0$, by Step~1 in the proof of Theorem~\ref{S-state} and bracketing, we get convergence of the auxiliary root and value. Because $|U_i|\le N_i$, the Lipschitz constant in \eqref{S-auxiliary} is bounded by
$P_n\{N_i(1+|Z_i|+|G_i|)\}+d_n(P_nN_i^2)^{1/2}=O_P(1).$
By finite meshes, we get uniform convergence down to $r=0$. Conditional on $(N,Z,Y)$ at fixed $(b,t)$, we apply \eqref{F-eq-cgmt2} on compact sets by convexity in the nuisance coefficients and concavity in the dual coordinates. By a finite $(b,t)$ mesh, we get convergence of the perturbed minimum for either sign of $\epsilon$.

\OptProofStep{2}{Two-sided derivatives of the population value}
For $x=(b,t,r)$, let $a_x(U)$ be the maximand in \eqref{S-popobjective} and let $\psi(U)$ denote the perturbation in \eqref{S-perturb}. On the common localization set define
\[
 F_\epsilon(x)=\sup_U\{a_x(U)-\epsilon\E\psi(U)\},\qquad
 v(\epsilon)=\min_x F_\epsilon(x).
\]
By the uniform bound $|F_\epsilon-F_0|\le|\epsilon|\|\psi\|_\infty$, compactness and uniqueness, we get $x_\epsilon\to x_0$. At these minimizers the dual gap is at most $2|\epsilon|\|\psi\|_\infty$. By \eqref{S-entropygap}, $|U_\epsilon-U_0|\le2W$ and $\E W^2<\infty$, we get $\|U_\epsilon-U_0\|_{L^2}\to0$, hence convergence of expected perturbations by bounded $\psi_p$. From the two minimizer values, we get
\begin{equation}
\label{S-valuebounds}
 -\epsilon\E\psi\{U_0(x_\epsilon)\}
 \le v(\epsilon)-v(0)
 \le-\epsilon\E\psi\{U_\epsilon(x_0)\}.
\end{equation}
After division by $\epsilon$ with opposite inequalities for $\epsilon<0$, we get $v'(0)=-\E\psi(W,Z,Y,s(T_B))$. Under the design and intercept constraints, we obtain the empirical value as the supremum of
$-\frac1n\sum_{i:N_i>0}\{N_iH(Y_i+U_i/N_i)+\epsilon\psi(N_i,Z_i,Y_i,Y_i+U_i/N_i)\}.$
Because the maximized dual value is convex in $\epsilon$ with derivative $-n^{-1}\sum_i\psi(N_i,Z_i,Y_i,s(\widehat\eta_i^*))$ at zero by uniqueness, we obtain empirical convergence from secant bounds for both signs of $\epsilon$ as $\epsilon\to0$.

\OptProofStep{3}{Count truncation and unused observations}
We extend this limit to bounded continuous functions of $(Z,Y,p)$ at positive counts by Bernstein approximation and remove $Z$ truncation by Gaussian tail bounds. Note that for count truncation, we have
\[
\frac1n\sum_iN_i\ind\{N_i>M\}\pto\E[W\ind\{W>M\}],
 \qquad \frac1n\sum_i\ind\{N_i>M\}\pto\Prob(W>M).
\]
We then let $M\to\infty$ to remove count truncation.

We observe that given counts and selected rows, omitted rows are independent of the fit with nuisance projections $N(0,\|\widehat w^*\|^2)$ and conditional variance at most $C/(n-M_n)$ for bounded tests. Because $(n-M_n)/n\pto e^{-1}$, we get the omitted-row limit by conditional Chebyshev and Theorem~\ref{S-state}, and hence unweighted convergence from \eqref{S-mixture}.

\OptProofStep{4}{The score coordinate and conditional convergence}
We apply the logit on $[\delta,1-\delta]$ and let $\delta\downarrow0$ to obtain score convergence. For unit counts, we get the ordinary empirical limit, and from Theorem~\ref{S-state} we obtain the independent-row external limit and \eqref{S-conditional}. This completes the proof.
\end{proof}

\subsection{Performance functionals and continuity}\label{S-reader-performance}
Let $Q$ be a joint probability measure for $Y\in\{0,1\}$ and a score $L\in\R$ with $0<\pi=QY<1$. For independent pairs $(Y_1,L_1)$ and $(Y_0,L_0)$ with distribution $Q$ and joint distribution $Q\otimes Q$, we define
\begin{align}
 B(Q)&=Q\{s(L)-Y\}^2,&R(Q)&=Q\ell_Y(L),\label{S-risks}\\
 A(Q)&=\frac{(Q\otimes Q)[Y_1(1-Y_0)\{\ind\{L_1>L_0\}+\tfrac12\ind\{L_1=L_0\}\}]}{\pi(1-\pi)}.\label{S-auc}
\end{align}
The mean log likelihood is $-R(Q)$. We define the intercept and slope for calibration by
\begin{equation}\label{S-calibration}
 (a(Q),c(Q))=\argmin_{a,c}Q\ell_Y(a+cL),\qquad
 Qs\{d(Q)+L\}=\pi.
\end{equation}
The second equation defines calibration-in-the-large $d(Q)$, which differs from the calibration intercept $a(Q)$.

For the fitted scores $\widehat\eta_i=\widehat b+X_i^\top\widehat\beta$, define $\widehat Q_A=n^{-1}\sum_i\delta_{(Y_i,\widehat\eta_i)}$, where $\delta_{(y,l)}$ is unit mass at $(y,l)$. Let $\widehat Q_E$ be the conditional distribution of $(Y,\widehat b+X^\top\widehat\beta)$ for an independent new observation from the same population given $\mathcal D$. Then $F_{\rm ext}=F(\widehat Q_E)$.

For $j\in\{0,B\}$, let $Q_{jE}$ be the distribution of $(Y,V_j)$, with $Q_{0E}=Q_E$ and $Q_{BE}=Q_E^*$. To characterize external calibration, let $H\sim N(0,1)$ and define
$\nu_j=t_j^2+r_j^2,\qquad A_0=\E s'(\theta+\gamma Z),\qquad j\in\{0,B\}.$
Let $D_\pi(\nu)$ solve $\E s\{D_\pi(\nu)+\sqrt\nu H\}=\pi$.

\begin{proposition}\label{S-functional}
Under the conditions of Theorem~\ref{S-empirical}, the empirical and conditional external evaluations of each $F\in\{A,B,R,a,c,d\}$ converge to $F(Q)$ at the corresponding $Q$ in \eqref{S-five}. For the ordinary fit, this holds throughout $\gamma>0$, $0<\kappa<h(\theta,\gamma)$. The calibration functionals satisfy
\begin{equation}\label{S-apparentcal}
 (a(Q_A),c(Q_A),d(Q_A))=(0,1,0)
 =(a(Q_A^*),c(Q_A^*),d(Q_A^*)).
\end{equation}
For $j\in\{0,B\}$ and $(a,c)=(a(Q_{jE}),c(Q_{jE}))$,
\begin{equation}\label{S-externalcal}
 \E s(q+c\sqrt{\nu_j}H)=\pi,\qquad
 c\nu_j\E s'(q+c\sqrt{\nu_j}H)=\gamma t_jA_0,\qquad q=a+cb_j,
\end{equation}
and
\[
 d(Q_{jE})=D_\pi(\nu_j)-b_j.
\]
\end{proposition}

\begin{proof}
\OptProofStep{1}{Bounded losses and AUC}
Note that by bounded continuity, we get Brier convergence. Because $r_j>0$ and the proximal map is strictly increasing, both $V_j$ and $T_j$ are continuous given $Y$ at each count. From class-specific weak convergence and zero tie mass, we get AUC convergence.

\OptProofStep{2}{Log loss and its tails}
On the localization event,
\begin{equation}
\label{S-scoreL2}
 \frac1n\sum_i(\widehat\eta_i^*)^2
 \le C\{1+\norm Z^2/n+\norm A_\op^2/n\}=O_P(1)
\end{equation}
by \eqref{F-eq-opbound}. We then combine $\ell_y(l)\le\log2+|l|$ with
\[
\frac1n\sum_i|\widehat\eta_i^*|\ind\{|\widehat\eta_i^*|>M\}
 \le\frac1{Mn}\sum_i(\widehat\eta_i^*)^2=O_P(M^{-1}),
\]
to obtain unweighted and OOB log-loss convergence via truncation, because the OOB fraction tends to $e^{-1}$.

Note that by the score equations, we have
\begin{equation}
\label{S-entropyidentity}
 \frac1n\sum_iN_i\ell_{Y_i}(\widehat\eta_i^*)
 =-\frac1n\sum_iN_iH\{s(\widehat\eta_i^*)\}\le\log2.
\end{equation}
This entropy average is bounded and converges in probability, and by the state equations and $T_B=V_B-\lambda_B U$, we have
\[
 \E UV_B=r_B\E GU=\kappa r_B^2/\lambda_B=\lambda_B\E U^2,
 \qquad \E UT_B=0.
\]
We thus get the limit $\E W\ell_Y(T_B)$, with $W=1$ for the ordinary fit.

\OptProofStep{3}{Calibration}
We observe that $c_Q=\inf_{a^2+c^2=1}Q[-(2Y-1)(a+cL)]_+>0$ by compactness, $Q_U^*\ge e^{-1}Q_E^*$, and positive conditional label probabilities at finite external scores. Next, we bound the empirical hinge loss below by $c_Q/2$ with probability $1-o(1)$ by monotone truncation, finite meshes, and \eqref{S-scoreL2}, and compare its minimum with zero to get
$(c_Q/2)\sqrt{a^2+c^2}\le\text{empirical calibration loss}\le\log2.$
On this fixed ball,
\[
Qs'(a+cL)\begin{pmatrix}1&L\\L&L^2\end{pmatrix}\succ0.
\]
By uniform convergence and monotone brackets, we get consistency of $(a,c)$ and the offset. Because the limiting radius is positive, apparent scores are finite and nonconstant with probability tending to one. From their score equations, we have $(a,c,d)=(0,1,0)$, and by \eqref{S-state-eq}, $\E UT=0$, and strict convexity we get \eqref{S-apparentcal}. For external scores, we apply integration by parts in $V_j=b_j+t_jZ+r_jG$ to get
\[
 \begin{aligned}
 \E s(a+cV_j)&=\pi,\qquad
 0=\E V_j\{s(a+cV_j)-Y\}
   =c\nu_j\E s'(a+cV_j)-\gamma t_jA_0 .
 \end{aligned}
\]
With $c=1$ in the first equation, we obtain the offset in \eqref{S-externalcal}. This completes the proof.
\end{proof}

\subsection{Correction and bootstrap averaging}\label{S-reader-bootstrap}
For unpenalized numerical fits, we compute bootstrap means and LS quantiles from successful refits, and the statistic is unavailable if no refit succeeds.

In the Gaussian bootstrap simulations, numerical failure meant a maximum absolute fitted score above 30, nonconvergence within 50 Newton steps, Hessian factorization failure, or 40 failed step-halving attempts. Here, the threshold of 30 was a numerical stopping rule because a finite unpenalized fit can exist with scores above 30. The released code also checks separation by linear programming, and we report the simulation results from the stored records.

Let $\widehat Q_A^*$ and $\widehat Q_U^*$ denote the weighted and unweighted empirical measures in \eqref{S-empiricalBA} and \eqref{S-empiricalBU}. For $F\in\{A,B,R,a,c,d\}$, define $F_{0A,n}=F(\widehat Q_A)$. With $B_n\ge1$ independent resamples conditional on $\mathcal D$, let $F_{BA,n}^{*(b)}=F(\widehat Q_A^{*(b)})$ and $F_{BU,n}^{*(b)}=F(\widehat Q_U^{*(b)})$, $1\le b\le B_n$. Harrell's estimator \citep{harrell2015regression,harrell2026rms} is
\begin{equation}\label{S-Hestimator}
 \widehat H_F=F_{0A,n}-\overline D_F,\qquad
 \overline D_F=\frac1{B_n}\sum_{b=1}^{B_n}
      \{F_{BA,n}^{*(b)}-F_{BU,n}^{*(b)}\}.
\end{equation}
By replacing the replicate average in \eqref{S-Hestimator} with $\E_*$, we define the conditional bootstrap expectation.

\begin{theorem}\label{S-Harrell}
Under \eqref{S-regime}, for $F\in\{A,B,R,a,c,d\}$ and fixed $B_n<\infty$,
\begin{align}
 \widehat H_F&\pto H_F:=F(Q_A)-F(Q_A^*)+F(Q_U^*),\label{S-Hlimit}\\
 \widehat H_F-F_{\rm ext}&\pto
 \Delta_F:=H_F-F(Q_E).\label{S-bias}
\end{align}
For $F\in\{A,B\}$ these conclusions hold for every deterministic $B_n\ge1$ and for the conditional bootstrap expectation. For calibration,
\begin{equation}\label{S-Hcal}
 H_a=a(Q_U^*),\qquad H_c=c(Q_U^*),\qquad H_d=d(Q_U^*).
\end{equation}
\end{theorem}

\begin{proof}
\OptProofStep{1}{A fixed number of replicates}
By Theorem~\ref{S-empirical}, Proposition~\ref{S-functional}, and a union over fixed $B_n$, we get $\overline D_F\pto F(Q_A^*)-F(Q_U^*)$ and \eqref{S-Hlimit}. From state convergence and the independent-row representation, we obtain $F_{\rm ext}\pto F(Q_E)$ and \eqref{S-bias}. Then, we substitute \eqref{S-apparentcal} to get \eqref{S-Hcal}.

\OptProofStep{2}{Bounded bootstrap averages}
Let $F_n^{*(b)}$ be either replicate evaluation of AUC or Brier loss with limit $F_*$. Boundedness gives $\E|F_n^{*(1)}-F_*|\to0$. For deterministic $B_n\ge1$,
\[
 \begin{aligned}
 \E_*\left|B_n^{-1}\sum_b(F_n^{*(b)}-F_*)\right|
 &\le B_n^{-1}\sum_b\E_*|F_n^{*(b)}-F_*|\\
 &=\E_*|F_n^{*(1)}-F_*|\pto0 ,
 \end{aligned}
\]
by conditional exchangeability and Markov's inequality. From conditional Jensen and Markov, we obtain the expectation and average limits. Then, we apply the bound to failure indicators to get the failure-fraction limit. This completes the proof.
\end{proof}

Let $I_{nb}$ indicate that the exact bootstrap refit and evaluations exist, with $p_n=\Prob(I_{n1}=0)\to0$ by Theorem~\ref{S-state} and Proposition~\ref{S-functional}. Under Section~\ref{S-model} assignments and a common value if the original fit fails, for fixed $B_n$ the successful-refit and all-replicate averages differ with probability at most $B_np_n$. For contributions and assigned averages of absolute value at most $M$, we bound their difference by $2Mf_n$ even if all refits fail, where $f_n=B_n^{-1}\sum_b(1-I_{nb})$ has $\E f_n=p_n$. Thus, this difference tends to zero in probability for all deterministic $B_n$. For numerical fits, we require single-replicate failure probability $o(1)$ and numerical errors $o_P(1)$ for fixed counts or $o_{L^1}(1)$ for bounded averages.

\subsection{Integral representations for AUC}
Here, $T_{\rm case}$ and $T_{\rm control}$ denote independent fitted scores for a case and control. Note that for $T_{\rm case}$ at $(z_1,g_1,w_1)$, by monotonicity of the control proximal equation we have
\begin{equation}\label{S-controlCDF}
 \Prob(T_{\rm control}<T_{\rm case}\mid z_0,w_0,T_{\rm case})
 =\Phi\left\{\frac{T_{\rm case}+\lambda w_0s(T_{\rm case})-b-tz_0}{r}\right\}.
\end{equation}
We integrate over $z_1,g_1,z_0$ with factor $s(\theta+\gamma z_1)\{1-s(\theta+\gamma z_0)\}/\{\pi(1-\pi)\}$ to compute AUC, where the two count variables have ordinary Poisson probabilities under $Q_U^*$ and size-biased $w\Prob(W=w)$ under $Q_A^*$. When we condition on the two independent signal coordinates, we have
\begin{equation}\label{S-externalauc}
 A(Q_{jE})=\frac{\E\left[p_1(1-p_0)
       \Phi\{t_j(Z_1-Z_0)/(\sqrt2r_j)\}\right]}{\pi(1-\pi)},
 \qquad p_y=s(\theta+\gamma Z_y),
\end{equation}
with independent standard normal $Z_1,Z_0$.

To avoid numerical rank comparisons, for $H\sim N(0,1)$, $p_H=\Prob(Y=1\mid H)$ and $0<\pi=\E p_H<1$, we define the unnormalized cumulative functions
$F_1(h)=\int_{-\infty}^h\phi(u)p_u\,du$ and $F_0(h)=\Phi(h)-F_1(h)$. When we rank by $H$, we obtain
\begin{equation}\label{S-gaussianauc}
 A=\frac{\int F_0\,dF_1}{\pi(1-\pi)}
  =\frac{\E[p_H\Phi(H)]-\pi^2/2}{\pi(1-\pi)}.
\end{equation}
Because $F_1$ is continuous with $F_1(-\infty)=0$ and $F_1(\infty)=\pi$, from $d(F_1^2)/2$ we get $\int F_1\,dF_1=\pi^2/2$. For an external score, we use $H=(tZ+rG)/\sqrt{t^2+r^2}$ to compute $p_H$ by conditional Gaussian integration, and from \eqref{S-controlCDF} we obtain the corresponding integral for training scores.

\subsection{Bias under weak signals}\label{S-signs}
We next expand the proportional-dimensional limits in \eqref{S-bias} as $\kappa\downarrow0$. At $\theta=\gamma=0$, we define $A(Q_A)$, $A(Q_A^*)$, and $\Delta_A$ by continuity as $\gamma\downarrow0$, and we let $\pi_{\rm circ}$ denote the circle constant.

Because we have $b=t=0$ by symmetry at $\theta=\gamma=0$, for $X=(2Y-1)G$ and $M=(2Y-1)T$, the variable $X$ is standard normal and independent of $W$, and we obtain
\begin{equation}\label{S-nullprox}
 M=rX+\lambda W h_0(M),\qquad h_0(m)=s(-m).
\end{equation}
We define $m_j=\E W^j$ for either count distribution, with $m_1=1$, $m_j=1$ for unit counts, and $m_2=2,m_3=5$ for Poisson counts.

\begin{proposition}\label{S-null}
At $\theta=\gamma=0$, as $\kappa\downarrow0$,
\begin{align}
 \lambda&=4\kappa+8m_2\kappa^2+O(\kappa^3),&
 r^2&=4m_2\kappa+(20m_2^2-8m_3)\kappa^2+O(\kappa^3),\label{S-nullstate}\\
 \Delta_A&=(1-2^{-1/2})\sqrt{\kappa/\pi_{\rm circ}}+o(\sqrt\kappa),&
 \Delta_B&=-\tfrac12\kappa^2+O(\kappa^3),\label{S-nullbias}\\
 c(Q_U^*)&=\tfrac12+O(\kappa),&c(Q_E)&=0.\label{S-nullslope}
\end{align}
For every sufficiently small fixed $\kappa>0$, there is $\epsilon(\kappa)>0$ such that
\begin{equation}\label{S-nearnull}
 \theta=0,\quad 0<\gamma<\epsilon(\kappa)
 \quad\Longrightarrow\quad
 \kappa<c_Bh(0,\gamma),\qquad \Delta_A>0,\quad\Delta_B<0,\quad\Delta_c>0.
\end{equation}
For $Q\in\{Q_A,Q_E,Q_A^*,Q_U^*,Q_E^*\}$,
\[
 \theta=0\quad\Longrightarrow\quad a(Q)=d(Q)=0.
\]
\end{proposition}

\begin{proof}
\OptProofStep{1}{Smooth expansions at zero ratio}
Let $\delta=\sqrt\kappa$, $r=\delta R$, and $\lambda=\delta^2L$. Next, we divide the trace and norm equations in \eqref{S-state-eq} by $\delta^2$ and $\delta^4$ to get $L/4=1$ and $L^2m_2/4=R^2$ at zero. The Jacobian at $(4,2\sqrt{m_2})$ is triangular and nonsingular, and by the implicit function theorem we obtain smooth $L,R$ that are even in $\delta$ by $(\delta,X)\mapsto(-\delta,-X)$. Because $1+\lambda Ws'(M)\ge1$, proximal derivatives have polynomial bounds in $W+|X|$. By finite moments, differentiation and the Taylor remainder bounds hold under polynomial count weights.

When we substitute $v=r^2$ into the trace and norm expansions, we obtain
\[
\kappa=\frac\lambda4-\frac{\lambda v}{16}-\frac{\lambda^2m_2}{16}+O(\kappa^3),\qquad
 \kappa v=\lambda^2\left\{\frac{m_2}4-\frac{\lambda m_3}8+
                              \frac{vm_2}{16}+O(\kappa^2)\right\}.
\]
When we solve successively for the coefficients, we obtain \eqref{S-nullstate}.

\OptProofStep{2}{Brier and logistic loss}
For $a=\lambda W$, by Gaussian integration in \eqref{S-nullprox}, we obtain
\begin{align*}
\E_X M&=a/2-a^2/8+O(\kappa^3),&
 \E_X M^2&=v+a^2/4-av/2+O(\kappa^3),\\
 \E_X M^3&=3av/2+O(\kappa^3),&
 \E_X M^4&=3v^2+O(\kappa^3).
\end{align*}
Next, we substitute $h_0(m)=1/2-m/4+m^3/48+O(m^5)$ into \eqref{S-nullprox}, and we use
\[
\E X=\E X^3=0,\qquad \E X^2=1,\qquad \E X^4=3,\qquad
 |M-rX|\le\lambda W.
\]
Under any fixed polynomial count weight, odd-power coefficients have zero expectation by Step~1 symmetry, and hence we obtain the stated remainders. When we expand $h_0(m)^2=\tfrac14-\tfrac m4+\tfrac{m^2}{16}+\tfrac{m^3}{48} -\tfrac{m^4}{96}+O(m^5)$, for $S=S(W)$ we get
\begin{equation}
\label{S-nullBweighted}
 \E[Sh_0(M)^2]=\frac{\E S}4-\frac{\lambda\E SW}8+
 \frac{v\E S}{16}+\frac{3\lambda^2\E SW^2}{64}
 -\frac{v^2\E S}{32}+O(\kappa^3).
\end{equation}
We use $(S,W)=(1,1),(1,W),(W,W)$ for $Q_A,Q_U^*,Q_A^*$, respectively, and $a=0$ for external scores. Substitution of \eqref{S-nullstate} gives the following expansions.\begin{center}
\begin{tabular}{@{}lcc@{}}
\toprule
Distribution & Brier loss & Logistic loss\\
\midrule
$Q_A$ & $\frac14-\frac\kappa4+O(\kappa^3)$
 & $\log2-\frac\kappa2-\frac{\kappa^2}4+O(\kappa^3)$\\
$Q_E$ & $\frac14+\frac\kappa4+\frac{\kappa^2}4+O(\kappa^3)$
 & $\log2+\frac\kappa2+\frac{5\kappa^2}4+O(\kappa^3)$\\
$Q_A^*$ & $\frac14-\frac\kappa2+\frac{\kappa^2}4+O(\kappa^3)$
 & $\log2-\kappa-\frac{\kappa^2}2+O(\kappa^3)$\\
$Q_U^*$ & $\frac14+O(\kappa^3)$
 & $\log2+\kappa^2+O(\kappa^3)$\\
$Q_E^*$ & $\frac14+\frac\kappa2+\frac{\kappa^2}2+O(\kappa^3)$
 & $\log2+\kappa+4\kappa^2+O(\kappa^3)$\\
\bottomrule
\end{tabular}
\end{center}
We use $\log(1+e^{-m})=\log2-m/2+m^2/8-m^4/192+O(m^6)$ with the same moments to obtain the log-loss expansions. From Harrell's combination, we get \eqref{S-nullbias}, and the $\kappa$ and $\kappa^2$ coefficients of $\Delta_R$ equal zero.

\OptProofStep{3}{AUC at the null}
For fitted case score $T_{\rm case}$ and case Gaussian noise $G_1\sim N(0,1)$ independent of evaluation case and control counts $w_1,w_0$, \eqref{S-controlCDF} and \eqref{S-nullprox} give
\[
A(Q)=\E\Phi\left\{G_1+\frac{\lambda}{2r}(w_1+w_0)+\mathcal R\right\},
 \qquad \E|\mathcal R|=O(\kappa).
\]
Here, we have $|s(T_{\rm case})-1/2|\le|T_{\rm case}|/4$, $\E|T_{\rm case}|^j=O(\kappa^{j/2})$ under polynomial count weights, and $\lambda/r=O(\sqrt\kappa)$, and hence $\E|\mathcal R|=O(\kappa)$. Because $\Phi''$ is bounded and $\E\phi(G_1)=1/(2\sqrt{\pi_{\rm circ}})$, with mean count $\E_{\rm eval}W$ under the evaluation measure, we obtain
\[
A(Q)=\tfrac12+\frac{\lambda\E_{\rm eval}W}{2r\sqrt{\pi_{\rm circ}}}
 +o(\sqrt\kappa).
\]
Because the mean count is $1$ under ordinary and original-sample bootstrap evaluation and $2$ under size-biased apparent bootstrap evaluation, we obtain
\[
\begin{gathered}
A(Q_A)=\tfrac12+\sqrt{\kappa/\pi_{\rm circ}}+o(\sqrt\kappa),\quad
 A(Q_U^*)=\tfrac12+\sqrt{\kappa/(2\pi_{\rm circ})}+o(\sqrt\kappa),\\
A(Q_A^*)=\tfrac12+2\sqrt{\kappa/(2\pi_{\rm circ})}+o(\sqrt\kappa),
 \qquad A(Q_E)=A(Q_E^*)=\tfrac12.
\end{gathered}
\]
Next, we combine these expansions in \eqref{S-Hlimit} to obtain the AUC expansion in \eqref{S-nullbias}.

\OptProofStep{4}{Calibration and continuation to positive signal}
When $\theta=0$, we obtain $b=0$ and zero calibration intercept and offset from uniqueness and symmetry under $(Z,G,Y)\mapsto(-Z,-G,1-Y)$. At the null, we have $M\ge rX$ with strict inequality on $W>0$, and hence $\E M>0$ and the derivative of the free-slope risk at zero equals $-\E M/2<0$. Its unique minimizer is positive, and under $Q_U^*$ we obtain
\[
 \E T^2=8\kappa+O(\kappa^2),\qquad
 \E\{T(Y-1/2)\}=\kappa+O(\kappa^2).
\]
After division by $\kappa$, the slope equation is $2c-1+O(\kappa)=0$ uniformly on compact $c$-intervals. Its signs at $1/2\pm\epsilon$ bracket the root, and thus we obtain \eqref{S-nullslope}. By independence of the null external score and $Y$, we get $c(Q_E)=0$.

Note that Lemma~\ref{S-NG-lem-localization} is uniform near $(0,0)$. By dominated convergence, we obtain continuity of $h$, with $h(0,0)=1/2$ by conditional Jensen. From Theorem~\ref{S-state}, Proposition~\ref{S-functional}, and uniqueness, we get continuity of the score and calibration functionals. Thus, the strict signs in \eqref{S-nullbias} and \eqref{S-nullslope} hold at each sufficiently small fixed $\kappa<c_B/2$ for small positive $\gamma$, and hence we have \eqref{S-nearnull}. This completes the proof.\end{proof}

\subsection{Cross-validation and bootstrap intervals}\label{S-alternatives}\label{S-reader-comparators}

Let $\widehat F_{\rm CV}$ and $\widehat F_{\rm OOB}$ denote the fixed-$K$ cross-validation and out-of-bag estimators in Section~\ref{subsec:optboot-procedures}. Define $a_K=1-1/K$ for $K\ge2$, and let $Q_E(\theta,\gamma,\tau)$ be the ordinary external distribution at aspect ratio $\tau$.

Define the .632 estimator \citep{efron1983estimating} by
$\widehat F_{.632}=(1-c_B)F_{0A,n}+c_B\widehat F_{\rm OOB}.$

\begin{theorem}\label{S-altlimit}
For fixed $K$ and $\kappa/a_K<h(\theta,\gamma)$, all fold fits exist with probability $1-o(1)$. For each $F$ in Proposition~\ref{S-functional},
\begin{equation}\label{S-CVlimit}
 \widehat F_{\rm CV}\pto F\{Q_E(\theta,\gamma,\kappa/a_K)\}.
\end{equation}
Under \eqref{S-regime}, for a fixed positive number of bootstrap replicates,
\begin{equation}\label{S-OOBlimit}
\begin{gathered}
 \widehat F_{\rm OOB}\pto F(Q_E^*),\qquad
 \widehat F_{.632}\pto (1-c_B)F(Q_A)+c_BF(Q_E^*).
\end{gathered}
\end{equation}
For AUC and Brier loss, \eqref{S-OOBlimit} also holds for conditional bootstrap expectations and any deterministic positive number of replicates.
\end{theorem}

\begin{proof}
\OptProofStep{1}{Independent held-out rows}
With $a_Kn+O(1)$ independent training observations per fold, by Theorem~\ref{F-thm-exist} and the unit-weight case of Theorem~\ref{S-state} we have a finite fit and the state at $\kappa/a_K$. Conditional on this fit, held-out scores are independent and Gaussian given $Z$, and from bounded-test averages, conditional second moments, and Proposition~\ref{S-functional} we obtain the performance limit in each fold. As fold weights tend to $1/K$, we use a union bound over fixed $K$ and get $\widehat F_{\rm CV}\pto F\{Q_E(\theta,\gamma,\kappa/a_K)\}$.

\OptProofStep{2}{Out-of-bag rows}
Given the counts and selected rows, the omitted observations are independent of the fit and their fraction tends to $e^{-1}$, and by Theorem~\ref{S-empirical} and Proposition~\ref{S-functional} we obtain \eqref{S-OOBlimit} for fixed replicate counts. For bounded AUC and Brier loss, from Step~2 of the proof of Theorem~\ref{S-Harrell} we get the same limit for arbitrary deterministic counts and conditional expectations.

\OptProofStep{3}{Bias relative to the original fit}
By Theorem~\ref{S-empirical} and Proposition~\ref{S-functional}, we have $F_{\rm ext}\pto F(Q_E)$, and from its difference with \eqref{S-OOBlimit} we obtain the limits of the estimation errors.

At $\theta=\gamma=0$, from \eqref{S-OOBlimit} and the Brier expansions in Step~2 of the Proposition~\ref{S-null} proof, we get
\[
 \begin{aligned}
 B(Q_E^*)-B(Q_E)
 &=\frac{\kappa}{4}+\frac{\kappa^2}{4}+O(\kappa^3),\\
 (1-c_B)B(Q_A)+c_B B(Q_E^*)-B(Q_E)
 &=\frac{3c_B-2}{4}\kappa
   +\frac{2c_B-1}{4}\kappa^2+O(\kappa^3).
 \end{aligned}
\]
Because $3c_B-2=1-3/e<0$ and $h(0,0)=1/2$, the two leading terms have opposite signs, and for each sufficiently small fixed $\kappa<c_B/2$, by continuity these signs and $\kappa<c_Bh(0,\gamma)$ hold on $0<\gamma<\epsilon(\kappa)$ for some $\epsilon(\kappa)>0$ at $\theta=0$. By \eqref{S-OOBlimit}, we obtain nonzero bias in this setting for every deterministic $B_n\ge1$. This completes the proof.
\end{proof}

\subsection{Location-shifted bootstrap intervals}\label{S-ls-intervals}
We use the empirical bootstrap quantiles $q^*_{u,F}$ and location-shifted interval $I_F^{\rm LS}$ defined in Section~\ref{subsec:optboot-procedures}.

\begin{proof}[Proof of Proposition~\ref{prop:optboot-ls}]
We define $x_F=F(Q_A^*)$, and by Theorem~\ref{S-empirical} and Proposition~\ref{S-functional}, we obtain convergence of one bootstrap apparent evaluation to $x_F$. For every deterministic $B_n\ge1$, the expected fraction outside its $\epsilon$-neighborhood is the single-refit probability and tends to zero, as does the expected failure fraction $p_n$. By Markov's inequality, we get the same limits among successful refits after division by $1-f_n$. Thus, both empirical and interpolated quantiles tend to $x_F$. From Theorem~\ref{S-Harrell} and its successful-refit comparison, we obtain the optimism limit $x_F-F(Q_U^*)$, and hence we have
$(\inf I_F^{\rm LS},\sup I_F^{\rm LS})\pto(F(Q_U^*),F(Q_U^*)).$
Because $F_{\rm ext}\pto F(Q_E)$, coverage tends to zero if $F(Q_U^*)\ne F(Q_E)$. We observe by continuity at fixed small $\kappa$ that the strict signs in Proposition~\ref{S-null} hold for Brier and AUC gaps at small positive signals. This completes the proof.
\end{proof}

\subsection{Bootstrap calibration shrinkage}\label{S-shrinkage}
Under bootstrap evaluation on the original sample, the free calibration slope converges to $S_B=c(Q_U^*)$. Note that when $S_B>0$ and we multiply fitted slopes by a consistent estimator of $S_B$, we get limiting external slope $c(Q_E)/S_B$ from the bijection $(a,c)\mapsto(a+c(1-S_B)b_0,cS_B)$ in the population calibration objective. At the null, we have zero external slope and $S_B=1/2+O(\kappa)$ by Proposition~\ref{S-null}, and hence this shrinkage coefficient is inconsistent. Specifically, for small positive signals in \eqref{S-nearnull}, we combine $S_B-c(Q_E)=\Delta_c>0$ in \eqref{S-Hcal} with continuity and \eqref{S-nullslope} to get $S_B>0$ and $c(Q_E)/S_B<1$.

\section{Single-fit estimation and joint inference}\label{S-onefit}
\subsection{Consistency of the adjusted scores}
We assume $\gamma>0$ and $0<\kappa<h(\theta,\gamma)$.

We use the ordinary fitted scores $\widehat\eta_i$ of the original fit to define
\begin{equation}\label{S-augmented}
 \frac1n\sum_i\frac{\widehat\lambda s'(\widehat\eta_i)}{1+\widehat\lambda s'(\widehat\eta_i)}
 =\kappa_n,\qquad
 \widetilde V_i=\widehat\eta_i+\widehat\lambda(s(\widehat\eta_i)-Y_i),\qquad
 \widehat Q_{\rm aug}=\frac1n\sum_i\delta_{(Y_i,\widetilde V_i)}.
\end{equation}
Let $\widehat F=F(\widehat Q_{\rm aug})$, with realized target $F_{\rm ext}$.

\begin{theorem}\label{S-augconsistent}
In the strict interior for the ordinary fit with $\gamma>0$, for every functional in Proposition~\ref{S-functional},
\begin{equation}\label{S-augconsistency}
 \widehat\lambda\pto\lambda_0,\qquad
 \widehat F\pto F(Q_E),\qquad
 \widehat F-F_{\rm ext}\pto0.
\end{equation}
\end{theorem}

\begin{proof}
\OptProofStep{1}{The trace equation}
We let $g_n(l)$ denote the empirical trace in \eqref{S-augmented}, and we obtain $g_n(l)\pto g(l)=\E\{l s'(T_0)/(1+l s'(T_0))\}$ for each $l>0$ from Theorem~\ref{S-empirical}. By strict monotonicity and $g(\lambda_0)=\kappa$, for $0<\epsilon<\lambda_0$ we get
\[
 \begin{aligned}
 g_n(\lambda_0-\epsilon)&\pto g(\lambda_0-\epsilon)<\kappa,\\
 g_n(\lambda_0+\epsilon)&\pto g(\lambda_0+\epsilon)>\kappa .
 \end{aligned}
\]
We have $g_n(\widehat\lambda)=\kappa_n\to\kappa$ on finite fits with $\Prob(\text{finite fit})\to1$, and hence we obtain $\widehat\lambda\pto\lambda_0$ from this bracket.

\OptProofStep{2}{The corrected distribution and its functionals}
Adjustment at $\lambda_0$ transforms $(Y,T_0)$ into $(Y,V_0)$ by the proximal identity, and substitution of $\widehat\lambda$ changes each score by at most $|\widehat\lambda-\lambda_0|=o_P(1)$. We then apply Theorem~\ref{F-thm-exist} and \eqref{S-scoreL2} to obtain
\[
 P_n|\widetilde V_i|\ind_{\{|\widetilde V_i|>M\}}
 \le M^{-1}P_n\widetilde V_i^2
 \le 2M^{-1}(P_n\widehat\eta_i^2+\widehat\lambda^2)=O_P(M^{-1}).
\]
We obtain log-loss convergence by truncation, and we get $\widehat F\pto F(Q_E)$ from bounded Brier loss and the AUC and calibration arguments in Proposition~\ref{S-functional}. Here, $F_{\rm ext}$ has the same limit by the independent-row representation, and hence \eqref{S-augconsistency} holds. This completes the proof.
\end{proof}

\subsection{Joint normal approximation for realized losses}\label{S-xa-section}

The independent rows in both designs have a fixed-dimensional component $Z_i\in\R^{d_0}$ and a Gaussian component $W_i\sim N_{p-d_0}(0,I)$ independent of $(Z_i,Y_i)$, where $\Prob(Y_i=1\mid Z_i)=s(\theta_n+\varpi_n^\top Z_i)$. In the Gaussian design \eqref{S-coordinates}, we have $d_0=1$, $Z_i\sim N(0,1)$, and $\varpi_n=\gamma_n>0$, and $\zeta_n=(\kappa_n,\theta_n,\gamma_n)$ belongs to a fixed compact set $\mathcal K$ in $0<\kappa<h(\theta,\gamma)$ with $\gamma>0$. On the other hand, in the design \eqref{S-NG-eq-head}--\eqref{S-NG-eq-finitehead}, we have $d_0=m$ and $Z_i$ has the fixed standardized distribution $P_Z$, and $\zeta_n=(\kappa_n,\theta_n,\varpi_n)$ belongs to a fixed compact set $\mathcal K$ in $0<\kappa<h_Z(\theta,\varpi)$ for $h_Z$ defined in \eqref{S-NG-eq-headboundary}.

Let $A$ and $W$ have rows $(1,Z_i^\top)$ and $W_i^\top$, respectively, and let $B=[A,W]$. For the unpenalized fit with these covariates, let $\widehat u_n=(\widehat b,\widehat t^\top)^\top$ and $\widehat a_n=(\widehat u_n^\top,\widehat w^\top)^\top$. We evaluate population quantities at $\zeta_n$, with $\kappa_n=p/n$. Let $(b,t,r,\lambda)$ be the unit-weight state, $u_*=(b,t^\top)^\top$, and $G\sim N(0,1)$ independent of $(Z,Y)$. Define
\begin{equation}\label{S-xa-state}
 V=b+t^\top Z+rG,\qquad T+\lambda\{s(T)-Y\}=V,\qquad
 q=s(T)-Y,\quad d=s'(T),\quad\omega=\frac{d}{1+\lambda d}.
\end{equation}
The state satisfies \eqref{S-state-eq} or \eqref{S-NG-eq-newstate}, respectively. For the first-order fit, score, and trace-root limits we use Theorems~\ref{S-state}, \ref{S-empirical}, and~\ref{S-augconsistent} in the Gaussian design and Theorem~\ref{S-NG-thm-head}(a) in the second design.
Define
$\ell_B(y,v)=\{s(v)-y\}^2$ and $\ell_L(y,v)=\ell_R(y,v)=\log(1+e^v)-yv$. Here $L$ indexes log loss, denoted by $R$ in Section~\ref{subsec:optboot-logistic}.
Let $z_a=\Phi^{-1}(a)$ denote the standard normal $a$-quantile.
For $F,H\in\{B,L\}$ define
\begin{equation}\label{S-xa-integrals}
 \begin{gathered}
 m_F=\E\ell_F(Y,V),\qquad
 a_F=\E\{q\ell_F'(Y,V)\},\qquad
 J=\E\frac{d}{(1+\lambda d)^2},\\
 \Sigma_{FH}=\operatorname{Cov}\{\ell_F(Y,V),\ell_H(Y,V)\}
                       +\frac{\lambda a_Fa_H}{J}.
 \end{gathered}
\end{equation}

With $\widehat\lambda$ and $\widetilde V_i$ from \eqref{S-augmented}, let $P_n$ denote the empirical average and define $q_i=s(\widehat\eta_i)-Y_i$, $d_i=s'(\widehat\eta_i)$, and $\widehat C_F=P_n\ell_F(Y_i,\widetilde V_i)$. The realized risk is
$R_{F,n}=\E_{Z,Y,G}\ell_F(Y,\widehat b+\widehat t^\top Z+\|\widehat w\|G),$
where the expectation is over an independent new observation. Thus $(\widehat C_B,\widehat C_L)=(\widehat B,\widehat R)$ and $(R_{B,n},R_{L,n})=(B_{\rm ext},R_{\rm ext})$ in Sections~\ref{subsec:optboot-logistic} and~\ref{sec:optboot-onefit}. For the finite-$n$ covariance $\Sigma=\Sigma(\zeta_n)$, a normal limit denotes Gaussian approximation in the bounded Lipschitz metric.

\begin{theorem}\label{S-xa-clt}
Uniformly for $\zeta_n\in\mathcal K$,
\[
 \sqrt n(\widehat C_B-R_{B,n},\widehat C_L-R_{L,n})^\top
       \Longrightarrow N_2(0,\Sigma),\qquad \Sigma\succ0.
\]
\end{theorem}

\subsection{Restricted fits and transformed losses}

Define the transformed losses
\begin{equation}\label{S-xa-tilts}
 \tau(x)=\frac{\lambda s'(x)}{1+\lambda s'(x)},\qquad
 \varrho_F=\frac{a_F}{J},\qquad
 H_F(y,x)=\ell_F\bigl(y,x+\lambda\{s(x)-y\}\bigr)-\varrho_F\tau(x).
\end{equation}
We hold $\lambda,\varrho_F$ in these functions fixed in every auxiliary parameter differentiation below, and define

\[
 \mathcal R_F(b',t',v)=\E\ell_F(Y,b'+(t')^\top Z+\sqrt vG),\qquad
 f_{F,v}=\partial_v\mathcal R_F(b,t,r^2)
         =\tfrac12\E\ell_F''(Y,V).
\]

Let $w_n^0$ be the original restricted fit at $u_*$, and we define
$\eta_i^0=b+t^\top Z_i+W_i^\top w_n^0$.
With a fixed radius bound larger than the localization radius, we assign zero to each restricted coefficient outside its own localization event. Then, the truncated coefficient agrees with the actual restricted coefficient with probability tending to one. Next, we define
\begin{equation}\label{S-xa-core}
 X_{F,n}=\frac1{\sqrt n}\sum_i\{H_F(Y_i,\eta_i^0)-m_F+\varrho_F\kappa_n\},\qquad
 T_n=\sqrt n(\|w_n^0\|^2-r^2).
\end{equation}

\begin{proof}[Proof of Theorem~\ref{S-xa-clt}]
By Proposition~\ref{S-xa-head-cancellation}, for $F=B,L$,
\[
 \sqrt n(\widehat C_F-R_{F,n})=X_{F,n}-f_{F,v}T_n+o_P(1).
\]
Lemma~\ref{S-xa-restricted-kernel} gives the joint normal approximation for these projections, with covariance \eqref{S-xa-integrals} by Proposition~\ref{S-xa-covariance-identity}.

Note that $V$ has positive density on $\R$ given $Y=0$. When $a s(v)^2+b\log(1+e^v)$ is constant for all $v$, we get $b=0$ by division as $v\to\infty$ and $a=0$ by nonconstancy of $s(v)^2$. Hence, both the loss covariance and its sum with $\lambda(a_Fa_H)/J$ are positive definite.

We obtain continuity of $\Sigma$ and a positive minimum eigenvalue on $\mathcal K$ from sub-Gaussian moments, state continuity, and the positive lower bounds on $\lambda,J$. Along each $\zeta_n\to\zeta_0$, both characteristic functions converge to $\exp\{-a^\top\Sigma(\zeta_0)a/2\}$, and we get uniformity from compactness. This completes the proof.
\end{proof}

\subsection{Gaussian variation of the restricted fit}

For the vector $\epsilon=(\epsilon_B,\epsilon_L)$, we next define
$L_\epsilon(y,x)=\ell_L(y,x)+\epsilon_B H_B(y,x)+\epsilon_L H_L(y,x).$
Define
$\xi=(u,\rho,\epsilon)$, where $u=(b',(t')^\top)^\top$ and $\rho\ge0$, and define
\begin{align}
 f_n(\xi)&=\min_w\left\{P_n L_\epsilon(Y_i,(1,Z_i^\top)u+W_i^\top w)
                                    +\tfrac12\rho\|w\|^2\right\},\notag\\
 \mathcal A(\xi,r',l)&=\E M_l L_\epsilon(Y,b'+(t')^\top Z+r'G)
                       -\frac{\kappa_n(r')^2}{2l}+\frac{\rho(r')^2}{2},
                       \label{S-xa-perturbed-value}
\end{align}
where $M_l L(y,v)=\min_x\{L(y,x)+(x-v)^2/(2l)\}$.
Let $w_n(\xi)$ denote the minimizer defining $f_n(\xi)$.
Let $(r(\xi),l(\xi))$ be the stationary solution of $\mathcal A$ near
$(r,\lambda)$, and define $\Phi(\xi)=\mathcal A(\xi,r(\xi),l(\xi))$.
When the loss perturbation is zero, define $w_n(u,\rho)=w_n(u,\rho,0)$.
At any parameter point in this fixed neighborhood, including at zero perturbation, define $q_\xi(y,v)=L_\epsilon'(y,T)$ by
$T+l(\xi)L_\epsilon'(y,T)=v$. Let $T_\xi+l(\xi)L_\epsilon'(Y,T_\xi)=b'+(t')^\top Z+r(\xi)G$. Let $R_i=(Z_i,Y_i)$ denote the $i$th sample row, and for a generic row $R=(z,y)$ define
\[
\begin{gathered}
 m_R(\xi)=\E_G M_{l(\xi)}L_\epsilon(y,b'+(t')^\top z+r(\xi)G),\\
 \mathcal D=(\partial_{\epsilon_B},\partial_{\epsilon_L},2\partial_\rho)^\top,
 \qquad h(R)=\mathcal Dm_R-\E\mathcal Dm_R.
\end{gathered}
\]
We evaluate derivatives at $\xi_*=(u_*,0,0)$ when no other evaluation point is specified, and we use right derivatives in $\rho$ at zero.

For two parameter points $\xi,\widetilde\xi$, let $(N,\widetilde N)$ be
centered Gaussian with variances $r(\xi)^2,r(\widetilde\xi)^2$ and covariance
$\eta k$, independent of $R$. Define
\[
 Q_\eta(k;\xi,\widetilde\xi)
 =\E\{q_\xi(Y,b'+(t')^\top Z+N)
          q_{\widetilde\xi}(Y,\widetilde b+\widetilde t^\top Z+\widetilde N)\}.
\]
Let $\mathcal D_{\widetilde\xi}=(\partial_{\widetilde\epsilon_B},\partial_{\widetilde\epsilon_L},2\partial_{\widetilde\rho})^\top$ denote the same derivative operator as $\mathcal D$ acting on the second parameter point $\widetilde\xi$. For $0\le\eta\le1$ let $k_\eta$ be the solution near the unperturbed solution
of $k_\eta=\eta l(\xi)l(\widetilde\xi)Q_\eta(k_\eta)/\kappa_n$, and define
\begin{equation}\label{S-xa-value-kernel}
 \mathcal H(\xi,\widetilde\xi)
   =\int_0^1 k_\eta Q_\eta(k_\eta;\xi,\widetilde\xi)\,d\eta,
 \qquad \Gamma=\mathcal D\mathcal D_{\!\widetilde\xi}^{\top}\mathcal H.
\end{equation}

To control the derivatives of the restricted fit, let $w=w_n(\xi)$ be its minimizer and define
\[
 \eta_i=(1,Z_i^\top)u+W_i^\top w,\qquad
 D=\operatorname{diag}L_\epsilon''(Y_i,\eta_i),\qquad
 \mathscr H=W^\top D W/n+\rho I.
\]
For a parameter direction $a$, we define $w_a=\partial_a w$ and $w_{aa}=\partial_a^2w$.
Let $h$ consist of the $d_0+1$ components of $a$ corresponding to $u$, and let $\nu$ denote a parameter multi-index.

\begin{lemma}\label{S-tilted-geometry}
Fix $M>0$. For all sufficiently large $n$, with probability at least $1-C_Mn^{-M}-Ce^{-cn}$, uniformly on a sufficiently small fixed neighborhood of $\xi_*$ with $\rho\ge0$, the minimizer $w$ is unique and, for every admissible unit direction $a$,
\[
\begin{aligned}
 &\|w\|\le C,\qquad \|\mathscr H^{-1}\|\le C,\qquad
   \|w_a\|\le C,\qquad \|w_{aa}\|\le C_M\sqrt{\log(en)},\\
 &\max_i|(1,Z_i^\top)h+W_i^\top w_a|
       +\max_i|W_i^\top\mathscr H^{-1}w|\le C_M\sqrt{\log(en)},\\
 &|\partial^\nu f_n(\xi)|\le
 \begin{cases}
 C_M\sqrt{\log(en)},&|\nu|=3,\\
 C_M\log(en),&|\nu|=4.
 \end{cases}
\end{aligned}
\]
These conclusions also hold after any fixed number of row deletions and during insertion with weights in $[0,1]$, with fixed normalization $n$.
\end{lemma}
\begin{proof}
We define $L_n=\log(en)$, and from $|s^{(k)}|\le C_ks'$ and $s'(x+v)\le e^{|v|}s'(x)$, we differentiate \eqref{S-xa-tilts} to get $|H_F'|\le C$ and $|H_F^{(j)}|\le C_js'$ for $j\ge2$. Here, the first two augmented-loss derivatives at $x+\lambda(s-y)$ are $\ell_F'(1+\lambda s')$ and $\ell_F''(1+\lambda s')^2+\lambda\ell_F's''$, and $\ell_F''$ contains $s'$. Thus, we have $|H_B|+|H_L-\ell_L|\le C$, and on a small perturbation box,
\[
\tfrac12s'(x)\le L_\epsilon''(y,x)\le C s'(x),\qquad
 |L_\epsilon'|\le C,
 \qquad |L_\epsilon^{(j)}|\le C_j s'(x)\quad(j\ge2).
\]
With bounded $L_\epsilon-(1+\epsilon_L)\ell_L$, we use Lemma~\ref{S-NG-lem-localization} to localize all fixed-deletion fits, and from \eqref{S-NG-eq-geo-curvature} and $L_\epsilon''\ge s'/2$ we get curvature during insertion. For $d_0>0$, we apply H\"older's inequality in \eqref{S-NG-eq-head} to obtain
\[
 \E e^{\|Z\|^2/(d_0K_Z^2)}\le2,\qquad
 \Prob\{\max_i\|Z_i\|>C_a\sqrt{L_n}\}\le C_an^{-a}.
\]
For each fixed polynomial $f$, we have $\E|P_n f(Z_i)-\E f(Z)|^{2j}=O(n^{-j})$ because each nonzero summand has at most $j$ distinct row indices. In the case $d_0=0$, $A$ consists only of the intercept column.

For $q_i=L_\epsilon'(Y_i,\eta_i)$ and a direction with component $h$ for $(b,t)$, penalty component $c$, and loss component $g=\epsilon'_BH_B+\epsilon'_LH_L$, we define $w_a=\partial_aw$ and $E_a=Ah+Ww_a$ and differentiate implicitly to get
\begin{align*}
\mathscr H w_a=-n^{-1}W^\top(DAh+g')-cw,\quad
 \mathscr H w_{aa}=-n^{-1}W^\top
       \{L_\epsilon''' E_a^2+2g''E_a\}-2cw_a
\end{align*}
with entrywise products, and by the design bounds we have $\|w_a\|\le C$ and $\|E_a\|_2\le C\sqrt n$. Here, subscript zero denotes deletion of the incoming row $(z,y,u)$ with $u\in\R^{p-d_0}$ at fixed divisor $n$, and by bounded deleted coefficients and conditional Gaussian tails, it holds that
\[
|u^\top w_0|+|u^\top w_{a,0}|+|u^\top\mathscr H_0^{-1}w_0|
 \le C_A\sqrt{L_n},\qquad
 |u^\top\mathscr H_0^{-1}W_j|\le C_A\sqrt{nL_n}.
\]
We apply a union bound over rows, row pairs, and at most $Cn^{C_{d_0}}$ grid points of mesh $n^{-4}$ to obtain a failure probability of $C_A n^{-A}$. Note that from Gaussian quadratic-form tails, we also bound $l_i=u^\top\mathscr H_0^{-1}u/n$ in a fixed compact interval.

\begingroup\setlength{\emergencystretch}{1em} Let $T$ solve $T+l_iq_T=b'+(t')^\top z+u^\top w_0$ with $q_T=L_\epsilon'(y,T)$, $d_T=L_\epsilon''(y,T)$, and $\Delta=-q_T\mathscr H_0^{-1}u/n$. We observe from the projection bounds that $\|W_0\Delta\|_2\le C$ and $\|W_0\Delta\|_\infty\le C_A\sqrt{L_n/n}$, and from curvature we get $\|w-w_0-\Delta\|\le C_A\sqrt{L_n}/n$ because the score has a quadratic remainder of at most $C_A\sqrt{L_n}/n$.\par\endgroup

Next, with $\mathscr H_i^0=\mathscr H_0+d_Tuu^\top/n$, we define
\[
 v_a^0=w_{a,0}-(\mathscr H_i^0)^{-1}\frac{u}{n}
 \{d_T[(1,z^\top)h+u^\top w_{a,0}]+g'(y,T)\}.
\]
Specifically, by rank-one inversion, it holds that
\[
 (1,z^\top)h+u^\top v_a^0
 =\frac{(1,z^\top)h+u^\top w_{a,0}-l_i g'(y,T)}{1+d_Tl_i}
 =O(\sqrt{L_n})
\]
on the projection event. The retained predictor increment, $\|Ah\|_2\le C\sqrt n$ and bounded derivatives of $g$ bound the Hessian and derivative-equation changes by $C_A\sqrt{L_n/n}$. The incoming changes multiplied by $u/n$ have the same order, and $\|c(w-w_0)\|\le C_An^{-1/2}$. Inverting the Hessian gives $\|w_a-v_a^0\|\le C_A\sqrt{L_n/n}$, hence $\max_i|E_{a,i}|\le C_A\sqrt{L_n}$. For $v=\mathscr H^{-1}w$,
\[
u^\top(\mathscr H_i^0)^{-1}w_0
 =\frac{u^\top\mathscr H_0^{-1}w_0}{1+d_Tl_i}.
\]
Because the inverse and coefficient errors are bounded by $C_A\sqrt{L_n/n}$, we have $\max_i|W_i^\top\mathscr H^{-1}w|\le C_A\sqrt{L_n}$ and obtain $\|w_{aa}\|\le C_A\sqrt{L_n}$ from the second derivative equation.

By repeated implicit differentiation, we have polynomial derivative bounds, and from a finer polynomial mesh and Gaussian union bounds we obtain uniform failure probability $C_An^{-A}$ through fixed deletions and insertions of weights in $[0,1]$.

In direction $(h,c,g)$, we differentiate the profile value and apply the nuisance score equation to obtain
\begin{align*}
f_n^{(3)}&=P_n\{L_\epsilon^{(3)}E_a^3+3g''E_a^2\}+3c\|w_a\|^2,\\
 f_n^{(4)}&=P_n\{L_\epsilon^{(4)}E_a^4+4g^{(3)}E_a^3
             +3L_\epsilon^{(3)}E_a^2Ww_{aa}+6g''E_aWw_{aa}\}
             +6c\,w_a^\top w_{aa}.
\end{align*}
From $\|E_a\|_2\le C\sqrt n$, $\|E_a\|_\infty\le C_M\sqrt{L_n}$, and $\|w_{aa}\|\le C_M\sqrt{L_n}$, we obtain the cubic and quartic bounds $C_M\sqrt{L_n}$ and $C_ML_n$. In particular, \[
P_n|E_a^2Ww_{aa}|\le n^{-1}\|E_a\|_\infty\|E_a\|_2\|Ww_{aa}\|_2
 \le C_M L_n.
\]
For a symmetric $j$-linear form $B$ and coordinate vectors $z_1,\ldots,z_j$, we use polarization to get
\[
j!B(z_1,\ldots,z_j)
 =\sum_{S\subseteq\{1,\ldots,j\}}(-1)^{j-|S|}
        B\Big[\sum_{i\in S}z_i\Big]^j
\]
where we sum only over terms that involve all $j$ indices. On a smaller neighborhood with nonnegative penalty increments, all evaluation points remain in the domain, and hence from the bounds on the pure derivatives we obtain those on the mixed derivatives. This completes the proof.\end{proof}

For the empirical-average bounds, we consider $F$ with at most linear growth and bounded positive-order predictor derivatives through order six. We also allow bounded functions that are smooth in their arguments and the deterministic $l(\xi)$, where the constants $C_F$ may depend on $F$.

\begin{lemma}\leavevmode\label{S-xa-restricted-kernel}
For the restricted quantities in \eqref{S-xa-core}, uniformly on $\mathcal K$,
\begin{equation}\label{S-xa-core-clt}
 (X_{B,n},X_{L,n},T_n)^\top
       \ \Longrightarrow\ N_3(0,\E hh^\top+\Gamma).
\end{equation}
For each fixed $a\in\R^3$, at the moving finite-$n$ parameters,
\[
\begin{aligned}
 &\left|\E\exp\{ia^\top(X_{B,n},X_{L,n},T_n)^\top\}
 -\exp\{-a^\top(\E hh^\top+\Gamma)a/2\}\right|\\
 &\qquad\le C_a\left\{n^{-1/12}\log(en)^{1/3}
          +n^{-1/8}\sqrt{\log(en)}+\frac{\log(en)^4}{\sqrt n}\right\}.
\end{aligned}
\]
For each deterministic nearby $\xi$ and each $F$ specified above,
\[
 \operatorname{Var}\{P_nF(Y_i,\eta_i)\}\le\frac{C_F}{n},\qquad
 \left|\E P_nF(Y_i,\eta_i)-\E F(Y,T_\xi)\right|
 \le\frac{C_F\log(en)^4}{n}.
\]
\end{lemma}
\begin{proof}
The constants $C,C_q$ depend on $\mathcal K$, the fixed perturbation neighborhood, and the derivative or moment order, and we define $L_n=\log(en)$.

\OptProofStep{1}{Geometry, derivatives and moment bounds}
From Lemma~\ref{S-tilted-geometry}, we obtain the coefficient, rowwise and profile bounds, and for $\rho\ge n^{-2}$, we have $\|\mathscr H^{-1}\|\le n^2$ globally. By objective comparison and implicit differentiation, we bound the coefficients and fixed derivatives polynomially in $n$ and the design norm. For any fixed $m$, we choose $A$ large and apply H\"older's inequality with failure probability $C_An^{-A}$ to get exceptional contributions $O(n^{-m})$.

On localization, from the integrated score equations we get coefficient error $Cn^{-2}$ between penalties zero and $n^{-2}$, and hence $L^2$ error $Cn^{-3/2}$ for scaled coordinates and conditional means. We truncate coefficients and radii on each array's localization event and integrate incoming rows before we condition on full-array localization. Then, we differentiate the ridge quantities before the unpenalized limit.

At the state, we have $\mathcal A_{ll}=-\E\{q^2/(1+\lambda d)\}/\lambda<0$, and its radial Schur complement is positive by \eqref{S-popHessian} or \eqref{S-NG-eq-populationhessian}. By the implicit function theorem, on a smaller box we obtain smooth $r(\xi),l(\xi)$ that have bounded fixed-order derivatives and satisfy
\begin{equation}
\label{S-xa-tilted-state}
 l(\E\omega+\rho)=\kappa_n,\qquad l^2\E q^2=\kappa_n r^2,
 \qquad \omega=\frac{L_\epsilon''(Y,T)}{1+lL_\epsilon''(Y,T)}.
\end{equation}
In Steps~2--6, $r,l,q,\omega$ denote this perturbed state.

\OptProofStep{2}{Insertion moments and concentration}
We delete $(z,y,u)$ at fixed divisor $n$ and let $B_0=\mathscr H_0^{-1}$, $l_i=u^\top B_0u/n$, $v=b'+(t')^\top z+u^\top w_0$, and $T+l_iq_T=v$ with $q_T=L_\epsilon'(y,T)$. Note that at $\Delta=-q_TB_0u/n$, we have incoming predictor $T$ and the incoming and linearized retained scores cancel with remainder at most
$\frac Cn\|W_0\|\left(\sum_{j\ne i}|W_j^\top\Delta|^4\right)^{1/2}.$
For deleted vectors $a_j$ with $\|a_j\|\le C\sqrt n$, it holds by Gaussian moments, the $L^{q/2}$ triangle inequality, and a Gaussian tail union bound that
\begin{equation}
\label{S-xa-gaussian-powers}
 \left\|\left(\sum_j|a_j^\top u/n|^{2k}\right)^{1/2}\right\|_{L^q}
       \le C_{q,k}n^{(1-k)/2},\qquad
 \left\|\max_j|a_j^\top u|\right\|_{L^q}\le C_q\sqrt{nL_n}.
\end{equation}
Specifically, by the first bound with $a_j=B_0W_j$, $k=2$, and curvature, it holds that
\[
 w_1=w_0+\Delta+e,\qquad
 \|\Delta\|_{L^q}\le C_qn^{-1/2},\quad
 \|e\|_{L^q}\le C_q/n,\quad
 \|g^\top\Delta\|_{L^q}\le C_q/n
\]
for bounded deleted $g$, because $|q_T|\le C$ and $g^\top B_0u$ is conditionally Gaussian. For insertion weight $s$, replace $l_iq_T,\Delta$ by $sl_iq_T,s\Delta$.

We define $M=\|w\|^2$, $Q_2=P_nq_i^2$, $\Lambda=\operatorname{tr}\mathscr H^{-1}/n$, $P=D-DW\mathscr H^{-1}W^\top D/n$, and $\tau_n=\operatorname{tr}P/n$. By expansion of the squared norm, we get row-replacement change $O_{L^q}(n^{-1})$ in $M$. For $F$ with two bounded predictor derivatives, we let $(F_i')$ denote its derivatives at the retained scores $\eta_{0j}=b'+(t')^\top Z_j+W_j^\top w_0$. Then, its retained average changes by $g^\top(\Delta+e)+R$, where
\[
g=W_0^\top(F_i')/n,\quad \|g\|\le C,\qquad
 |R|\le C\|W_0(\Delta+e)\|^2/n.
\]
By the insertion bounds, we have $\|g^\top\Delta\|_{L^q}+\|g^\top e\|_{L^q}+\|R\|_{L^q}\le C_q/n$. From $|T|\le|v|+Cl_i$ and $\|u^\top e\|_{L^q}\le\|u\|_{L^{2q}}\|e\|_{L^{2q}}\le C_qn^{-1/2}$, we also get bounded incoming moments for linearly growing $F$.

For $d_i=L_\epsilon''(Y_i,\eta_i)$ and $B_*=(\mathscr H-d_iuu^\top/n)^{-1}$ at a fixed coefficient vector, when we apply rank-one insertion, the inverse changes by
$-\frac{d_iB_*uu^\top B_*}{n(1+d_i u^\top B_*u/n)},$
with normalized trace $O_{L^q}(n^{-1})$, and from \eqref{S-xa-gaussian-powers} we obtain $\|E\|_{L^q}\le C_q\sqrt{L_n/n}$ for the Hessian perturbation on the remaining rows. In
$(\mathscr H_0+E)^{-1}=B_0-B_0EB_0+B_0E(\mathscr H_0+E)^{-1}EB_0,$
the quadratic trace is at most $C\|E\|^2=O_{L^q}(L_n/n)$, and the linear trace is $-g^\top(\Delta+e)$ for $g=n^{-2}\sum_{j\ne i}L_\epsilon'''(Y_j,\eta_{0j})(W_j^\top B_0^2W_j)W_j$, which we bound via the design norm. Because the Taylor remainder is no more than $CP_n|W_j^\top(\Delta+e)|^2=O_{L^q}(n^{-1})$, we get change $O_{L^q}(L_n/n)$ in $\Lambda$. For $M_*=W\mathscr H^{-1}W^\top/n$, $L_*=I-M_*D$ and diagonal $E,F$,
\[
P'[E]=L_*^\top E L_*,\qquad L_*'[E]=-M_*EL_*,\qquad
 |\operatorname{tr}P''[E,F]|/n\le C\|E\|_F\|F\|_F/n.
\]
From Frobenius Cauchy--Schwarz, we obtain the last inequality, and from $D$ we bound the retained-row predictor-gradient entries and Hessian norm of $\tau_n$ by $C/n$. Because $\operatorname{tr}P$ changes by $O_{L^q}(1)$ under rank-one insertion, we bound the row-replacement change in $\tau_n$ by $O_{L^q}(n^{-1})$.

For $0\le\eta\le1$, we define $\widetilde W=\eta W+\sqrt{1-\eta^2}W'$, where $W'$ is standard Gaussian and independent of $W,(Z_i,Y_i)_{i\le n}$. At $\widetilde\xi$, we use the same fixed coordinates and labels, and each incoming marginal is independent of both deleted arrays. By H\"older's inequality and the coefficient bounds, we bound the row-replacement change in $K=w^\top\widetilde w$ by $O_{L^q}(n^{-1})$, and from orthogonal decomposition by independent row subsets, we obtain
$\operatorname{Var}U\le\tfrac12\sum_i\E|U-U^{(i)}|^2,$
where in $U^{(i)}$ we replace row $i$ by an independent copy. Because each nonconstant subset variance appears at least once on the right, we obtain
\begin{equation}
\label{S-xa-replacement-bounds}
 \operatorname{Var}M+\operatorname{Var}Q_2+\operatorname{Var}\tau_n
       +\operatorname{Var}K\le C/n,\qquad
 \operatorname{Var}\Lambda\le CL_n^2/n.
\end{equation}

\OptProofStep{3}{Signed insertion and finite-$n$ centering}
Let $a_j=B_0W_j$ and $c_j=L_\epsilon'''(Y_j,\eta_{0j})$. For $Z_3=n^{-3}\sum_{j\ne i}c_j(a_j^\top u)^3$, \eqref{S-xa-gaussian-powers} implies $\|Z_3\|_{L^q}\le C_qn^{-1/2}$. The full $n$-normalized penalized score $g_0$ at $w_0+\Delta$ has second-order remainder $R_0$, with
\[
g_0=\frac1{2n}W_0^\top\{c_j(W_j^\top\Delta)^2\}_{j\ne i}+R_0,
 \qquad \|R_0\|_{L^q}\le C_qn^{-3/2}
\]
by \eqref{S-xa-gaussian-powers} with $k=3$, and for $k=2$, we have $\|g_0\|_{L^q}\le C_q/n$. When we replace the integrated Hessian by $\mathscr H_0+d_Tuu^\top/n$ with $d_T=L_\epsilon''(y,T)$, the error is at most $C_q\sqrt{L_n/n}\|g_0\|_{L^{2q}}\le C_q\sqrt{L_n}n^{-3/2}$. Because the term quadratic in $e$ has the same order, by rank-one inversion, we obtain
\begin{equation}
\label{S-xa-signed-predictor}
 u^\top e=-\frac{q_T^2}{2(1+d_Tl_i)}Z_3+R_\eta,
 \qquad \|R_\eta\|_{L^q}\le C_qL_n/n
\end{equation}
because by H\"older's inequality, we bound the corresponding predictor error by $C_q\sqrt n\,\sqrt{L_n}n^{-3/2}\le C_qL_n/n$.

For admissible $F(v,l_i)$ with a fixed trace argument, we apply three Gaussian integrations by parts and obtain
\begin{equation}\label{S-xa-signed-gaussian}
 \E\{(a^\top u)^3F(c+w^\top u)\}
   =3\|a\|^2(a^\top w)\E F'+(a^\top w)^3\E F'''
\end{equation}
We define $l_0=\operatorname{tr}B_0/n$, and from $\|l_i-l_0\|_{L^q}\le C_qn^{-1/2}$, the cubic moment bound, and Cauchy--Schwarz, we replace $l_i$ by $l_0$ with error at most $C/n$. Conditional on the deleted array and $(z,y)$, we use \eqref{S-xa-signed-gaussian} and
\[
\max_j\|a_j\|^2\le Cn,\qquad
 \sum_j|a_j^\top w_0|\le Cn,\qquad
 \sum_j|a_j^\top w_0|^3\le Cn^{3/2}
\]
to obtain $|\E\{F(v,l_i)Z_3\}|\le Cn^{-3}(n^2+n^{3/2})+C/n\le C/n$. Here, implicit derivatives have denominator at least one and polynomial numerators in $l$, and by quadratic-form tails and H\"older we bound the complement of the compact trace interval.

By expansion of $\eta_i=T+u^\top e$ in \eqref{S-xa-signed-predictor}, the signed bound, and $\E|u^\top e|^2\le C/n$, we get mean error $CL_n/n$. For the trace argument, we use two Gaussian integrations by parts to obtain
\[
\E\{(u^\top B_0u-\operatorname{tr}B_0)F(c+w_0^\top u)\}
 =w_0^\top B_0w_0\,\E F''.
\]
We thus get $|\E\{(l_i-l_0)F(v)\}|+\E(l_i-l_0)^2\le C/n$, and on the incoming row this conditional mean is smooth in $m_0=\|w_0\|^2,l_0,z,y$ and at $m=0$, where $\partial_m\E F(c+\sqrt mG)=\E F''(c+\sqrt mG)/2$. We next use \eqref{S-xa-replacement-bounds} at $(\E m_0,\E l_0)$ to bound the expansion by $CL_n^2/n$ and deletion changes by $CL_n/n$. For linearly growing tests and polynomial factors in $z$, we apply the positive-order derivative bounds to get
\begin{equation}\label{S-xa-signed-test}
\begin{gathered}
 \left|\E\{\chi(Z_i)F(Y_i,\eta_i,l_i)\}
    -\E\{\chi(Z)F(Y,T(b'+(t')^\top Z+\sqrt mG;l),l)\}\right|
       \le CL_n^4/n,\\
 m=\E M,\qquad l=\E\Lambda,
\end{gathered}
\end{equation}
for fixed polynomially growing $\chi$.

For $\ell_i=W_i^\top\mathscr H^{-1}W_i/n$ and $d_i=L_\epsilon''(Y_i,\eta_i)$, we use the Step~2 inverse expansion to get
\[
\ell_i=\frac{l_i}{1+d_i l_i}
          +\frac{q_T}{(1+d_i l_i)^2}Z_3+R_{\ell,i},
 \qquad \E|R_{\ell,i}|\le CL_n/n
\]
where \eqref{S-xa-gaussian-powers} and the insertion moments bound the four remainder terms by
\[
\begin{aligned}
 \E[C\|u\|^2\|E\|^2/n]&\le CL_n/n,\quad
 \E[Cn^{-2}\max_j|a_j^\top u|^2\sum_j|W_j^\top e|]\le CL_n/n,\\
 \E[Cn^{-4}\sum_j|a_j^\top u|^4]&\le C/n,\quad
 \E[Cn^{-2}\max_j|a_j^\top u|^2\|W_0e\|^2]\le CL_n/n.
\end{aligned}
\]
We replace $d_i$ by $d_T$ in the cubic coefficient with error $C/n$ in expectation. For $T+lL_\epsilon'(Y,T)=b'+(t')^\top Z+\sqrt mG$, we define $q(m,l)=L_\epsilon'(Y,T)$ and $\omega(m,l)=L_\epsilon''(Y,T)/(1+lL_\epsilon''(Y,T))$ and suppress their arguments at $(m,l)=(\E M,\E\Lambda)$. From the signed bound for $j=1,2$, we get
\[
\begin{gathered}
\E(d_i^j\ell_i)=l\E\{d_T^j/(1+ld_T)\}+O(L_n^4/n),\\
\E\tau_n=\E\omega+O(L_n^4/n),\qquad
 \E Q_2=\E q^2+O(L_n^4/n),\qquad (m,l)=(\E M,\E\Lambda).
\end{gathered}
\]

\begin{equation}
\label{S-xa-exact-overlap}
 \E\{(\rho+\tau_n)K\}
  =\eta\E(\widetilde\Lambda Q_{12})+E_n,\qquad
 Q_{12}=P_nq_i\widetilde q_i,\quad |E_n|\le C/n,
\end{equation}
with
\[
E_n=\frac1{n^2}\E\sum_i
 \{d_iq_iW_i^\top\mathscr H^{-1}\widetilde w
      +\eta q_i\widetilde d_i\widetilde W_i^\top
                          \widetilde{\mathscr H}^{-1}\widetilde w\}.
\]
Let $\partial_{ij},\widetilde\partial_{ij}$ differentiate the separate arguments $W_{ij},\widetilde W_{ij}$ and let $e_j$ be the coordinate vector in $\R^{p-d_0}$. Score differentiation gives
\[
\begin{gathered}
 \partial_{ij}w=-\mathscr H^{-1}(e_jq_i+d_iW_iw_j)/n,\qquad
 \partial_{ij}q_i=d_i\{(1-d_i\ell_i)w_j-q_i(\mathscr H^{-1}W_i)_j/n\},\\
 \sum_j\widetilde\partial_{ij}\widetilde w_j
 =-\widetilde q_i\widetilde\Lambda
   -\widetilde d_i\widetilde W_i^\top\widetilde{\mathscr H}^{-1}\widetilde w/n.
\end{gathered}
\]
We apply integration by parts to $q_i\widetilde w_j$ in the correlated Gaussian entries and sum over $i,j$ with $\sum_{ij}W_{ij}q_i\widetilde w_j=-n\rho K$ to obtain \eqref{S-xa-exact-overlap}, because $\|dq\|\|W\|\|\mathscr H^{-1}\widetilde w\|\le Cn$. This identity also holds conditionally on all $(Z_i,Y_i)$.

From the exact trace equation $(p-d_0)/n=\rho\Lambda+P_nd_i\ell_i$, we apply \eqref{S-xa-exact-overlap} at $\eta=1$ to two identical fits and factor expectations via \eqref{S-xa-replacement-bounds}. Then, we combine the signed estimates to obtain
\[
l\{\rho+\E\omega(m,l)\}-\kappa_n=O(L_n^4/n),\qquad
 l^2\E q(m,l)^2-\kappa_n m=O(L_n^4/n)
\]
with the dimension correction $-d_0/n$. At zero perturbation, from curvature and the score equation we bound the full-to-restricted nuisance difference by $C\|\widehat u_n-u_*\|=o_P(1)$, and hence the restricted radius converges. By bounded radius and compact positive inverse trace, every subsequential expectation limit solves these equations, and we get $\E\Lambda\to\lambda$ from the unique norm root.

By Step~1 and the resolvent identity, we obtain uniformly Lipschitz coefficients and bound trace changes by $C\{|\rho-\widetilde\rho|+P_n|d_i-\widetilde d_i|\}$, and each trace is Lipschitz in the parameters and in expectation after truncation. Hence, on a smaller box, $(m,l)$ and the population stationary pair stay near the unperturbed state, and we define
\[
F_1(m,l)=l\{\rho+\E\omega(m,l)\}-\kappa_n,\qquad
 F_2(m,l)=l^2\E q(m,l)^2-\kappa_nm.
\]
When we apply Gaussian integration by parts and the Moreau derivative, we obtain
\[
\mathcal A_r=(r/l)F_1(r^2,l),\qquad
 \mathcal A_l=-F_2(r^2,l)/(2l^2).
\]
At $F_1=F_2=0$ with $r,l>0$, from the nonsingularity of the stationary Hessian, we obtain nonsingularity of the residual Jacobian and apply the uniform inverse-function bound to get
\begin{equation}\label{S-xa-mean-bounds}
 \begin{gathered}
 |\E M-r^2|+|\E\Lambda-l(\xi)|\le CL_n^4/n,\\
 \|M-r^2\|_{L^2}\le Cn^{-1/2}+CL_n^4/n,\qquad
 \|\Lambda-l(\xi)\|_{L^2}\le CL_n n^{-1/2}+CL_n^4/n.
 \end{gathered}
\end{equation}
When we substitute this bound into \eqref{S-xa-signed-test} and apply Step~2, we obtain the bounds on the empirical averages. Because $\partial_{\epsilon_F}f_n=P_nH_F(Y_i,\eta_i)$ and $2\partial_\rho f_n=M$, we get $\|\mathcal D(\E f_n-\Phi)\|\le CL_n^4/n$.

\OptProofStep{4}{Conditional Gaussian approximation}
Given $\mathcal B=\sigma\{(Z_i,Y_i):1\le i\le n\}$, the entries of $W$ are independent Gaussian. For scalar $G_n=\sqrt n P_ng(Y_i,\eta_i)$, let $j=-\mathscr H^{-1}W^\top g'/n$ and $e=g'+DWj$. Note that for $E$ with $\|E\|_F=1$, we differentiate to get
\begin{align*}
w_E&=-\mathscr H^{-1}(E^\top q+W^\top DEw)/n,\qquad
       \eta_E=Ew+Ww_E,\\
 \nabla_WG_n&=(ew^\top+qj^\top)/\sqrt n,\qquad
 j_E=-\mathscr H^{-1}\left[E^\top e+
 W^\top\{(g''+L_\epsilon'''Wj)\eta_E+DEj\}\right]/n.
\end{align*}
Here, $j$ in $Wj$ is the coefficient derivative for the loss, and by the row bounds in Step~1 it holds that
\[
\|w_E\|\le Cn^{-1/2},\quad \|\eta_E\|_2\le C,\quad
 \|j_E\|\le C\sqrt{L_n/n},\quad \|e_E\|_2\le C\sqrt{L_n}.
\]
When we differentiate $\nabla_WG_n$ and use $\|e\|_2+\|q\|_2\le C\sqrt n$, we obtain
\[
\|\nabla_WG_n\|_F\le C,\qquad
 \|\nabla_W^2G_n\|\le C\sqrt{L_n/n}
\]
from derivatives of $g=H_B$ or $H_L$.

For $G_n=\sqrt n\|w\|^2$ and $v=\mathscr H^{-1}w$,
$\nabla_WG_n=-2\{qv^\top+(DWv)w^\top\}/\sqrt n.$
For unit perturbations $E,F$,
\[
\mathscr H w_{EF}=-\{E^\top D\eta_F+F^\top D\eta_E
 +W^\top L_\epsilon'''\eta_E\eta_F+W^\top D(Ew_F+Fw_E)\}/n.
\]
Here, the radius Hessian equals $2\sqrt n(w_E^\top w_F+v^\top\mathscr H w_{EF})$ with $|w_E^\top w_F|\le C/n$. From the inner product of the equation with $v$, we bound the first, second and fourth terms by $C/n$ and the third term by $C\|Wv\|_\infty\|\eta_E\|_2\|\eta_F\|_2/n\le C\sqrt{L_n}/n$. Hence, we obtain the same derivative bounds for the radius, and by Step~1 they also hold for $L^4$ norms and fixed linear combinations.

For standard Gaussian $X$, an independent copy $X'$, and centered $g(X)$ with $v=\E g^2$ (allowing zero variance), we define $X_t=tX+\sqrt{1-t^2}X'$ and differentiate $\E f(X)g(X_t)$ by parts to obtain
\[
\operatorname{Cov}(f(X),g(X))
  =\int_0^1\E\{\nabla f(X)^\top\nabla g(X_t)\}\,dt
\]
after we cancel second-derivative terms and integrate from $t=0$ to $t=1$. Then, we define
\[
\begin{gathered}
T(X)=\int_0^1\nabla g(X)^\top\E(\nabla g(X_t)\mid X)\,dt,\quad
k_1=(\E\|\nabla g\|^4)^{1/4},\quad
 k_2=(\E\|\nabla^2g\|^4)^{1/4}.
\end{gathered}
\]
When we substitute $f=F\circ g$ and differentiate $T$, we obtain
\[
\E gF(g)=\E TF'(g),\quad \E T=v,\quad
 \|\nabla T\|_{L^2}\le3k_1k_2/2,\quad
 \operatorname{Var}T\le\E\|\nabla T\|^2.
\]
For $\phi(a)=\E e^{iag}$, we solve the differential equation to obtain
\[
\phi'(a)+va\phi(a)=-a\E\{e^{iag}(T-v)\},\qquad
 |\phi(a)-e^{-va^2/2}|\le3a^2k_1k_2/4
\]
By polynomial growth and dominated convergence after truncation, these identities hold for ridge perturbations. Let $U_n=\sqrt n\mathcal D(f_n-\Phi)$ and $C_n=\Cov_W(U_n\mid\mathcal B)$. From conditional H\"older's inequality and $L^4$ derivative bounds, we obtain
\[
 \E\left|\E_W e^{ia^\top(U_n-\E_WU_n)}-e^{-a^\top C_na/2}\right|
 \le C_a\sqrt{L_n/n}.
\]

\OptProofStep{5}{The conditional covariance and its derivatives}
Conditional on $\mathcal B$, we use the joint Gaussian design representation, \eqref{S-xa-exact-overlap}, and the bounds \eqref{S-xa-replacement-bounds}--\eqref{S-xa-mean-bounds}. Then, for $\overline k=\E K$, we get
\begin{equation}
\label{S-xa-pair-bound}
 \|Q_{12}-Q_\eta(\overline k;\xi,\widetilde\xi)\|_{L^2}
       \le Cn^{-1/4},
\end{equation}
where we project $\overline k$ onto the covariance interval. For deleted row $i$, the predictor pair is conditionally Gaussian with variances $\|w_0\|^2,\|\widetilde w_0\|^2$ and covariance $\eta w_0^\top\widetilde w_0$, and under insertion the residual product changes by $O_{L^2}(n^{-1/2})$.

For a second deleted row $j$, the coefficients change independently of row $i$, with $L^2$ norm $Cn^{-1/2}$ also for predictor and residual changes by Step~2. Note that incoming pairs are conditionally independent, and their means are smooth in radii, traces, and overlap. Specifically, by \eqref{S-xa-replacement-bounds}--\eqref{S-xa-mean-bounds}, deterministic replacements cost $CL_n/\sqrt n+CL_n^4/n$, and we get
\[
\E|Q_{12}-Q_\eta(\overline k;\xi,\widetilde\xi)|^2
 \le C/n+Cn^{-1/2}+CL_n^2/n+CL_n^8/n^2
\]
by diagonal boundedness, conditional independence, and Cauchy--Schwarz off the diagonal. We include the expectation change under two deletions in $Cn^{-1/2}$ to get \eqref{S-xa-pair-bound} from square roots. Gaussian differentiation holds at singular covariances by bounded proximal derivatives, and we combine \eqref{S-xa-mean-bounds} with Cauchy--Schwarz to bound the projection shift by $CL_n^4/n$.

Note that for $\overline k=\eta l\widetilde l Q_\eta(\overline k)/\kappa_n+O(n^{-1/4})$, we have
\[
\left|\partial_k\{\eta l\widetilde l Q_\eta(k)/\kappa_n\}\right|
 \le\frac{\eta^2l\widetilde l}{\kappa_n}\E(\omega\widetilde\omega)
 \le\chi<1
\]
by $l\omega\le lC/(1+lC)<1$, $l\E\omega\le\kappa_n$ in \eqref{S-xa-tilted-state}, and Cauchy--Schwarz. We observe from $|Q_\eta|\le\sqrt{\E q^2\E\widetilde q^2} =\kappa_n r\widetilde r/(l\widetilde l)$ that this map sends the covariance interval to itself. We then apply contraction with the residual and concentration bounds to get a unique fixed point that satisfies
$\|K-k_\eta\|_{L^2}+\|Q_{12}-Q_\eta(k_\eta)\|_{L^2}\le Cn^{-1/4}.$
By implicit and Gaussian differentiation, we bound the fixed-order parameter derivatives of the population kernel up to $\eta=1$.

Because the design gradient of the optimal value is $qw^\top/n$, we apply the covariance identity in Step~4 to obtain
\[
n\operatorname{Cov}_W(f_n(\xi),f_n(\widetilde\xi)\mid\mathcal B)
     =\int_0^1\E_W(KQ_{12}\mid\mathcal B)\,d\eta.
\]
Conditional Jensen's inequality and the $L^2$ bounds give $L^1$ error at most $Cn^{-1/4}$ from \eqref{S-xa-value-kernel}. For any unit parameter direction, implicit differentiation gives
\[
\|\nabla_W\partial_a f_n\|_{L^2}\le Cn^{-1/2},\qquad
 \|\nabla_W\partial_{aa}f_n\|_{L^2}
                  \le C\sqrt{L_n/n}
\]
by differentiating $qw^\top/n$ with $q_a=DE_a+g'$, $q_{aa}=L_\epsilon'''E_a^2+2g''E_a+DWw_{aa}$ and the Step~1 bounds. From the Gaussian covariance identity, we then get the $L^1$ bound $C\sqrt{L_n}$ on mixed third derivatives. For $h=n^{-1/12}L_n^{-1/6}$, mixed forward differences have error $4Cn^{-1/4}/h^2+Ch\sqrt{L_n}\le Cn^{-1/12}L_n^{1/3}$. The population kernel has bounded third derivatives, and hence we obtain
\begin{equation}
\label{S-xa-conditional-covariance}
 \|C_n-\Gamma\|_{L^1}\le Cn^{-1/12}L_n^{1/3}
\end{equation}
where penalty derivatives use one-sided forward quotients in $\rho\ge n^{-2}$ and the Step~1 $L^2$ bound to pass to zero penalty.

\OptProofStep{6}{The conditional mean}
We delete the $i$th loss at fixed divisor $n$ and penalty and denote the minimum by $f_{n,-i}$. For $A_n^{\rm ins}=n(f_n-f_{n,-i})$, by the Step~2 trial we obtain
\[
\|A_n^{\rm ins}-M_{l_i}L_\epsilon(y,b'+(t')^\top z+u^\top w_0)\|_{L^2}
 \le Cn^{-1/2}.
\]
Note that the deleted score is zero at $w_0$, and by \eqref{S-xa-gaussian-powers} its quadratic change plus incoming loss equals $l_iq_T^2/2+L_\epsilon(y,T)$ with cubic remainder $O_{L^2}(n^{-1/2})$. From Step~3 and curvature, we bound further minimization by $Cn\|g_0\|^2=O_{L^2}(n^{-1})$. By conditional Jensen, Gaussian integration and \eqref{S-xa-mean-bounds}, we replace $l_i$ and the deleted radius by $l(\xi),r(\xi)$ at cost $CL_n/\sqrt n+CL_n^4/n\le Cn^{-1/4}$. Then, from linear Moreau growth and polynomial derivative bounds, we get
$\|\E_W A_n^{\rm ins}-m_{R_i}(\xi)\|_{L^2}\le Cn^{-1/4}.$

For incoming weight $s\in[0,1]$, we fix the divisor $n$, penalty, and other weights, and let $w_s,e_s,\mathscr H_s$ be the fit, insertion remainder, and Hessian. We define $v_s=\partial_aw_s$, $z_s=\partial_{aa}w_s$, $A_{-i}$ as $A$ with row $i$ removed, $R_s=A_{-i}h+W_0v_s$, and $E_s=(1,z^\top)h+u^\top v_s$ on the incoming row, where subscripts $s,0$ denote inserted and deleted fits. When we differentiate the score equation, we obtain
\begin{align*}
\mathscr H_s(v_s-v_0)
  ={}&-n^{-1}W_0^\top\{(D_s-D_0)R_0+(g_s'-g_0')\}\\
    &-sn^{-1}u\{d_i[(1,z^\top)h+u^\top v_0]+g_i'\}
      -c(w_s-w_0).
\end{align*}
From \eqref{S-xa-gaussian-powers}, $n^{-2}\sum_jR_{0j}^2\|B_0W_j\|^2\le C$, and $n\|e_s\|=O_{L^q}(1)$, we bound the retained increment and its product with $R_0$ before we multiply by $W_0^\top/n$. Because the incoming projection of the derivative of the deleted fit has bounded conditional Gaussian moments, we obtain $\|v_s-v_0\|_{L^q}\le C_qn^{-1/2}$ and $\|E_s\|_{L^q}\le C_q$.

From the second-order score derivatives, we obtain
\begin{align*}
\mathscr H_s(z_s-z_0)={}&-(\mathscr H_s-\mathscr H_0)z_0-2c(v_s-v_0)-sn^{-1}u\{L_i'''E_s^2+2g_i''E_s\}\\
 &-n^{-1}W_0^\top\{(L_s'''-L_0''')R_0^2
                    +L_s'''(R_s^2-R_0^2)+2(g_s''-g_0'')R_0+2g_s''(R_s-R_0)\}.
\end{align*}
Here, the change in the retained Hessian is $O_{L^{2q}}(\sqrt{L_n/n})$. Because $\|z_0\|_{L^{2q}}+\|u^\top z_0\|_{L^{2q}}\le C_q\sqrt{L_n}$, its product with $z_0$ is $O_{L^q}(L_n/\sqrt n)$ and $sd_i uu^\top z_0/n$ is $O_{L^q}(\sqrt{L_n/n})$. From the $L^{2q}$ increment bounds and $\|R_0\|_\infty+\|R_s\|_\infty=O_{L^{2q}}(\sqrt{L_n})$, we bound the four retained terms by
\[
C_qL_n/\sqrt n,\quad C_q\sqrt{L_n/n},\quad
 C_q\sqrt{L_n/n},\quad C_qn^{-1/2}.
\]
Here, the incoming term is at most $C_qn^{-1/2}$, and by uniform curvature we obtain
\[
\|z_s-z_0\|_{L^q}\le C_qL_n/\sqrt n,\qquad
 \|u^\top z_s\|_{L^q}\le C_qL_n.
\]

We differentiate the exact insertion integral to get
\[
\partial_a A_n^{\rm ins}=\int_0^1\{q_iE_s+g(Y_i,\eta_i)\}\,ds,
 \qquad
 \partial_{aa} A_n^{\rm ins}
       =\int_0^1\{d_iE_s^2+2g_i'E_s+q_i u^\top z_s\}\,ds
\]
where the second integral has $L^2$ norm at most $CL_n$. With $h=n^{-1/8}L_n^{-1/2}$, we use forward differences that are one-sided in the penalty to get
\[
 \|\partial_a\E_W A_n^{\rm ins}-\partial_a m_{R_i}\|_{L^2}
 \le Cn^{-1/4}/h+ChL_n\le Cn^{-1/8}\sqrt{L_n}.
\]
In a row replacement, the deleted value cancels, and by this derivative bound the centered residual changes by $O_{L^2}(n^{-1/2}n^{-1/8}\sqrt{L_n})$. By the replacement variance inequality in Step~2 and the ridge error, we get
\begin{equation}
\label{S-xa-conditional-mean}
 \left\|\E_W U_n-\E U_n-\frac1{\sqrt n}\sum_i h(R_i)\right\|_{L^2}
       \le Cn^{-1/8}\sqrt{L_n}+Cn^{-3/2}.
\end{equation}

\OptProofStep{7}{Combining the two sources of randomness}
By Step~1 and Gaussian integration, $m_R$ and its derivatives have polynomial growth, and thus the independent centered $h(R_i)$ have bounded third moments. For each factor of the characteristic function, the quadratic Taylor error is at most $\E|a^\top h(R_i)|^3/(6n^{3/2})$ for the data and $O(n^{-2})$ for the Gaussian comparison with covariance $\E hh^\top$. Telescoping these errors yields the bound $C_a/\sqrt n$.

We substitute \eqref{S-xa-conditional-covariance}--\eqref{S-xa-conditional-mean} in Step~4 and use $|e^{ix}-e^{iy}|\le|x-y|$, conditional Jensen's inequality, and $\|\E U_n\|\le CL_n^4/\sqrt n$. At $\xi_*$, we have $\mathcal D\Phi=(m_B-\varrho_B\kappa_n,m_L-\varrho_L\kappa_n,r^2)^\top$ and $U_n=(X_{B,n},X_{L,n},T_n)^\top$ up to the ridge errors, and hence we obtain \eqref{S-xa-core-clt} uniformly. This completes the proof.\end{proof}

\subsection{Expansion at the full fit}
Let $H_i=(1,Z_i^\top)^\top$, $a_n^0=(u_*^\top,(w_n^0)^\top)^\top$, $S_n=n^{-1/2}\sum_i H_i\{s(\eta_i^0)-Y_i\}$, $F_n=\sqrt n(\widehat u_n-u_*)$, and $g_n=\nabla_u r(u,0,0)^2|_{u_*}$.
At the population $\lambda$, define
\[
\begin{aligned}
 S&=P_n\frac{\lambda d_i}{1+\lambda d_i}-\kappa_n,
 &D&=P_n\frac{d_i}{(1+\lambda d_i)^2},\\
 A_F&=P_nq_i\ell_F'(Y_i,\widehat\eta_i+\lambda q_i),
 &C_F^0&=P_n\ell_F(Y_i,\widehat\eta_i+\lambda q_i).
\end{aligned}
\]
\begin{lemma}\label{S-xa-multiplier}
Under the conditions of Theorem~\ref{S-xa-clt}, uniformly on $\mathcal K$, $\E\|S_n\|^2\le C$, $\|\widehat a_n-a_n^0\|=O_P(n^{-1/2})$, $F_n=O_P(1)$, and
\[
 \sqrt n(\|\widehat w\|^2-r^2)=T_n+g_n^\top F_n+o_P(1).
\]
The multiplier and loss estimates satisfy
\begin{gather}
 S,D-J,A_F-a_F=O_P(n^{-1/2}),\notag\\
 \widehat\lambda-\lambda=-S/J+O_P(n^{-1}),\label{S-xa-root-remainder}\\
 \widehat C_F-C_F^0=-(a_F/J)S+O_P(n^{-1}).\label{S-xa-loss-remainder}
\end{gather}
\end{lemma}
\begin{proof}
For a coordinate $h$ of $(1,Z^\top)^\top$, let $h_i=h(Z_i)$ and $A_h=P_nh_i\{s(\eta_i^0)-Y_i\}$. Use Step~2 of the proof of Lemma~\ref{S-xa-restricted-kernel} at deleted scores $\eta_j^{(-i)}$. The retained gradient satisfies
\[
 g=n^{-1}W_0^\top\{h_js'(\eta_j^{(-i)})\}_{j\ne i},\qquad
 \|g\|\le C(P_nh_i^2)^{1/2},\qquad
 \|g^\top\Delta\|_{L^q}+\|g^\top e\|_{L^q}\le C_q/n
\]
by conditional Gaussian moments and the bound on $e$. Let $R_\Delta,R_e$ be the quadratic terms and let $\mathcal D_{-i}$ denote the deleted data. Then, we have
\[
 \begin{aligned}
 \|R_\Delta\|_{L^q\mid\mathcal D_{-i}}
 &\le C_qn^{-3}\sum_{j\ne i}|h_j|\|B_0W_j\|^2
 \le C_qP_n|h_i|/n,\\
 |R_e|&\le C\max_j|h_j|\|e\|^2=O_{L^q}(\sqrt{\log(en)}/n^2).
 \end{aligned}
\]
By the Cauchy--Schwarz inequality, we bound the cross term, the added row is $O_{L^q}(n^{-1})$, and we get the variance bound from two insertions. From \eqref{S-xa-signed-test}--\eqref{S-xa-mean-bounds} with $\chi=h$ and $\E h(Z)q=0$, we get the mean bound and sum over coordinates to obtain
\[
 \Var A_h\le C/n,\qquad |\E A_h|\le C\log(en)^4/n,\qquad
 \E\|S_n\|^2=n\sum_h\{\Var A_h+(\E A_h)^2\}\le C .
\]

Note that the restricted score is $(S_n^\top/\sqrt n,0)^\top$ and the full score is zero, and from \eqref{S-NG-eq-geo-curvature} we get $\|\widehat a_n-a_n^0\|\le C\|S_n\|/\sqrt n$. Let $R_n(u)=\|w_n(u,0,0)\|^2$, $R(u)=r(u,0,0)^2$ and $L_n=\log(en)$. At deterministic nearby $u$, we obtain $R_n(u)-R(u)=O_P(n^{-1/2}+L_n^4/n)$ from \eqref{S-xa-mean-bounds} and $\sup_u\|\nabla^2R_n(u)\|=O_P(\sqrt{L_n})$ from Lemma~\ref{S-tilted-geometry}. Then, we use centered differences of step $h=n^{-1/8}$ to get
\[
 \|\nabla R_n(u_*)-g_n\|
 =O_P\{(n^{-1/2}+L_n^4/n)/h+h\sqrt{L_n}\}=o_P(1).
\]
Taylor expansion at $u_*$ gives the radius expansion with scaled remainder $O_P(\sqrt{L_n/n})$ from the second derivative.

Lemma~\ref{S-xa-restricted-kernel} applies to the bounded-derivative integrands of $S,D,A_F$. Their deviations at the restricted fit are $O_P(n^{-1/2})$. The coefficient bound and $\|B\|_{\op}=O_P(\sqrt n)$ give the same order at the full fit. For $f(x)=P_n\{xd_i/(1+xd_i)\}-\kappa_n$, $f(\lambda)=S$, $f'(\lambda)=D$ and $|f''|\le1/8$ for $x\ge0$. Root consistency and $\inf J>0$ give $|\widehat\lambda-\lambda|\le2|S|/J$ with probability tending to one. Taylor expansion, $|q_i|\le1$ and bounded second loss derivatives give
\[
\begin{aligned}
 \widehat\lambda-\lambda&=-S/J+O_P(S^2+|S|\,|D-J|),\\
 \widehat C_F-C_F^0&=A_F(\widehat\lambda-\lambda)+O_P(n^{-1}).
\end{aligned}
\]
Because we have $A_F-a_F=O_P(n^{-1/2})$, we then obtain \eqref{S-xa-loss-remainder} by substitution. This completes the proof.
\end{proof}

\begin{proposition}\label{S-xa-head-cancellation}
For $F=B,L$, jointly and uniformly on $\mathcal K$,
\begin{equation}\label{S-xa-realized-expansion}
 \sqrt n(\widehat C_F-R_{F,n})=X_{F,n}-f_{F,v}T_n+o_P(1).
\end{equation}
\end{proposition}
\begin{proof}
At zero penalty let $u=(b',(t')^\top)^\top$, with $T_u+l(u)\{s(T_u)-Y\}=b'+(t')^\top Z+r(u)G$. Evaluate derivatives at $u_*$ and let $g=\nabla_u[r(u)^2]|_{u_*}$. Differentiating the trace equation, with $\lambda$ fixed in $\tau$, gives
\[
 0=\nabla_u\E\frac{l(u)s'(T_u)}{1+l(u)s'(T_u)}
   =\nabla_u\E\tau(T_u)+J\nabla_ul(u).
\]
The multiplier derivative of the loss expectation at fixed score is $a_F$. Differentiation of the proximal identity and $\varrho_FJ=a_F$ give
\begin{equation}\label{S-xa-head-identity}
 \begin{aligned}
 \nabla_u\E H_F(Y,T_u)
 &=\nabla_u\mathcal R_F(u_*,r^2)+f_{F,v}g+(\varrho_FJ-a_F)\nabla_ul(u)\\
 &=\nabla_u\mathcal R_F(u_*,r^2)+f_{F,v}g .
 \end{aligned}
\end{equation}

We define $j_F(u)=P_nH_F(Y_i,(1,Z_i^\top)u+W_i^\top w_n(u,0))$, and for deterministic nearby $u$ we get $j_F(u)-\E H_F(Y,T_u)=O_P(n^{-1/2})$ by Lemma~\ref{S-xa-restricted-kernel}. From bounded $H_F',H_F''$, the design norm and Lemma~\ref{S-tilted-geometry}, we have $\sup_u\|\nabla^2j_F(u)\|=O_P(\sqrt{\log(en)})$. With bounded population second derivatives, we use forward differences of step $h=n^{-1/4}$ to obtain
\[
 \big\|\nabla j_F(u_*)-\left.\nabla_u\E H_F(Y,T_u)\right|_{u_*}\big\|
 =O_P\{n^{-1/2}/h+h\sqrt{\log(en)}\}=o_P(1).
\]
By Lemma~\ref{S-xa-multiplier}, we have $\widehat u_n-u_*=O_P(n^{-1/2})$, and the scaled quadratic remainder is $O_P(\sqrt{\log(en)/n})$. From \eqref{S-xa-head-identity}, we see that the two expansions at $u_*$ are
\[
 \begin{aligned}
 \sqrt n(\widehat C_F-m_F)
 &=\sqrt n\{j_F(\widehat u_n)-m_F+\varrho_F\kappa_n\}+o_P(1)\\
 &=X_{F,n}+(\nabla_u\mathcal R_F+f_{F,v}g)^\top F_n+o_P(1),\\
 \sqrt n(R_{F,n}-m_F)
 &=(\nabla_u\mathcal R_F+f_{F,v}g)^\top F_n+f_{F,v}T_n+o_P(1).
 \end{aligned}
\]
In the difference, the common $F_n$ term cancels and \eqref{S-xa-realized-expansion} holds uniformly. This completes the proof.
\end{proof}

\subsection{Evaluation of the covariance}

\begin{proposition}\label{S-xa-covariance-identity}
For $F,H\in\{B,L\}$, the limiting covariance of
$X_{F,n}-f_{F,v}T_n$ and $X_{H,n}-f_{H,v}T_n$ is
$\Sigma_{FH}$ in \eqref{S-xa-integrals}.
The covariance of the conditional mean is
\[
 \operatorname{Cov}_{Z,Y}\{\E_G\ell_F(Y,V),\E_G\ell_H(Y,V)\}.
\]
The remaining covariance equals
\begin{equation}\label{S-xa-design-correction}
 \E_{Z,Y}\operatorname{Cov}_G\{\ell_F(Y,V),\ell_H(Y,V)\}
           +\lambda a_Fa_H/J.
\end{equation}
\end{proposition}
\begin{proof}
Let $\mathcal D_F=\partial_{\epsilon_F}-2f_{F,v}\partial_\rho$ with $\widetilde{\mathcal D}_H$ acting on the second point. From Lemma~\ref{S-xa-restricted-kernel} and the right penalty derivative, we obtain $\sqrt n\mathcal D_F(f_n-\Phi)=X_{F,n}-f_{F,v}T_n+o_{L^2}(1)$ at $\xi_*$. Because $q_v=\omega$ and $q_{vv}=s''(T)/(1+\lambda d)^3$, we use differentiation at fixed $V,\lambda$ to get
\begin{equation}\label{S-xa-residual-direction}
 \frac{H_F'(Y,T)}{1+\lambda d}=\ell_F'(Y,V)-\varrho_F\lambda q_{vv}.
\end{equation}
The state derivatives in direction $\mathcal D_F$ are
\begin{equation}\label{S-xa-nuisance-response}
 \dot r_F=\varrho_F\lambda/r,\qquad \dot\lambda_F=0 .
\end{equation}
With these derivatives, $\dot q_F=\ell_F'(Y,V)+\varrho_F\lambda(G\omega/r-q_{vv})$. Gaussian integration by parts gives
\[
 \begin{aligned}
 \E\{q(G\omega/r-q_{vv})\}
   &=r^{-1}\E\partial_G(q\omega)-\E(q q_{vv})=\E\omega^2,\\
 \E\{G(G\omega/r-q_{vv})\}
   &=r^{-1}\E(\omega+rGq_{vv})-\E(Gq_{vv})=\E\omega/r .
 \end{aligned}
\]
In \eqref{S-xa-perturbed-value}, $\mathcal A_r=\E Gq-\kappa_n r/\lambda+\rho r$ and $\mathcal A_\lambda=-\E q^2/2+\kappa_n r^2/(2\lambda^2)$. Using $\E G\ell_F'=r\E\ell_F''=2rf_{F,v}$ gives
\[
 \begin{aligned}
 \mathcal D_F\mathcal A_r
 &=2rf_{F,v}+\varrho_F(\lambda\E\omega-\kappa_n)/r-2rf_{F,v}=0,\\
 \mathcal D_F\mathcal A_\lambda
 &=-a_F+\varrho_F(\kappa_n-\lambda^2\E\omega^2)/\lambda=0 ,
 \end{aligned}
\]
by $\lambda\E\omega=\kappa_n$, $\varrho_FJ=a_F$, and
\begin{equation}\label{S-xa-gap-identity}
 \kappa_n-\lambda^2\E\omega^2
 =\E\!\left\{\frac{\lambda d}{1+\lambda d}
             -\left(\frac{\lambda d}{1+\lambda d}\right)^2\right\}
 =\lambda J .
\end{equation}
The nonsingular Hessian in Step~1 of the proof of Lemma~\ref{S-xa-restricted-kernel} proves \eqref{S-xa-nuisance-response}. Conditional Gaussian integration gives $\E_GGq=r\E_G\omega$. Using $\tau(T)=\lambda\omega$,
\[
 \begin{aligned}
 \mathcal D_Fm_R
 &=\E_G H_F(Y,T)+\dot r_F\E_G Gq-\tfrac12\dot\lambda_F\E_G q^2\\
 &=\E_G\ell_F(Y,V)-\varrho_F\lambda\E_G\omega
                    +\varrho_F\lambda\E_G\omega
 =\E_G\ell_F(Y,V).
 \end{aligned}
\]
This gives the covariance of the conditional means.

Let $V,\widetilde V$ have offset $b+t^\top Z$ and centered Gaussian noises independent of $(Z,Y)$ with variance $r^2$ and correlation $x\in[0,1]$. For $\widetilde q=q(Y,\widetilde V)$, define
\[
 Q(x)=\E(q\widetilde q),\quad P_F(x)=\E\{\ell_F'(Y,V)\widetilde q\},\quad
 R_{FH}(x)=\E\{\ell_F'(Y,V)\ell_H'(Y,\widetilde V)\}.
\]
Gaussian differentiation gives $Q'=r^2\E(\omega\widetilde\omega)$, continuous at $0,1$ by bounded derivatives. The common strict residual sign gives $Q>0$. With $C=\lambda^2/\kappa_n$, the equations $k=\eta C Q(x)$ and $x=\eta k/r^2$ imply
\[
 \eta(x)=\sqrt{\frac{x r^2}{C Q(x)}},\qquad
 \frac{d\eta}{dx}=\frac{\eta(x)D(x)}{2x},\qquad D(x)=1-\frac{xQ'(x)}{Q(x)},\quad x>0 .
\]
Step~5 of the proof of Lemma~\ref{S-xa-restricted-kernel} bounds $D$ away from zero. Because $Q(1)=r^2/C$, $\eta(0)=0$ and $\eta(1)=1$.

Differentiate $Q_\eta$ at fixed covariance $\eta k$ before equating the parameter points. The variance derivative $2r\dot r_F=2\varrho_F\lambda$ and $\partial_{r^2}\E f=\E f_{vv}/2$ give
\[
 \begin{aligned}
 \mathcal D_F Q_\eta\big|_k
 &=\E\{(\ell_F'(Y,V)-\varrho_F\lambda q_{vv})\widetilde q\}
     +\varrho_F\lambda\E(q_{vv}\widetilde q)=P_F(x),\\
 \widetilde{\mathcal D}_H\mathcal D_F Q_\eta\big|_k&=R_{FH}(x).
 \end{aligned}
\]
The second line applies \eqref{S-xa-residual-direction} and \eqref{S-xa-nuisance-response} to the other argument. Differentiating the overlap at fixed $\eta$ gives
\[
 \begin{aligned}
 k_F&=\eta C P_F/D,\qquad k_H=\eta C P_H/D,\\
 k_{FH}&=\frac{\eta C}{D}
 \left\{R_{FH}+\frac{\eta}{r^2}(P_F'k_H+P_H'k_F)
                   +\frac{\eta^2}{r^4}Q''k_Fk_H\right\},
 \end{aligned}
\]
where subscripts denote parameter derivatives, primes $x$ derivatives, and both multiplier derivatives are zero. We substitute into the mixed derivative of $kQ_\eta=k^2/(\eta C)$ and use $(Q-xQ')'=-xQ''$ to obtain
\begin{equation}\label{S-xa-boundary-integral}
 \begin{aligned}
 \frac{2(k_Fk_H+kk_{FH})}{\eta C}\frac{d\eta}{dx}
 &=r^2 R_{FH}+r^2\frac{P_FP_H+x(P_F'P_H+P_FP_H')}{Q-xQ'}\\
 &\quad +r^2\frac{x^2P_FP_HQ''}{(Q-xQ')^2}\\
 &=r^2 R_{FH}+r^2\frac{d}{dx}\left(\frac{xP_FP_H}{Q-xQ'}\right).
 \end{aligned}
\end{equation}
The quotient equals zero at $x=0$, with $P_F=a_F$ and $Q-Q'=r^2J/\lambda$ at $x=1$ by \eqref{S-xa-gap-identity}. By Step~4 of the proof of Lemma~\ref{S-xa-restricted-kernel}, we integrate \eqref{S-xa-boundary-integral} to obtain
\[
 \begin{aligned}
 \mathcal D_F\widetilde{\mathcal D}_H\mathcal H
 &=r^2\int_0^1 R_{FH}(x)\,dx+\lambda a_Fa_H/J
 =\E_{Z,Y}\Cov_G\{\ell_F(Y,V),\ell_H(Y,V)\}+\lambda a_Fa_H/J .
 \end{aligned}
\]
Hence, \eqref{S-xa-design-correction} holds, and when we add the conditional-mean covariance, we obtain $\Sigma_{FH}$. This completes the proof.
\end{proof}

\subsection{Covariance estimation and marginal coverage}\label{S-xb-section}

From one fitted model, we define
\begin{equation}\label{S-xa-j-estimator}
 \widehat a_F=P_nq_i\ell_F'(Y_i,\widetilde V_i),\quad F\in\{B,L\},
 \qquad \widehat J=P_n\frac{d_i}{(1+\widehat\lambda d_i)^2}.
\end{equation}
For $\widehat L_i=(\ell_B(Y_i,\widetilde V_i),\ell_L(Y_i,\widetilde V_i))^\top$, we define
\begin{equation}\label{S-xa-variance-estimator}
 \widehat\Sigma=P_n(\widehat L_i-P_n\widehat L_i)
                         (\widehat L_i-P_n\widehat L_i)^\top
                +\frac{\widehat\lambda}{\widehat J}(\widehat a_F\widehat a_H)_{F,H\in\{B,L\}}.
\end{equation}

\begin{proposition}\label{S-xa-variance-consistency}
Uniformly for $\zeta_n\in\mathcal K$,
$
 \widehat\Sigma-\Sigma(\zeta_n)\xrightarrow{P}0.
$
\end{proposition}
\begin{proof}
\OptProofStep{1}{Fourth moments of the predictors}
At zero perturbation, we have $\sup_i\E|\eta_i^0|^4\le C$ by Steps~1--2 of Lemma~\ref{S-xa-restricted-kernel} and $\max_i|\widehat\eta_i-\eta_i^0|=O_P(\sqrt{\log(en)/n})$ by Lemmas~\ref{S-tilted-geometry} and~\ref{S-xa-multiplier}. By Markov, $|q_i|\le1$ and $\widehat\lambda=O_P(1)$, we get
\[
 \begin{aligned}
 P_n|\widetilde V_i|^4
 &=P_n|\eta_i^0+(\widehat\eta_i-\eta_i^0)+\widehat\lambda q_i|^4\\
 &\le C\{P_n|\eta_i^0|^4+\max_i|\widehat\eta_i-\eta_i^0|^4+\widehat\lambda^4\}
 =O_P(1).
 \end{aligned}
\]

\OptProofStep{2}{Means and covariances}
Note that by the fourth-moment bounds, it holds that the empirical and population second-moment tails above $M$ are $O_P(M^{-2})$, and hence by truncation we get bounded-test convergence for the loss means and covariances. We then combine bounded continuous integrands and root consistency to obtain $\widehat a_F-a_F,\widehat J-J=o_P(1)$, and because $\inf_{\mathcal K}J>0$ we get covariance consistency. This completes the proof.
\end{proof}

Let $\mathcal S_B=[0,1]$ and $\mathcal S_L=[0,\infty)$.

For any $0<\alpha<1$ and a valid fit, we define the confidence interval
\begin{equation}\label{S-xb-interval}
 I_{F,n}=\left[\widehat C_F\pm z_{1-\alpha/2}
          \sqrt{\widehat\Sigma_{FF}/n}\right]\cap\mathcal S_F.
\end{equation}
If the fit does not exist, the root is undefined, or the variance is nonpositive or nonfinite, we record a failure and define $I_{F,n}=\mathcal S_F$.

\begin{corollary}\label{S-xb-clt}
For $F\in\{B,L\}$ and $0<\alpha<1$, uniformly on $\mathcal K$,
\[
 \Prob\{R_{F,n}\in I_{F,n}\}\longrightarrow1-\alpha.
\]
This coverage limit also holds under numerical errors $o_P(n^{-1/2})$ in the root and fitted losses, errors $o_P(1)$ in the variance estimate, and probability $o(1)$ that evaluation fails.
\end{corollary}
\begin{proof}
Let $Z_{F,n}=\sqrt n(\widehat C_F-R_{F,n})/\sqrt{\widehat\Sigma_{FF}}$ on success and zero on failure. The failure probability is $o(1)$ by existence, root consistency, and positive definiteness. By Theorem~\ref{S-xa-clt}, Proposition~\ref{S-xa-variance-consistency}, and the Slutsky theorem, we get $Z_{F,n}\dto N(0,1)$ uniformly, and because $R_{F,n}\in\mathcal S_F$, we have
\[
 \begin{aligned}
 \Prob\{R_{F,n}\in I_{F,n}\}
 &=\Prob\{|Z_{F,n}|\le z_{1-\alpha/2}\}+o(1)
 =2\Phi(z_{1-\alpha/2})-1+o(1)=1-\alpha+o(1).
 \end{aligned}
\]
Note that the stated numerical errors change $Z_{F,n}$ by $o_P(1)$ on success, and numerical failure has probability $o(1)$. With $z_{1-\alpha/4}$, the two marginal noncoverage probabilities sum to $\alpha+o(1)$, and thus we obtain the simultaneous bound. This completes the proof.
\end{proof}

We define $I_F^{\rm iid}$ for $F\in\{B,L\}$ by \eqref{S-xb-interval} with only the empirical variance of the adjusted losses. Because $q=s(T)-Y$ and $\ell_F'(Y,V)$ share a nonzero sign, we have $a_F>0$, and positive continuous $\lambda a_F^2/J$ on compact $\mathcal K$ implies a uniform gap below nominal. For fixed $\alpha$, by the uniform expansion in \eqref{S-xa-integrals} and Corollary~\ref{S-xb-clt}, we obtain
\begin{equation}\label{eq:independent-coverage}
 \Prob(R_{F,n}\in I_F^{\rm iid})=
 2\Phi\!\left(z_{1-\alpha/2}\sqrt{\operatorname{Var}\{\ell_F(Y,V)\}/\Sigma_{FF}}\right)-1
 +o(1).
\end{equation}

\subsection{Comparison with an intercept-only fit}\label{S-reference-gain}
Here we let $R$ replace $L$ as the log-loss index. When both response classes appear, let $\widehat\pi=n^{-1}\sum_iY_i$. For an independent new response $Y$ and $F\in\{B,R\}$, define $G_F=\E[\ell_F(Y,\log\{\widehat\pi/(1-\widehat\pi)\})\mid\mathcal D]-F_{\rm ext}$. Let $g_B(u)=u(1-u)$, $g_R(u)=-u\log u-(1-u)\log(1-u)$, $\widehat G_F=g_F(\widehat\pi)-\widehat F$, and $U_{F,i}=\ell_F(Y_i,\widetilde V_i)-g_F'(\widehat\pi)(Y_i-\widehat\pi)$. Define $\widehat\Omega=(\widehat\Omega_{FG})_{F,G\in\{B,R\}}$ as the sample covariance of $(U_{B,i},U_{R,i})$ with divisor $n$, plus $(\widehat\lambda\widehat a_F\widehat a_G/\widehat J)$.

Let $\Omega$ denote the population version of this matrix at $(\kappa_n,\theta_n,\gamma_n)$ and define $L_F=\widehat G_F-z_{1-\alpha}(\widehat\Omega_{FF}/n)^{1/2}$. We declare improvement over the intercept-only model when $L_F>0$.

\begin{corollary}\label{cor:reference-gain}
\rightskip=0pt plus .4em Under the conditions of Theorem~\ref{thm:optboot-onefitci}, for each fixed loss $F$, uniformly on $\mathcal K$, we have $\sqrt n(\widehat G_B-G_B,\widehat G_R-G_R)^\top\dto N_2(0,\Omega)$, $\widehat\Omega-\Omega\pto0$ with $\Omega\succ0$, and $\limsup_n\sup_{(\kappa_n,\theta_n,\gamma_n)\in\mathcal K}\Prob\{L_F>0,\ G_F\le0\}\le\alpha$.
\end{corollary}

\begin{proof}[Proof of Corollary~\ref{cor:reference-gain}]
Let $\pi_n=\Prob(Y_i=1)$ and $\delta_n=\widehat\pi-\pi_n$. Then, we have the conditional reference risks
\[
R_{{\rm ref},B,n}=\pi_n(1-\pi_n)+\delta_n^2,\qquad
 R_{{\rm ref},R,n}=-\pi_n\log\widehat\pi-(1-\pi_n)\log(1-\widehat\pi).
\]
We observe that $g_F(\widehat\pi)-R_{{\rm ref},F,n}=g_F'(\widehat\pi)\delta_n$ for both losses. Because $\delta_n=O_P(n^{-1/2})$ and $\pi_n$ stays in a compact subset of $(0,1)$, it holds uniformly that
\[
 \sqrt n(\widehat G_F-G_F)
 =g_F'(\pi_n)\sqrt n\delta_n-\sqrt n(\widehat C_F-R_{F,n})+o_P(1).
\]
We append $Y-\pi_n$ in Step~7 of the proof of Lemma~\ref{S-xa-restricted-kernel}, where for $\epsilon Y$ we add $\epsilon P_nY$ with fixed minimizer. We observe from Proposition~\ref{S-xa-covariance-identity} that $\E_G\ell_F(Y,V)-m_F$ has covariance $\Cov\{\ell_F(Y,V),Y\}$ with $Y-\pi_n$, and by third moments of bounded labels it holds that
\[
\Omega_{FG}=\operatorname{Cov}\{\ell_F(Y,V)-g_F'(\pi_n)(Y-\pi_n),
               \ell_G(Y,V)-g_G'(\pi_n)(Y-\pi_n)\}
                  +\lambda a_Fa_G/J.
\]
We combine Proposition~\ref{S-xa-variance-consistency} with $g_F'(\widehat\pi)\to g_F'(\pi_n)$ to get $\widehat\Omega\to\Omega$, and because the label adjustment is constant given $Y=0$, we obtain positive definiteness with a uniform lower eigenvalue bound by compactness.

The probability of an absent response class is $\pi_n^n+(1-\pi_n)^n\to0$ uniformly, we use $\widehat\pi=1/2$ with no declaration on that event, and on valid fits we have
\[
\{L_F>0,\ G_F\le0\}
 \subseteq
 \left\{\frac{\sqrt n(\widehat G_F-G_F)}{\sqrt{\widehat\Omega_{FF}}}>z_{1-\alpha}\right\}
\]
and by the normal limit we obtain the bound. This completes the proof.
\end{proof}

\section{Single-fit inference with non-Gaussian components}\label{S-NG-section}
\subsection{Standardized predictors and state equations}\label{S-NG-model}

As in Sections~\ref{S-model} and~\ref{S-xa-section}, we use the two losses, adjusted scores, and conventions for undefined fits, and exclude the intercept from $\kappa_n=p/n$. For fixed $m<\infty$, let $P_Z$ be a fixed distribution such that
\begin{equation}\label{S-NG-eq-head}
 \E Z=0,\qquad \E ZZ^\top=I_m,\qquad
 \sup_{\norm{v}=1}\E\exp\{(v^\top Z)^2/K_Z^2\}\le2
\end{equation}
for some $K_Z<\infty$. This distribution may have atoms and nonzero
third moments.
For deterministic $x_{0,n}$ and invertible $C_n$, assume independent rows
\begin{equation}\label{S-NG-eq-finitehead}
 \begin{gathered}
 X_i=x_{0,n}+C_n\binom{Z_i}{A_i},\qquad Z_i\sim P_Z,\qquad
 A_i\sim N_{p-m}(0,I),\qquad A_i\perp Z_i,\\
 C_n^\top\beta_n=\binom{\varpi_n}{0},\qquad
 \Prob(Y_i=1\mid Z_i,A_i)=s(\theta_n+\varpi_n^\top Z_i).
 \end{gathered}
\end{equation}
Let $(\widehat b,\widehat t,\widehat w)$ minimize the empirical loss $L_n(b,t,w)=P_n\ell_{Y_i}(b+t^\top Z_i+A_i^\top w)$. For augmented covariates $z_X=Lz_U$ with invertible $L$, we have $z_X^\top u_X=z_U^\top L^\top u_X$, and thus Algorithm~1 and its covariance estimate are invariant under this change of coordinates. In case (a) of Theorem~\ref{thm:optboot-nongaussian}, $X_1$ has fixed dimension $m_0$, fixed distribution, mean $\mu_1$, nonsingular covariance $\Sigma_1$, and sub-Gaussian projections, and the conditional model is
\begin{equation}\label{S-NG-eq-conditionalclass}
 X_2\mid X_1\sim N_{p-m_0}(\mu_n+\Psi_nX_1,\Omega_n),\qquad
 \Omega_n\succ0,\qquad
 \Prob(Y=1\mid X)=s(\theta_n^{\rm orig}+X^\top\beta_n).
\end{equation}
Let $\beta_n=(\beta_{1,n}^\top,\beta_{2,n}^\top)^\top$. Using symmetric
positive square roots, we define
\begin{equation}\label{S-NG-eq-affineparameters}
 \begin{gathered}
 U=\Sigma_1^{-1/2}(X_1-\mu_1),\qquad
 G_n=\Omega_n^{-1/2}(X_2-\mu_n-\Psi_nX_1),\\
 c_n=\beta_{1,n}+\Psi_n^\top\beta_{2,n},\qquad
 a_n=\Sigma_1^{1/2}c_n,\qquad
 \gamma_{G,n}=(\beta_{2,n}^\top\Omega_n\beta_{2,n})^{1/2},\\
 \theta_n=\theta_n^{\rm orig}+\beta_{2,n}^\top\mu_n+c_n^\top\mu_1,
 \qquad \varpi_n=(a_n^\top,\gamma_{G,n})^\top.
 \end{gathered}
\end{equation}
Note that the conditional distribution of $G_n$ given $U$ is standard normal. We choose an orthogonal $Q_n$ with first column $\Omega_n^{1/2}\beta_{2,n}/\gamma_{G,n}$ for $\gamma_{G,n}>0$ and $Q_n=I$ for $\gamma_{G,n}=0$. Let $(G_0,A^\top)^\top=Q_n^\top G_n$ and $Z=(U^\top,G_0)^\top$. Then, $P_Z=P_U\otimes N(0,1)$ is fixed and satisfies \eqref{S-NG-eq-head} by Cauchy--Schwarz and the projection bounds, and the inverse transformation is
\begin{equation}\label{S-NG-eq-affinemap}
 X=\binom{\mu_1}{\mu_n+\Psi_n\mu_1}
 +\begin{pmatrix}
   \Sigma_1^{1/2}&0\\
   \Psi_n\Sigma_1^{1/2}&\Omega_n^{1/2}Q_n
  \end{pmatrix}\begin{pmatrix}U\\G_0\\A\end{pmatrix}.
\end{equation}
Its invertible matrix gives coefficient $(\varpi_n^\top,0)^\top$ for either value of $\gamma_{G,n}$, with $m=m_0+1\le p$ for all large $n$.

For independent $G\sim N(0,1)$, we define $D=2Y-1$ and
\begin{equation}\label{S-NG-eq-headboundary}
 h_Z(\theta,\varpi)=\inf_{b\in\R,t\in\R^m}
   \E\{D(b+t^\top Z)-G\}_+^2,\qquad
 Y\mid Z\sim\operatorname{Bernoulli}\{s(\theta+\varpi^\top Z)\}.
\end{equation}
We observe from Lemma~\ref{S-NG-lem-localization} that we have $0<h_Z\le1/2$ as well as continuity and existence below this threshold. For Gaussian $Z$, we apply rotation and conditional Jensen's inequality to get $h_Z(\theta,\varpi)=h(\theta,\norm{\varpi})$, and we let $h_P=h_Z$ when $Z\sim P$. Note that under~\eqref{S-NG-eq-affineparameters}, the compactness condition below means that the transformed parameters are bounded, $p/n$ is bounded away from zero, and $h_{P_U\otimes N(0,1)}(\theta_n,\varpi_n)-p/n$ is bounded away from zero, and thus we do not need any separate bound on $\Psi_n$ or on the eigenvalues of $\Omega_n$. Next, for $r,\lambda>0$, we define the ordinary state $(b,t,r,\lambda)$ by
\begin{equation}\label{S-NG-eq-newstate}
 \begin{gathered}
 V=b+t^\top Z+rG,\qquad T+\lambda\{s(T)-Y\}=V,\qquad
 U=s(T)-Y,\qquad d=s'(T),\\
 \E U=0,\qquad \E ZU=0,\qquad
 \E\frac{\lambda d}{1+\lambda d}=\kappa_n,\qquad
 \lambda^2\E U^2=\kappa_n r^2.
 \end{gathered}
\end{equation}
The subscript $0$ denotes this state solution. For all finite-$n$ population
quantities, we use $(\kappa_n,\theta_n,\varpi_n)$ and $P_Z$. A prime on
$\ell_F$ denotes a score derivative.

\begin{theorem}\label{S-NG-thm-head}
Assume that \eqref{S-NG-eq-head}--\eqref{S-NG-eq-finitehead} hold and
that $(\kappa_n,\theta_n,\varpi_n)$ lies in a fixed compact subset
$\mathcal K$ of
$\{(\kappa,\theta,\varpi):0<\kappa<h_Z(\theta,\varpi)\}$.
The ordinary state in \eqref{S-NG-eq-newstate} is unique and smooth
on this open set. The maximum likelihood estimator exists uniquely
with probability tending to one, and its fitted coefficients in the
standardized coordinates satisfy
$(\widehat b,\widehat t,\norm{\widehat w})-(b_0,t_0,r_0)\pto0$.
All the following conclusions hold uniformly on $\mathcal K$.

(a) We use Algorithm~1 without modification. Its root satisfies
$\widehat\lambda-\lambda_0\pto0$, and we have
\[
 \widehat F-F_{\rm ext}\pto0,\qquad F\in\{A,B,R,a,c,d\}.
\]

(b) For $F,H\in\{B,R\}$, define
\begin{equation}\label{S-NG-eq-cov}
 \begin{gathered}
 q=s(T_0)-Y,\qquad a_F=\E\{q\ell_F'(Y,V_0)\},\qquad
 J=\E\frac{s'(T_0)}{\{1+\lambda_0s'(T_0)\}^2},\\
 \Sigma_{FH}=\Cov\{\ell_F(Y,V_0),\ell_H(Y,V_0)\}
                 +\lambda_0a_Fa_H/J.
 \end{gathered}
\end{equation}
Let $\Sigma_n$ denote this matrix at the finite-$n$ parameters. Then
\begin{equation}\label{S-NG-eq-normal}
 \begin{gathered}
 d_{\rm BL}\left(
  \mathcal L\left\{\sqrt n
   \binom{\widehat B-B_{\rm ext}}{\widehat R-R_{\rm ext}}\right\},
  N_2(0,\Sigma_n)\right)\longrightarrow0,\\
 \widehat\Sigma-\Sigma_n\pto0,\qquad
 \inf_{(\kappa,\theta,\varpi)\in\mathcal K}
       \lambda_{\min}\{\Sigma(\kappa,\theta,\varpi)\}>0.
 \end{gathered}
\end{equation}
Here $d_{\rm BL}$ is the bounded Lipschitz distance, and
$\widehat\Sigma$ is given by the exact expression in
\eqref{S-xa-variance-estimator} with the quantities in
\eqref{S-xa-j-estimator} and $\ell_R=\ell_L$.
In \eqref{S-NG-eq-normal}, when we apply $\widehat\Sigma^{-1/2}$ to the
error vector, the resulting vector converges in distribution to
$N_2(0,I_2)$. For each fixed $0<\zeta<1$, the marginal normal intervals
of Theorem~\ref{thm:optboot-onefitci} and the $\chi^2_2$ Wald ellipse
have limiting coverage $1-\zeta$ for the realized conditional losses.
\end{theorem}

In Theorem~\ref{thm:optboot-nongaussian}, $\theta_n$ and $\vartheta_n$ correspond to $\theta_n^{\rm orig}$ and $\theta_n$ in \eqref{S-NG-eq-affineparameters}. 

\subsection{Existence and curvature under a conditionally Gaussian design}

\begin{lemma}\label{S-NG-NG-mixed-geometry}
Let $d<\infty$ be fixed and let $a_i\in\R^d$ be independent and
identically distributed with $\E a_i a_i^\top=I_d$ and
$\sup_{\norm v=1}\E\exp\{(v^\top a_i)^2/K^2\}\le2$ for fixed $K$.
Independently, let $A\in\R^{n\times q}$ be standard Gaussian with
$q/n\le\bar\kappa<1$, and define $D=(a_i^\top)_i$, $B=[D,A]$.
There are $\varepsilon,c_*,M_B,c,C>0$, depending only on
$d,K,\bar\kappa$, such that, with probability at least $1-Ce^{-cn}$,
\begin{equation}\label{S-NG-eq-geo-singular}
 \norm B_{\op}\le M_B\sqrt n,\qquad
 \sigma_{\min}(B_I)\ge c_*\sqrt n
 \quad(|I^c|\le\varepsilon n).
\end{equation}
For each fixed $R<\infty$, on the same event and for some $c_R>0$,
\begin{equation}\label{S-NG-eq-geo-curvature}
 c_RI\preceq n^{-1}B_J^\top\diag\{s'(B_Ju)\}B_J
 \preceq M_B^2I/4
 \quad(\norm u\le R,\ |J^c|\le\varepsilon n/2).
\end{equation}
\end{lemma}

\begin{proof}
Let $\delta=1-\bar\kappa$, choose $L^2\ge4\E\|a_i\|^4$ with $L\ge1$, and define $T_i=a_i a_i^\top\ind\{\|a_i\|\le L\}$. Then, we have $\|I_d-\E T_i\|\le\E\|a_i\|^4/L^2\le1/4$, and from Hoeffding on a $1/4$-net of size $9^d$ we get $P_nT_i\succeq I_d/2$ with failure probability $Ce^{-cn}$. For $\varepsilon L^2\le1/4$, we also have
\[
 n^{-1}D_I^\top D_I\succeq P_nT_i-\varepsilon L^2I_d
 \succeq(1/2-\varepsilon L^2)I_d\succeq I_d/4 .
\]
H\"older gives $\E e^{\|a_i\|^2/(dK^2)}\le2$. Exponential Markov for $\sum_i\|a_i\|^2$ gives $\|D\|_{\op}\le M_D\sqrt n$ with exponential failure. Denote this joint event by $\mathcal E_D$.

For fixed $I$, let $k=|I|\ge(1-\varepsilon)n$, $\varepsilon\le\delta/2$. For $q\ge1$, \eqref{F-eq-cgmt} on unit spheres gives
\[
 \Prob\{\sigma_{\min}(A_I)<x\}
 \le2\Prob\{\|g_k\|-\|g_q\|\le x\},
\]
where the Gaussian vectors are independent. Optimizing $\E e^{t\chi_j^2}=(1-2t)^{-j/2}$ gives
\begin{equation}\label{S-NG-eq-geo-chi}
 \Prob(\norm{g_j}>\sqrt j+t)\le e^{-t^2/2},\qquad
 \Prob(\norm{g_j}<\sqrt j-t)\le e^{-t^2/2}.
\end{equation}
Using $\sqrt{k/n}-\sqrt{q/n}\ge\delta/4$ gives $\sigma_{\min}(A_I)\ge(\delta/16)\sqrt n$ with failure probability $4e^{-\delta^2n/512}$. If $P_I$ projects off $A_I$, rotational invariance gives, for nonzero $x$ fixed given $D$,
\[
 \frac{\|P_Ix\|^2}{\|x\|^2}\ \stackrel{d}{=}\ \frac{U}{U+V},
 \qquad U\sim\chi^2_{k-q},\quad V\sim\chi^2_q,\quad U\perp V.
\]
We obtain $U\ge(k-q)/2$ and $U+V\le2k$ with exponential failure from \eqref{S-NG-eq-geo-chi}. On $\mathcal E_D$, each unit $v$ then satisfies $\|P_ID_Iv\|\ge b_0\sqrt n$, where $b_0=\sqrt{\delta/32}$. We then use a $b_0/(2M_D)$-net to get $\sigma_{\min}(P_ID_I)\ge b_0\sqrt n/2$ with the same failure order.

We choose $\varepsilon$ with $-\varepsilon\log\varepsilon-(1-\varepsilon)\log(1-\varepsilon)$ below half the smaller tail exponent. By a union bound over $\sum_{j\le\varepsilon n}\binom nj$ subsets, we get both singular-value bounds simultaneously. For $y=D_Iv+A_Iw$, we have
\[
 (b_0/2)\sqrt n\|v\|\le\|P_Iy\|\le\|y\|,\qquad
 (\delta/16)\sqrt n\|w\|\le\|y\|+M_D\sqrt n\|v\|.
\]
Thus, we obtain the lower bound in \eqref{S-NG-eq-geo-singular}, where the bound on $D_I$ suffices for $q=0$. From \eqref{F-eq-opbound}, we also get the upper bound for $A$.

We have $\|Bu\|^2\le nM_B^2R^2$ for $\|u\|\le R$, and hence at most $\varepsilon n/2$ absolute scores exceed $T=1+M_BR\sqrt{2/\varepsilon}$ and at least $(1-\varepsilon)n$ weights exceed $s'(T)$ after deletions. From their singular-value bound, we obtain \eqref{S-NG-eq-geo-curvature} with $c_R=c_*^2s'(T)$, and the upper bound follows from $s'\le1/4$. This completes the proof.
\end{proof}

Note that the second-moment matrix of $a_i=(1,Z_i^\top)^\top$ is $I_{m+1}$. When $b^2+\norm t^2=1$, we combine $(b+t^\top Z)^2\le 2b^2+2(t^\top Z)^2$ with H\"older's inequality in \eqref{S-NG-eq-head} to obtain $\E e^{(b+t^\top Z)^2/K^2}\le2$ for $K^2=2\max\{K_Z^2,1/\log2\}$.

\begin{lemma}\label{S-NG-lem-localization}
Assume \eqref{S-NG-eq-head}--\eqref{S-NG-eq-finitehead}.
Then $0<h_Z(\theta,\varpi)\le1/2$, and $h_Z$ is continuous.
On each compact subset of
$\{(\theta,\varpi,\kappa):0<\kappa<h_Z(\theta,\varpi)\}$,
for each fixed nonnegative integer $k$, there are $c,C,R>0$ such
that, uniformly for all sufficiently large $n$, with probability
at least $1-Ce^{-cn}$,
\begin{equation}\label{S-NG-eq-fo-localization}
 L_n(b,t,w)\ge c\sqrt{b^2+\norm t^2+\norm w^2}.
\end{equation}
The ordinary maximum likelihood estimate and each estimate with
at most $k$ rows deleted exist uniquely and have norm at most $R$.
Their standardized Hessians, normalized by $n$, are bounded below
uniformly on each fixed coefficient ball. Restricted estimates with
$(b,t)$ in a fixed compact set have the same localization.
\end{lemma}

\begin{proof}
Let $H=(1,Z^\top)^\top$, $H_i=(1,Z_i^\top)^\top$. Positive probabilities of both labels and $\E HH^\top=I$ give
\[
 m_0^2=\min_{\|v\|=1}\E(DH^\top v)_+^2>0,\qquad
 \|(DH^\top v-G)_+\|_2\ge m_0\|v\|-1 .
\]
The latter is uniform on compact parameter sets by dominated convergence. Thus, the infimum defining $h_Z$ lies on a common compact ball, and we get continuity. By independent Gaussian noise, we have positivity, and by evaluation at zero, we obtain $h_Z\le1/2$.

On $\|v\|^2+r^2=1$ with $r\ge0$, we define $L=rG-DH^\top v$ and $N=\|L_+\|_2$. By Gaussian symmetry, we have $N-r\sqrt\kappa\ge r(\sqrt{h_Z}-\sqrt\kappa)$ for $r>0$ and $N\ge m_0$ at zero. Hence, the gap has positive compact minimum $\delta$. With $C_4=\sup\E L^4$, we choose $M^2\ge2C_4/\delta^2$ and $a=\min(L_+/M,1)$ to get
\[
 M\{\E(aL)-r\sqrt\kappa\|a\|_2\}
 \ge N(N-r\sqrt\kappa)-C_4/M^2
 \ge\delta^2/2 ,
\]
proving
\begin{equation}\label{S-NG-eq-fo-witness}
 \E(aL)-r\sqrt\kappa\,\|a\|_{L^2}\ge \delta^2/(2M).
\end{equation}

Use $\ell_Y(x)\ge(-Dx)_+=\max_{0\le a\le1}(-aDx)$ and the conditional Gaussian matrix $-\diag(D_i)A$. At fixed $v,r$, minimize the auxiliary objective in \eqref{F-eq-cgmt} in direction $-g_{p-m}$ to obtain
\[
 \max_{a_i\in[0,1]}\{P_na_i(rG_i-D_iH_i^\top v)
          -r(\|g_{p-m}\|/\sqrt n)(P_na_i^2)^{1/2}\}.
\]
We choose $a_i=\min\{(rG_i-D_iH_i^\top v)_+/M,1\}$ and note that the functions $aL,a^2$ have Lipschitz constants at most $C(1+\|Z\|+|G|)$ and centered moments at most $j!C^j$. From the exponential series, we get the tail $2e^{-cnu^2}$ for small fixed $u$. By a finite mesh and \eqref{S-NG-eq-geo-chi}, we transfer \eqref{S-NG-eq-fo-witness} to the auxiliary value with failure probability $Ce^{-cn}$.

Apply \eqref{F-eq-cgmt} on the nuisance spheres and dual box. Between mesh points, the primary hinge value changes by at most
$\|\Delta v\|P_n\|H_i\|+|\Delta r|\|A\|_{\op}/\sqrt n .$
By sub-Gaussian moments and \eqref{F-eq-opbound}, we bound these coefficients with exponential failure, and we get \eqref{S-NG-eq-fo-localization} by homogeneity. Note that deleting fixed $k$ rows costs at most $k\|[\mathbf1,Z,A]\|_{\op}/n=O(n^{-1/2})$ on the unit sphere. By comparison at zero, we bound the fit, and by comparison at $w=0$, we bound restricted norms by $(\log2+\|v\|P_n\|H_i\|)/c$. By Lemma~\ref{S-NG-NG-mixed-geometry}, we get curvature and uniqueness. This completes the proof.
\end{proof}

\subsection{Consistency under a conditionally Gaussian design}
\begin{proof}[Proof of Theorem~\ref{S-NG-thm-head}(a)]
Consider a convergent parameter sequence in $\mathcal K$.

\OptProofStep{1}{Existence and uniqueness of the state}
Let $H(v)=v\log v+(1-v)\log(1-v)$ on $(0,1)$ with its continuous extension to $[0,1]$, and for $r\ge0$ we define
\begin{equation}\label{S-NG-eq-populationobjective}
 \mathcal F(b,t,r)=\sup_{-Y\le U\le1-Y}
 \left[\E\{U(b+t^\top Z+rG)-H(Y+U)\}
       -r\sqrt\kappa\norm{U}_{L^2}\right].
\end{equation}
Here, the supremum is over measurable functions of $(Z,Y,G)$, and from $|U|\le1$ and $-H\le\log2$ we get a finite locally Lipschitz convex objective. We obtain coercivity from the choice $U=-Da$ in \eqref{S-NG-eq-fo-witness} with $G$ replaced by $-DG$.

Equation~\eqref{S-popobjective} reduces to this objective at unit weight and offset $b+t^\top Z$. We obtain a unique $\lambda>0$ for $r>0$ from the monotonicity proof of \eqref{S-normroot}. From \eqref{S-entropygap}, we have the dual gap $2\E(U'-U)^2$ and hence $L^2$ continuity. We next apply comparison of optimizers as in \eqref{S-popgradient} to get
\[
 \nabla\mathcal F=(\E U,\E ZU,\E GU-\sqrt\kappa\|U\|_{L^2}),\qquad
 \partial_{r+}\mathcal F(b,t,0)=-\sqrt\kappa\|s(b+t^\top Z)-Y\|_{L^2}<0.
\]
Every minimum has $r>0$, and by Gaussian integration by parts we obtain $\E GU=r\E\{d/(1+\lambda d)\}$ and hence equivalence of stationarity and \eqref{S-NG-eq-newstate}.

For two minimizers, the maximizing $U$ at their midpoint also maximizes at both ends by affinity for fixed $U$, and we subtract their proximal equations to obtain
$\Delta b+\Delta t^\top Z+\Delta r(G-cU)=0,\qquad c=\sqrt\kappa/\|U\|_{L^2}.$
If $\Delta r\ne0$, then $G-cU$, $T=b+t^\top Z+r(G-cU)$, and $U=s(T)-Y$ are measurable in $(Z,Y)$, which contradicts independence of $G$. Hence, we have $\Delta r=0$, and from the second-moment identity and norm equation we obtain $\Delta b=\Delta t=0$ and a unique $\lambda$.

\OptProofStep{2}{Smoothness}
We first obtain continuity from coercivity, dominated convergence, and uniqueness. Let $a_T=d/(1+\lambda d)$, $D_0=\E\{U^2/(1+\lambda d)\}$, and $L_h=h_b+h_t^\top Z+h_rG-(\lambda h_r/r)U$. Then, we differentiate the norm equation as in \eqref{S-popHessian} to get
\begin{equation}\label{S-NG-eq-populationhessian}
 h^\top\mathcal F''h
 =\E(a_TL_h^2)+\frac{\lambda}{D_0}\{\E(a_TUL_h)\}^2>0
 \qquad(h\ne0).
\end{equation}
Here, positivity follows from the second-moment matrix of $(1,Z^\top)^\top$ when $h_r=0$. When $h_r\ne0$, we have $\partial_G L_h=h_r/(1+\lambda d)\ne0$ and norm derivative $2\lambda D_0>0$. All fixed-order and response-probability derivatives satisfy integrable polynomial bounds in $\|Z\|+|G|$. We then apply the implicit function theorem and compactness to obtain smoothness and positive lower bounds on $r,\lambda,D_0$ and the least Hessian eigenvalue.

\OptProofStep{3}{Auxiliary values}
Conditional on $(Z,Y)$, we apply \eqref{S-auxiliary} with offset $b+t^\top Z_i$ and $\xi_n=\|g_{p-m}\|^2/n\pto\kappa$. Let $G_i$ be independent standard normals and define $e_y(v;\lambda)=\min_x\{\ell_y(x)+(x-v)^2/(2\lambda)\}$ with continuous extension $e_y(v;0)=\ell_y(v)$. We then apply Fenchel conjugacy at $(\lambda_n^{\rm aux})^2P_nU_i^2=\xi_nr^2$ to get
\begin{equation}\label{S-NG-eq-fo-finite-value}
 \mathcal F_n^{\rm aux}(b,t,r)
 =P_ne_{Y_i}(b+t^\top Z_i+rG_i;\lambda_n^{\rm aux})
       -\frac{r^2\xi_n}{2\lambda_n^{\rm aux}}.
\end{equation}
With $\E,\kappa$ and its root for population values, from sub-Gaussian moments, $\Var(P_nf)=\Var(f)/n$, and finite meshes we get uniform integrand convergence and polynomial local Lipschitz bounds. For moving parameters, we apply dominated convergence to $Y=\ind\{U_0\le s(\theta+\varpi^\top Z)\}$ with independent uniform $U_0$.

As $r\to0$, the brackets in Step~5 of the proof of Theorem~\ref{S-state} follow from $\lambda\le z(1+e^{C+z})$ for $z=|\lambda U_\lambda|$ on a fixed positive-probability set with $|V|\le C$. From the norm equation, we get $\lambda\to0$, and for $\lambda=rl$ we have
\[
 l(0)=\frac{\sqrt\kappa}{\|s(b+t^\top Z)-Y\|_{L^2}},\qquad
 \partial_l(l^2\E U_{rl}^2)=2l\E\{U_{rl}^2/(1+rls'(T_{rl}))\}>0 .
\]
By compactness and these brackets, we obtain uniform root convergence and, on each compact $K$,
\begin{equation}\label{S-NG-eq-fo-uniform-values}
 \sup_{(b,t,r)\in K}
 |\mathcal F_n^{\rm aux}(b,t,r)-\mathcal F(b,t,r)|\pto0
\end{equation}

\OptProofStep{4}{Fitted coordinates}
By Lemma~\ref{S-NG-lem-localization} and uniqueness, we bound the optimum and obtain a positive population gap outside any minimizer neighborhood. At the population $(b,t)$, we get the upper bound on the nuisance ball and dual box from \eqref{F-eq-cgmt2} and the lower bound on excluded radii from \eqref{F-eq-cgmt}. Next, we use a fixed-dimensional mesh with Lipschitz bound $P_n(1+\|Z_i\|)=O_P(1)$ to extend the comparison, and by \eqref{S-NG-eq-fo-uniform-values} we get
\begin{equation}\label{S-NG-eq-fo-coordinate-limit}
 (\widehat b,\widehat t,\norm{\widehat w})\pto(b_0,t_0,r_0).
\end{equation}
State continuity permits replacing $(p-m)/n$ by $p/n$.

\OptProofStep{5}{Empirical scores}
We apply \eqref{S-perturb} at unit weight for bounded continuous $\psi(Z,Y,p)$ of compact $Z$ support with two bounded continuous $p$ derivatives. For $|\epsilon|\|\psi_{pp}\|_\infty<2$, entropy curvature is at least two and its conjugate changes by at most $|\epsilon|\|\psi\|_\infty$, and hence localization holds. For $g_\epsilon(u)=H(Y+u)+\epsilon\psi(Z,Y,Y+u)$, by Step~1 of Theorem~\ref{S-empirical} we get $\partial_\lambda|\lambda U|=|U|g_\epsilon''/(g_\epsilon''+\lambda)>0$ and $|\lambda U|\to\infty$. With auxiliary Lipschitz bound $P_n(1+\|Z_i\|+|G_i|)+\sqrt{\xi_n}=O_P(1)$, by Steps~3--4 the perturbed minimum converges for either sign of $\epsilon$.

From \eqref{S-entropygap} and~\eqref{S-valuebounds}, we get population derivative $-\E\psi(Z,Y,s(T_0))$ at zero, and the strictly concave empirical dual on its box under $[\mathbf1,Z,A]^\top U=0$ has derivative $-P_n\psi(Z_i,Y_i,s(\widehat\eta_i))$. By two-sided secants and convexity in $\epsilon$, we get derivative convergence, and because $s(T_0)\in(0,1)$, by Bernstein approximation, $Z$ truncation, and the logit on $[\delta,1-\delta]$ as $\delta\downarrow0$, for bounded continuous $f$ we obtain
\begin{equation}\label{S-NG-eq-fo-scorelimits}
 P_nf(Z_i,Y_i,\widehat\eta_i)\pto\E f(Z,Y,T_0).
\end{equation}

\OptProofStep{6}{Independent scores}
Note that conditional on the fit, we have
\begin{equation}\label{S-NG-eq-fo-test-score}
 \widehat b+\widehat t^\top Z+\norm{\widehat w}G,
\end{equation}
with $G$ independent of $(Z,Y)$, and by \eqref{S-NG-eq-fo-coordinate-limit} we get convergence with $(Z,Y)$ in distribution to $(Z,Y,V_0)$.

\OptProofStep{7}{Performance measures}
Note that $V_0$ and $T_0$ have positive class-conditional densities because $r_0>0$, and hence we get zero AUC tie mass and a positive minimum of $Q[-(2Y-1)(a+cL)]_+$ on the unit circle. We bound empirical and conditional second moments by Lemmas~\ref{S-NG-lem-localization} and~\ref{S-NG-NG-mixed-geometry} and \eqref{S-NG-eq-fo-test-score} to get Brier convergence and $O_P(M^{-1})$ log-loss tails. We apply Step~3 of the proof of Proposition~\ref{S-functional} for calibration localization, strict convexity, uniform convergence, and offset brackets.

\OptProofStep{8}{Adjusted scores}
Note that $g(l)=\E\{ls'(T_0)/(1+ls'(T_0))\}$ increases strictly with $g(\lambda_0)=\kappa$, and hence $\widehat\lambda\pto\lambda_0$ by \eqref{S-NG-eq-fo-scorelimits} and bracketing. Each adjusted score then changes by at most $|\widehat\lambda-\lambda_0|$, and we use the proximal identity, $|q_i|\le1$, and compact-set uniform continuity to get, for bounded continuous $f$,
\begin{equation}\label{S-NG-eq-fo-adjusted}
 P_nf(Z_i,Y_i,\widetilde V_i)\pto\E f(Z,Y,V_0),
 \qquad P_n\widetilde V_i^2=O_P(1)
\end{equation}
We apply Step~7 to both the adjusted and external distributions, where every parameter sequence in $\mathcal K$ has a convergent subsequence and we obtain uniformity at the finite-$n$ parameters by continuity of the state. This completes the proof.
\end{proof}

\subsection{Inference under a conditionally Gaussian design}

\begin{proof}[Proof of Theorem~\ref{S-NG-thm-head}, part (b)]
Given $(Z_i,Y_i)_{i\le n}$, the nuisance coordinates in \eqref{S-NG-eq-finitehead} are independent standard normals, and by Part~(a), Lemmas~\ref{S-NG-lem-localization} and~\ref{S-NG-NG-mixed-geometry}, and \eqref{S-NG-eq-populationhessian}, the limits and curvature in Theorem~\ref{S-xa-clt} hold with $d_0=m$. Because the covariance \eqref{S-NG-eq-cov} has a positive uniform lower eigenvalue bound and $\widehat\Sigma\pto\Sigma_n$ by Proposition~\ref{S-xa-variance-consistency}, we obtain $N_2(0,I_2)$ after studentization from the Slutsky theorem and hence the marginal normal and joint $\chi_2^2$ coverage. This completes the proof.
\end{proof}

\subsection{Existence and score limits for independent null coordinates}\label{S-NG-U-section}
In case (b), we use $L_n$ for the lower-right block of \eqref{S-NG-eq-affinemap}. With $\beta_{2,n}=0$, the transformed signal is $a_n=\Sigma_1^{1/2}\beta_{1,n}$ on $U=\Sigma_1^{-1/2}(X_1-\mu_1)$, the intercept is $\vartheta_n=\theta_n^{\rm orig}+\mu_1^\top\beta_{1,n}$, and we have $\xi_i\perp(U_i,Y_i)$. Note that these are the coordinates of Theorem~\ref{S-NG-U-main}. For $Z^+=(U^\top,G_0)^\top$ with independent standard normal $G_0$ and zero additional coefficient, by conditional Jensen's inequality, we obtain
\[
 \E[\{D(b+t^\top U+cG_0)-G\}_+^2\mid U,Y,G]\ge\{D(b+t^\top U)-G\}_+^2.
\]
From equality at $c=0$, we obtain the equality of thresholds for $(U,a)$ and $(Z^+,(a^\top,0)^\top)$, with equivalent compactness conditions. The added state equation $\E[G_0\{s(T)-Y\}]=0$ holds by independence. By uniqueness, we then get the same multiplier and covariance in \eqref{S-NG-eq-cov}.

For $m<\infty$ and a fixed distribution $P_Z$ under \eqref{S-NG-eq-head}, we replace the Gaussian component in \eqref{S-NG-eq-finitehead} by independent null coordinates, and in canonical coordinates the independent and identically distributed rows satisfy
\begin{equation}\label{S-NG-U-model}
 \begin{gathered}
 X_i=C_n\binom{Z_i}{A_i},\qquad
 C_n^\top\beta_n=\binom{\varpi_n}{0},\qquad
 \Prob(Y_i=1\mid Z_i,A_i)=s(\theta_n+\varpi_n^\top Z_i),\\
 A_i\perp(Z_i,Y_i),\qquad
 \E A_{ik}=0,\qquad \E A_{ik}^2=1,\qquad
 \sup_{n,k}\E\exp(A_{ik}^2/K^2)\le2 .
 \end{gathered}
\end{equation}
Here $C_n$ is deterministic and invertible, $K<\infty$ is fixed, and $A_i\in\R^{p-m}$ has mutually independent coordinates with distributions that may depend on $n$ and $k$, where an independent test observation follows the same row distribution. By the affine invariance after \eqref{S-NG-eq-finitehead}, we also allow a deterministic translation of $X_i$ in the fitted intercept without restricting the singular values of $C_n$.

Let $B$ have rows $b_i^\top$ for $b_i=(1,Z_i^\top,A_i^\top)^\top$. In these standardized coordinates, we define $u=(b,t,w)$ and the empirical log loss $L_n(u)=P_n\ell_R(Y_i,b_i^\top u)$. Here, $\mathcal D$ denotes the training data, and we define the boundary, states, and covariance by \eqref{S-NG-eq-headboundary}, \eqref{S-NG-eq-newstate}, and \eqref{S-NG-eq-cov} at the finite-$n$ parameters unless limiting parameters are specified. Let $q=p-m$ denote the null dimension, and note that when $m=0$, we omit expressions in $Z$ and $t$.

\begin{theorem}\label{S-NG-U-main}
Assume that \eqref{S-NG-eq-head} and \eqref{S-NG-U-model} hold and
that, for all sufficiently large sample sizes,
$(\kappa_n,\theta_n,\varpi_n)$ lies in a fixed compact subset
$\mathcal K$ of
$\{(\kappa,\theta,\varpi):0<\kappa<h_Z(\theta,\varpi)\}$,
where $\kappa_n=p/n$. We use the unpenalized logistic fit with an
intercept and Algorithm~1 without modification.

(a) There is $R<\infty$ such that
\[
 \Prob\{\widehat u\text{ exists uniquely and }
                \norm{\widehat u}\le R\}\longrightarrow1.
\]
For each fixed $R'<\infty$, there are $c_{R'}>0$ and
$\varepsilon>0$ such that
\[
 \Prob\left\{
 \inf_{|J|\le\varepsilon n}\inf_{\norm u\le R'}
 \lambda_{\min}\left(n^{-1}\sum_{i\notin J}
           s'(b_i^\top u)b_ib_i^\top\right)\ge c_{R'}
          \right\}\longrightarrow1.
\]

(b) For the ordinary state $(b_0,t_0,r_0,\lambda_0)$ in
\eqref{S-NG-eq-newstate},
\begin{equation}\label{S-NG-U-firstorder}
 (\widehat b-b_0,\widehat t-t_0,
   \norm{\widehat w}-r_0,\widehat\lambda-\lambda_0)\pto0,
 \qquad
 \norm{\widehat w}_\infty=O_P\{\sqrt{\log n/n}\}.
\end{equation}
Let $\widehat Q_T$ and $\widehat Q_V$ be the empirical distributions
of $(Z_i,Y_i,\widehat\eta_i)$ and $(Z_i,Y_i,\widetilde V_i)$.
For an independent test observation, let
$\widehat V_{\rm new}=\widehat b+\widehat t^\top Z+A^\top\widehat w$.
Then
\begin{equation}\label{S-NG-U-scores}
 \begin{gathered}
 d_{\rm BL}\{\widehat Q_T,\mathcal L(Z,Y,T_0)\}\pto0,
 \qquad
 d_{\rm BL}\{\widehat Q_V,\mathcal L(Z,Y,V_0)\}\pto0,\\
 d_{\rm BL}\{\mathcal L(Z,Y,\widehat V_{\rm new}\mid\mathcal D),
                      \mathcal L(Z,Y,V_0)\}\pto0,\\
 \widehat F-F_{\rm ext}\pto0\quad(F\in\{A,B,R,a,c,d\}),
 \qquad \widehat\Sigma-\Sigma_n\pto0 .
 \end{gathered}
\end{equation}
Here we evaluate $F_{\rm ext}$ under the actual distribution of the
predictors, and $\Sigma_n$ is given by \eqref{S-NG-eq-cov}. We
compute $\widehat\Sigma$ precisely from
\eqref{S-xa-variance-estimator} with the quantities in
\eqref{S-xa-j-estimator} and $\ell_R=\ell_L$.

(c) For the realized conditional Brier and log losses,
\begin{equation}\label{S-NG-U-clt}
 d_{\rm BL}\left(
  \mathcal L\left\{\sqrt n
       \binom{\widehat B-B_{\rm ext}}{\widehat R-R_{\rm ext}}\right\},
       N_2(0,\Sigma_n)\right)\longrightarrow0,
 \qquad
 \sqrt n\widehat\Sigma^{-1/2}
       \binom{\widehat B-B_{\rm ext}}{\widehat R-R_{\rm ext}}
       \dto N_2(0,I_2).
\end{equation}
For each fixed $0<\zeta<1$, the marginal normal intervals in
Theorem~\ref{thm:optboot-onefitci} and the $\chi^2_2$ Wald ellipse
have limiting coverage $1-\zeta$. No symmetry, zero-third-moment
condition or fourth-moment matching is required for these results.
All conclusions are uniform on $\mathcal K$ and over the coordinate
distributions in \eqref{S-NG-U-model} with the fixed bound $K$.
\end{theorem}

Let $\underline\kappa=\inf_{\mathcal K}\kappa$, $\overline\kappa=\sup_{\mathcal K}\kappa$, and $K_*=\max(K,2)$, under which Gaussian comparison entries are allowed. We prove convergence along any subsequence with convergent parameters and denote a limiting objective value by $f$. Here, the bounds depend only on $P_Z,m,K_Z,K,\mathcal K$ and fixed derivative orders. Using compactness and continuity of the Gaussian state, we obtain convergence at the finite-$n$ parameters uniformly over allowed coordinate distributions.

\begin{lemma}\label{S-NG-U-geometry-lemma}
Assume that \eqref{S-NG-eq-head} and \eqref{S-NG-U-model} hold and
$(p-m)/n$ remains in a compact subset of $(0,1/2)$.
There are $\varepsilon,a,M,c,C>0$ such that
\begin{equation}\label{S-NG-U-geometry}
 \Prob\left\{\norm B_{\op}\le M\sqrt n,
 \min_{|I^c|\le\varepsilon n}\sigma_{\min}(B_I)\ge a\sqrt n\right\}
 \ge1-Ce^{-cn}.
\end{equation}
For each fixed $R$, on this event,
\begin{equation}\label{S-NG-U-curvature}
 \inf_{|J|\le\varepsilon n/2}\inf_{\norm u\le R}
 \lambda_{\min}\left\{n^{-1}\sum_{i\notin J}
 s'(b_i^\top u)b_ib_i^\top\right\}\ge c_R>0.
\end{equation}
In this lemma, the distributions of the coordinates may differ across
rows.
\end{lemma}

\begin{proof}
\OptProofStep{1}{Row differences and the independent component}
We choose $\varepsilon_0>0$ with $2(p-m)/\{n(1-\varepsilon_0)\}$ bounded strictly below one for large $n$. For fixed $|I|\ge(1-\varepsilon_0)n$, we define $(D_Iv)_j=(v_{2j-1}-v_{2j})/\sqrt2$ for $1\le j\le N=\lfloor|I|/2\rfloor$ on rows $I$ in order to get $D_IB_I=[0,\widetilde Z,\widetilde A]$. The entries of $\widetilde A$ are independent centered unit-variance differences $T=(U-V)/\sqrt2$ of disjoint entries in the same column, and
$\E e^{T^2/(2K_*^2)}\le(\E e^{U^2/K_*^2}\E e^{V^2/K_*^2})^{1/2}\le2.$
We apply Theorem~1.2 of \citet{litvak2012smallest} with $r=4$, fixed moment bound $\mu>1$, and $a_3=a_4=1$ from unit variances, which meet the theorem's strict lower bound on $a_4$. From Proposition~2.2, we get the exponential norm bound, and because $N/q$ is bounded away from one in $(1,\infty)$, the logarithmic aspect-ratio condition holds for large $n$. By Equation~(22), we obtain uniform constants and hence $\Prob\{\sigma_{\min}(\widetilde A)<a_A\sqrt n\}\le e^{-c_An}$.

\OptProofStep{2}{The remaining coordinates and all deletions}
Let $P$ project off the columns of $\widetilde A$, with $\operatorname{tr}P\ge\delta n$ for fixed $\delta>0$. Given $\widetilde A$, $\widetilde Zv$ has independent centered sub-Gaussian entries of unit variance for each unit $v\in\R^m$. By Theorem~1.1 of \citet{rudelson2013hanson} with threshold $\operatorname{tr}P/2$,
\[
 \|P\|_{\op}=1,\qquad \|P\|_F^2=\operatorname{tr}P,\qquad
 \Prob\{\|P\widetilde Zv\|^2<\delta n/2\mid\widetilde A\}\le2e^{-c_Zn}.
\]
We have $\|\widetilde Z\|_{\op}\le M_Z\sqrt n$ with exponential failure by Lemma~\ref{S-NG-NG-mixed-geometry} at $q=0$ and $a_i=(1,Z_i^\top)^\top$. With mesh $\eta=\min\{1,\sqrt\delta/(4M_Z)\}$ and net size $(1+2/\eta)^m$, we get $\sigma_{\min}(P\widetilde Z)\ge a_Z\sqrt n$, and for $m=0$ omit this projection with $M_Z=0,a_Z=1$.

By Proposition~2.2 of \citet{litvak2012smallest} for $A$ and Lemma~\ref{S-NG-NG-mixed-geometry} at $q=0$ for the intercept and $Z$ columns, we get $\|B\|_{\op}\le M\sqrt n$. For $y=D_IB_Iu=\widetilde Zt+\widetilde A w$ we have
\[
 \begin{aligned}
 a_Z\sqrt n\|t\|&\le\|y\|\le\|B_Iu\|,\\
 a_A\sqrt n\|w\|&\le\|y\|+M_Z\sqrt n\|t\|,\\
 |b|\sqrt{|I|}&\le\|B_Iu\|+M\sqrt n\|(t,w)\|.
 \end{aligned}
\]
We thus have $\sigma_{\min}(B_I)\ge a\sqrt n$ with uniform exponential failure. Next, we choose $\varepsilon\le\varepsilon_0$ with binary entropy below half this exponent and apply the union bound in Lemma~\ref{S-NG-NG-mixed-geometry} to get \eqref{S-NG-U-geometry}.

When $M^2R^2/T^2\le\varepsilon/2$ for $T$ and $P_n(b_i^\top u)^2\le M^2R^2$, at least $(1-\varepsilon)n$ rows have weight at least $s'(T)$ after deleting $J$. We thus get \eqref{S-NG-U-curvature} with $c_R=a^2s'(T)$ from the singular-value bound. This completes the proof.
\end{proof}

\begin{lemma}\label{S-NG-U-ridge}
Let $\rho>0$, and let the convex $C^3$ losses $\ell_i$ depend
only on $(Z_i,Y_i)$ and satisfy
$|\ell_i'|+|\ell_i''|+|\ell_i'''|\le D_0<\infty$.
We define $v=(b,t^\top)^\top$. When $|\delta|\le\rho/2$, let
\[
 \mathcal V_n(A)=\min_{u=(v,w)}
 \{P_n\ell_i(b_i^\top u)+\rho\norm u^2/2
                  +\zeta^\top v+\delta\norm w^2/2\}.
\]
For an independent standard Gaussian matrix $G$ and each bounded
$C^3$ function $\varphi$ with bounded derivatives,
\begin{equation}\label{S-NG-U-value}
 |\E\varphi\{\mathcal V_n(A)\}-\E\varphi\{\mathcal V_n(G)\}|
 \le C_{\rho,D_0,\varphi}(\log n)^{3/2}n^{-1/2}+o(1).
\end{equation}
This bound is uniform over the allowed distributions of the coordinates,
including different distributions across rows.
\end{lemma}

\begin{proof}
\OptProofStep{1}{Conditional linear forms}
We observe that the minimizer $u^{(-k)}$ at $w_k=0$ is measurable in $\mathcal F_{-k}=\sigma\{(Z_j,Y_j)_{j\le n},A_{-k}\}$ and has gradient $g_k(x)e_k$ for $\rho_*=\rho/2$ and $x=A_{ik}$, where
\[
 \begin{aligned}
 g_k(x)&=n^{-1}\sum_{j\ne i}A_{jk}\ell_j'(b_j^\top u^{(-k)})
             +x\ell_i'(b_i^\top u^{(-k)})/n,\\
 \rho_*|w_k(x)|^2&\le\rho_*\|u(x)-u^{(-k)}\|^2
             \le-g_k(x)w_k(x),
 \end{aligned}
\]
by strong convexity, and hence $|w_k(x)|\le|g_k(x)|/\rho_*$. For centered $T$ with independent copy $T'$, by symmetry and $\E|T-T'|^{2r}\le2^{2r+1}K_*^{2r}r!$, we obtain
\[
 \begin{aligned}
 \E e^{tT}
 &\le\E e^{t(T-T')}
 \le1+\sum_{r\ge1}\frac{2^{2r+1}K_*^{2r}r!t^{2r}}{(2r)!}
 \le e^{CK_*^2t^2}.
 \end{aligned}
\]
The first inequality is Jensen's. The last uses $(2r)!\ge2^r(r!)^2$. Independence and exponential Markov give
\[
 \Prob\{|\textstyle\sum_j a_jT_j|>z\}
 \le2\exp\{-cz^2/(K_*^2\textstyle\sum_j a_j^2)\}.
\]
Given $\mathcal F_{-k}$, the coefficients of the random entries in $ng_k$ are bounded by $D_0$. Thus, for each $D>0$, an event independent of $x$ with probability at least $1-n^{-D}$ gives
$\sup_{|x|\le C_D\sqrt{\log n}}|w_k(x)|\le C_{\rho,D_0,D}\sqrt{\log n/n}.$
Proposition~2.2 of \citet{litvak2012smallest} and Lemma~\ref{S-NG-U-geometry-lemma} give $\|B(0)\|_{\op}\le C\sqrt n$ with exponentially small failure probability. Varying $x$ adds at most $C_D\sqrt{\log n}$ to the norm.

\OptProofStep{2}{The value derivatives and replacement}
Let $L(x,u)$ be the objective, $H=L_{uu}$, $z=b_i(x)$, $w=w_k(x)$, $q_i=\ell_i'(z^\top u)$ and $d_i=\ell_i''(z^\top u)$. For the coordinate vector $e_k$ of $w_k$, differentiation of $L_u=0$ gives
\[
 \begin{aligned}
 g=L_{xu}&=(q_i e_k+d_iwz)/n,\qquad u_x=-H^{-1}g,\\
 \mathcal V_x&=q_iw/n,\qquad
 \mathcal V_{xx}=d_iw^2/n-g^\top H^{-1}g,\\
 \mathcal V_{xxx}
 &=L_{xxx}+3L_{xxu}[u_x]+3L_{xuu}[u_x,u_x]+L_{uuu}[u_x,u_x,u_x]\\
 &=\frac{\ell_i'''w^3}{n}
 +\frac{3\{\ell_i'''w^2z^\top u_x+2d_iw(e_k^\top u_x)\}}n\\
 &\quad+\frac{3\{\ell_i'''w(z^\top u_x)^2+2d_i(e_k^\top u_x)(z^\top u_x)\}}n
 +P_n\ell_j'''(b_j^\top u_x)^3 .
 \end{aligned}
\]
On the event in Step~1, $\|u_x\|\le C\sqrt{\log n}/n$ and $P_n|b_i^\top a|^3\le n^{-1}\|Ba\|_\infty\|Ba\|^2\le C\sqrt n\|a\|^3$, and by substitution we obtain
\[
 |\mathcal V_x|\le C\sqrt{\log n}\,n^{-3/2},\quad
 |\mathcal V_{xx}|\le C\log n\,n^{-2},\quad
 |\mathcal V_{xxx}|\le C(\log n)^{3/2}n^{-5/2}.
\]
We observe that $f(x)=\varphi\{\mathcal V_n(A(x))\}$ satisfies this third-derivative bound by the chain rule, and given $(Z,Y)$ and other entries, its quadratic Taylor terms match on the $x$-independent Step~1 event. Note that on $|x|\le C_D\sqrt{\log n}$, the expected remainder is $O((\log n)^{3/2}n^{-5/2})$ by bounded third moments, and outside it $2\|\varphi\|_\infty+|f'(0)x|+|f''(0)x^2|/2$ has expectation $O(n^{-D})$ by sub-Gaussian tails. Off this event, we bound the expectation by $2\|\varphi\|_\infty n^{-D}$ and sum over $nq=O(n^2)$ for $D>4$ to get \eqref{S-NG-U-value}. This completes the proof.
\end{proof}

\begin{proof}[Proof of Theorem~\ref{S-NG-U-main}(a)]
\OptProofStep{1}{Values for fixed ridge penalties}
For fixed $\rho>0$, we have $\|u\|^2\le2\log2/\rho$ by comparison at zero, and with $\rho(b^2+\|t\|^2+r^2)/2$ in \eqref{S-NG-eq-populationobjective}, the fixed-parameter dual root and uniform limit \eqref{S-NG-eq-fo-finite-value} hold with mesh Lipschitz bound $P_n(1+\|Z_i\|)+\rho R$. We then obtain the Gaussian penalized limit from Steps~3--4 of the proof of Theorem~\ref{S-NG-thm-head}(a), and apply Lemma~\ref{S-NG-U-ridge} to a smooth test equal to zero within $\epsilon/2$ of this limit and one beyond $\epsilon$ to get the same limit for the original design.

\OptProofStep{2}{Removing the penalty}
Let $R_0$ be the Gaussian localization radius in Lemma~\ref{S-NG-lem-localization}. Penalized optimality gives $\|\widehat u_\rho^G\|\le\|\widehat u_0^G\|$ and
\[
 \frac\rho4\norm{\widehat u_\rho^U}^2
 \le\mathcal V_n^U(\rho)-\mathcal V_n^U(\rho/2),\qquad
 \mathcal V_n^G(\rho)-\mathcal V_n^G(\rho/2)
 \le\frac\rho4\norm{\widehat u_{\rho/2}^G}^2.
\]
We have $\|\widehat u_\rho^U\|^2\le R_0^2+o_P(1)$ for fixed $\rho$ by common deterministic limits, and let $R=R_0+2$ and $\rho<c_R/[2(R_0+1)]$. Specifically, by Lemma~\ref{S-NG-U-geometry-lemma} with $h_Z\le1/2$, on $\|v-\widehat u_\rho^U\|=1$ we get
$\nabla L_n(v)^\top(v-\widehat u_\rho^U)\ge c_R-\rho\|\widehat u_\rho^U\|>0.$
We thus obtain a unique interior global minimum from convexity, full rank, and positive weights.

\OptProofStep{3}{Deleted fits}
We observe under deletion of $J$ that the gradient at the full fit changes by at most $\|B_J\|_{\op}\sqrt{|J|}/n\le M\sqrt{|J|/n}$, and for $\eta\le\varepsilon/2$ with $M\sqrt\eta<c_{R+1}/2$ we thus get bounded unique minima simultaneously for $|J|\le\eta n$ from Step~2. Next, we integrate curvature to bound their differences from the full fit by $c_{R+1}^{-1}M\sqrt{|J|/n}$, and $C\|b_i\|/n$ for one deletion, uniformly on $\mathcal K$. This completes the proof.
\end{proof}

\begin{proof}[Proof of Theorem~\ref{S-NG-U-main}(b)]
\OptProofStep{1}{Perturbed values and empirical scores}
We add the linear and quadratic auxiliary terms in Lemma~\ref{S-NG-U-ridge} for $\rho>0$, and let $H(p)=p\log p+(1-p)\log(1-p)$. Next, for bounded continuous $\psi(Z,Y,p)$ with compact $Z$ support and three bounded continuous $p$ derivatives, we define
\[
 g_{i,\epsilon}(U)=H(Y_i+U)+\epsilon\psi(Z_i,Y_i,Y_i+U),
 \qquad -Y_i\le U\le1-Y_i,
\]
For $|\epsilon|\|\psi_{pp}\|_\infty\le2$, we have $g_{i,\epsilon}''\ge H''/2\ge2$, and its conjugate differs from logistic loss by at most $|\epsilon|\|\psi\|_\infty$. Its first three derivatives are $U,1/g''$ and $-g'''/(g'')^3$, which are bounded because $|H'''|\le(H'')^2$, and thus we apply Lemma~\ref{S-NG-U-ridge} with fixed $D_0$.

In the Gaussian objective \eqref{S-NG-eq-populationobjective}, we add $\rho(b^2+\|t\|^2+r^2)/2+\zeta^\top(b,t)+\delta r^2/2$ and perturb the entropy by $\epsilon\psi$. From comparison at zero, we get $\rho\|u\|^2/4-\|\zeta\|\|u\|\le\log2+2|\epsilon|\|\psi\|_\infty$. Then, we apply Part~(a) on this ball with Lipschitz constant $P_n(1+\|Z_i\|)+\|\zeta\|+\rho R$. From Step~5 of the proof of Theorem~\ref{S-NG-thm-head}(a), we obtain the entropy bounds, and convexity holds for $|\delta|\le\rho/2$. By Lemma~\ref{S-NG-U-ridge}, we obtain a common limiting minimum $f(\rho,\zeta,\delta,\epsilon)$.

We have unique optimizers $(v_\rho,r_\rho,U_\rho)$ at zero auxiliary parameters by the quadratic and entropy gaps. Next, we combine \eqref{S-valuebounds} and \eqref{S-entropygap} at $W=1$ to get
\[
 \partial_{\zeta_j}f=(v_\rho)_j,\qquad \partial_\delta f=r_\rho^2/2,\qquad
 \partial_\epsilon f=-\E\psi(Z,Y,Y+U_\rho).
\]
When we minimize over the empirical primal quadratic, we obtain
\[
 \max_{-Y_i\le U_i\le1-Y_i}
 \left[-P_ng_{i,\epsilon}(U_i)
 -\frac12(B^\top U/n+\widetilde\zeta)^\top
 R_{\rho,\delta}^{-1}(B^\top U/n+\widetilde\zeta)\right],
\]
where $R_{\rho,\delta}=\diag\{\rho I_{m+1},(\rho+\delta)I_q\}$ and $\widetilde\zeta=(\zeta^\top,0_q^\top)^\top$. The value is convex in $\epsilon$ and concave in $\zeta_j,\delta$, with derivatives $-P_n\psi(Z_i,Y_i,s(\widehat\eta_{i,\rho}))$, $(\widehat v_\rho)_j$ and $\|\widehat w_\rho\|^2/2$ at zero. From two-sided secants, we get derivative convergence, and by Bernstein approximation, $Z$ truncation and the restricted logit as in \eqref{S-NG-eq-fo-scorelimits}, we obtain the coefficient, nuisance-norm and empirical-score limits.

\OptProofStep{2}{Unpenalized and independent scores}
Integrating the curvature bound on the localization ball, we obtain
\[
 \norm{\widehat u_\rho-\widehat u_0}\le C\rho,\qquad
 P_n\{b_i^\top(\widehat u_\rho-\widehat u_0)\}^2\le C\rho^2.
\]
Note that the Gaussian ridge fits are uniformly bounded for small $\rho$. As $n\to\infty$ and then $\rho\downarrow0$, the penalty converges uniformly to zero, and by uniqueness in Steps~1--2 of Theorem~\ref{S-NG-thm-head}(a) we obtain convergence of minimizers, and hence of their fitted-score distributions by the dual gap.

For \eqref{S-NG-U-firstorder}, let $u^{(-k)}$ be the minimum-norm minimizer on $\{\|u\|\le R,w_k=0\}$ with $R$ above localization. When we add $a\|u\|^2$, its unique minimizer is measurable without column $k$ by grid approximation, with norm at most that of any loss minimizer and excess loss at most $aR^2$. We thus get this limit as $a\downarrow0$ by compactness.

Let $\Delta=\widehat u-u^{(-k)}$ and $g_k=P_n A_{ik}\{s(b_i^\top u^{(-k)})-Y_i\}$, with constrained score $-a_k u^{(-k)}+g_ke_k$ for $a_k\ge0$ and $a_k(\|u^{(-k)}\|-R)=0$. Note that when $a_k>0$, we have $(u^{(-k)})^\top\widehat u\le R^2=\|u^{(-k)}\|^2$, and we integrate curvature between the two fits to get
\[
 c_R|\widehat w_k|^2\le c_R\|\Delta\|^2
 \le a_k(u^{(-k)})^\top\Delta-g_k\widehat w_k\le |g_k\widehat w_k|.
\]
The residuals in $g_k$ are bounded and measurable without column $k$. By Step~1 of the proof of Lemma~\ref{S-NG-U-ridge}, we obtain $\Prob(|g_k|>t\mid\mathcal F_{-k})\le2e^{-cnt^2}$. From a union bound over $q=O(n)$ and $|\widehat w_k|\le|g_k|/c_R$, we get the claimed rate.
For a new observation, by replacing the independent coordinates and
applying a Taylor expansion through order two, we obtain
\[
 \left|\E_A f(Z,Y,b+t^\top Z+A^\top w)
 -\E_G f(Z,Y,b+t^\top Z+\norm w G)\right|
 \le C_f\sum_k|w_k|^3
 \le C_f\norm w_\infty\norm w^2
\]
for tests with bounded third score derivative. The first two moments match and third absolute moments are uniformly bounded, and hence the right side tends to zero conditionally. We obtain the result for bounded Lipschitz tests by convolution, and from coefficient convergence we get the external-score limit. Finally, from Step~8 of the proof of Theorem~\ref{S-NG-thm-head}(a), we have $\widehat\lambda-\lambda_0=o_P(1)$, from which we also bound the adjusted-score error.

\OptProofStep{3}{Performance and covariance}
Bounded coefficients and \eqref{S-NG-U-geometry} give $P_n\widehat\eta_i^2+P_n\widetilde V_i^2=O_P(1)$ and conditional external second moment at most $C(1+\|\widehat u\|^2)$. On a compact box, rounding values on a width-$a$ mesh and extending by the Lipschitz bound gives a finite uniform $3a$-net of the unit bounded Lipschitz class. The moment bounds remove tails, proving \eqref{S-NG-U-scores}. Because $r_0>0$, both limiting class densities are positive. Step~7 of the proof of Theorem~\ref{S-NG-thm-head}(a) with these tail bounds gives all six performance limits.

On a fixed ball containing all localized row-deleted fits, we choose the minimum-norm row-$i$-deleted minimizer $\widehat u_{-i}$, which is independent of row~$i$. On the Step~3 event in part~(a), it equals the unrestricted fit, and we obtain
\[
 |\widehat\eta_i-b_i^\top\widehat u_{-i}|
 \le\|b_i\|\,\|\widehat u-\widehat u_{-i}\|
 \le C\|b_i\|^2/n\le CM^2.
\]
By~\eqref{S-NG-eq-head} and Lemma~\ref{S-NG-U-ridge}, we bound the conditional fourth moments of $b_i^\top\widehat u_{-i}$ and obtain $P_n\widetilde V_i^4=O_P(1)$ from the Markov bound, root consistency, and bounded residuals. Because the loss products have quadratic growth, their empirical and population tails have order $O_P(M^{-2})$, and we obtain covariance convergence by truncation. From the bounded integrands in~\eqref{S-xa-j-estimator}, we have $\widehat a_F-a_F,\widehat J-J=o_P(1)$, and from~\eqref{S-xa-variance-estimator}, the positive lower bound on $J$, and compactness we establish uniform consistency. This completes the proof.
\end{proof}

\subsection{Joint inference for independent null coordinates}\label{S-null-inference}
Let $R_0$ be larger than the original and Gaussian uniform localization radii, and let $\mathcal J$ be smooth, convex, and nondecreasing, equal to zero on $[0,R_0^2]$ with at least linear growth, such as $\mathcal J(a)=\int_{R_0^2}^a(a-z)e^{-1/(z-R_0^2)}\,dz$ for $a>R_0^2$ and zero otherwise. With $\rho_n=n^{-2}$, for the normal approximation we define
\begin{equation}\label{S-NG-U-eq1}
 u=\argmin_v\{L_n(v)+\mathcal J(\norm v^2)
                           +\rho_n\norm v^2/2\}.
\end{equation}
We let $\eta_i=b_i^\top u$ and $d_i=s'(\eta_i)$ at this auxiliary fit $u$, and define $\lambda^\dagger$ by
\begin{equation}\label{S-NG-U-eq2}
 P_n\frac{\lambda^\dagger d_i}{1+\lambda^\dagger d_i}
                       +n^{-2}\lambda^\dagger=\kappa_n.
\end{equation}

\begin{proof}[Proof of Theorem~\ref{S-NG-U-main}(c)]
Under Gaussian null coordinates, from Theorem~\ref{S-NG-thm-head}(b) we get the normal limit at $R_F^G$. By Lemma~\ref{S-NG-U-smooth}, the scaled errors for either design change by $o_P(1)$. We then apply Lemma~\ref{S-NG-U-risk} to their distributions and substitute the original risk for $R_F^G$. By the bounded Lipschitz triangle inequality, the first limit in \eqref{S-NG-U-clt} follows.

Let $E_n=\sqrt n(\widehat B-B_{\rm ext},\widehat R-R_{\rm ext})^\top$, where $cI_2\preceq\Sigma_n\preceq CI_2$ by Theorem~\ref{S-NG-thm-head}(b) and compactness. When $\Sigma_n\to\Sigma_*$, by part~(b) and Slutsky, we have
\[
 E_n\dto N_2(0,\Sigma_*),\qquad \widehat\Sigma\pto\Sigma_*\succ0,\qquad
 \widehat\Sigma^{-1/2}E_n\dto N_2(0,I_2).
\]
We thus get normal and $\chi_2^2$ coverage from marginal quantiles and the squared norm after restriction to the target ranges. Note that uniformity follows by compactness along every allowed parameter and coordinate sequence. This completes the proof.
\end{proof}

\subsection{Derivative bounds for entry replacement}\label{S-null-derivatives}
\begin{lemma}\label{S-NG-U-smooth}
For all sufficiently large $n$ and each real null-coordinate
array, the problem in \eqref{S-NG-U-eq1} has a unique minimizer that
varies smoothly with the array and satisfies $\norm u\le R_1<\infty$,
and at this minimizer the inverse Hessian of the criterion has norm at
most $n^2$. Equation~\eqref{S-NG-U-eq2} also has a unique root in
$(0,n^2)$, and the inverse of the derivative of its left-hand side is
at most $n^2$. For the original and Gaussian designs,
$\norm{u-\widehat u}+|\lambda^\dagger-\widehat\lambda|
=O_P(n^{-2})$. When we substitute these auxiliary quantities, the
scaled error of the realized risk in \eqref{S-NG-U-clt} changes by
$o_P(1)$.
\end{lemma}
\begin{proof}
We let $\mathcal J(R_1^2)>4\log2+1$ and get $\|u\|\le R_1$ from $\log2$ at zero, with uniqueness and smoothness from Hessian lower bound $n^{-2}I$. In \eqref{S-NG-U-eq2}, the trace starts at zero with derivative at least $n^{-2}$ and exceeds $\kappa_n$ at $n^2$.

On the localization event, $\|\widehat u\|<R_0$. Comparing the ridge objective at its minimizer and at the unpenalized fit $\widehat u$ bounds the ridge minimizer's norm by $\|\widehat u\|$. Its radial penalty equals zero, and it minimizes \eqref{S-NG-U-eq1}. We then integrate the score equations with part~(a) to get $\|u-\widehat u\|\le n^{-2}R_0/c$.

We observe that $P_n\eta_i^2\le M^2R_1^2$ for either fit when $\|B\|_{\op}\le M\sqrt n$, and let $1-\delta>\overline\kappa$ with $M_1$ large enough that $|\eta_i|\le M_1$ for at least $(1-\delta)n$ scores. For $d_*=s'(M_1)$ and $\lambda_+$ with $(1-\delta)\lambda_+d_* /(1+\lambda_+d_*)>\overline\kappa$, both roots lie in $[\lambda_-,\lambda_+]$ with $\lambda_-=\underline\kappa/(1/4+n^{-2})$, and both trace derivatives are at least $c_\lambda=(1-\delta)d_* /(1+\lambda_+/4)^2$. Because the score derivative is bounded, it holds that
$|\lambda^\dagger-\widehat\lambda|\le C\{\norm B_{\op}\norm{u-\widehat u}/\sqrt n+n^{-2}\}.$
We bound the empirical and conditional external means of absolute score changes by $\|B\|_{\op}\|u-\widehat u\|/\sqrt n$ and $\|u-\widehat u\|$ from $\E b_{\rm new}b_{\rm new}^\top=I$, and get $o_P(1)$ scaled error under bounded loss derivatives and residuals. This completes the proof.
\end{proof}

We consider a comparison array with independent null entries from their original or standard normal distributions, independent of $(Z,Y)$. Let $L=\log(en)$, where up to two entries may be fixed in $[-C_0\sqrt L,C_0\sqrt L]$ and the random entries have sub-Gaussian bound $K_*$. On the root interval $[\lambda_-,\lambda_+]$ from the proof of Lemma~\ref{S-NG-U-smooth} at its deterministic radius and common operator bound, we define
\begin{equation}\label{S-NG-U-eq3}
 g_{\lambda,F}(y,x)=\ell_F(y,x+\lambda\{s(x)-y\}),\qquad
 \tau_\lambda(x)=\frac{\lambda s'(x)}{1+\lambda s'(x)}.
\end{equation}
Note that on this interval, $|s^{(r)}|\le C_rs'$, $s'(x+h)/s'(x)\le e^{|h|}$, and $|\lambda(s-y)|\le\lambda_+$, and by Lemma~\ref{S-tilted-geometry} we bound first score derivatives by $C$ and those of order $r\ge2$ by $C_rs'$. We retain these bounds under any fixed number of $\lambda$ derivatives because $x+\lambda(s-y)$ is affine in $\lambda$ with bounded derivative.

We define the empirical perturbation $\mathcal F_\epsilon(u)=P_nL_\epsilon(Y_i,b_i^\top u)+P_\epsilon(u)$ with $L_\epsilon=\ell_R+\epsilon g$ and $P_\epsilon=\mathcal J(\norm u^2)+\rho_n\norm u^2/2$, where $g$ is a function in~\eqref{S-NG-U-eq3} or a required $\lambda$ derivative. Note that for small fixed $|\epsilon|$, we have $L_\epsilon''\ge s'/2$, and the log-loss perturbation differs from log loss by a bounded term with the others bounded. On a smaller interval with $\mathcal J(\norm{u_\epsilon}^2)<4\log2+1$ by comparison at zero, the radius $R_1$ still holds, and for population perturbations we add $\epsilon R_F^G(u)$ to the penalty, where
\begin{equation}\label{S-NG-U-eq4}
 R_F^G(u)=\E\ell_F(Y,b+t^\top Z+\norm wG)
        =\E\ell_F(Y,b+t^\top Z+w^\top G_q),
\end{equation}
where $G$ and $G_q$ are standard normal in one and $q$ dimensions, independent of $(Z,Y)$, and this vector representation is differentiable at $w=0$. Note that for $b_{\rm new}=(1,Z^\top,G_q^\top)^\top$, its derivative in directions $a_1,\ldots,a_r$ equals $\E\{\ell_F^{(r)}\prod_j b_{\rm new}^\top a_j\}$, which we bound by $C_r\prod_j\norm{a_j}$ from H\"older's inequality and projection moments. We also get zero in that gradient coordinate at $w_k=0$ by independence and centering of $G_k$, as for the radial penalty. For arbitrary arrays, we use the local stationary solution through the fit at $\epsilon=0$ by invertibility of the Hessian, and on the geometry event at radius $R_1+1$ we obtain a common interval from the bounded population gradient and Hessian. When $|\epsilon|C<c_{R_1+1}/4$, we have positive outward derivative on the unit sphere about the fit and positive curvature. Below, we evaluate all perturbation derivatives at zero, where constants and fixed powers of $L$ may depend on derivative order.

\begin{lemma}\label{S-NG-U-mixed}
Let $S$ be a labeled multiset of $r$ null-entry directions with
$1\le r\le6$, and define $u_S=\partial_Su$,
$\eta_S=\partial_S(Bu)$ and
$u_{\epsilon,S}=\partial_\epsilon\partial_Su|_0$.
Also define $\eta_{\epsilon,S}=\partial_\epsilon\partial_S(Bu)|_0$.
Let $v=\partial_\epsilon u_\epsilon|_0$ and let $G_n$ be an empirical average of a fixed $\lambda$ derivative of $g_{\lambda,F}$ or $\tau_\lambda$ in \eqref{S-NG-U-eq3}, or the risk $R_F^G(u)$. For $T_F^G=\sqrt n\{P_ng_{\lambda^\dagger,F}(Y_i,\eta_i)-R_F^G(u)\}$ and $\Phi=\phi(T_B^G,T_R^G)$ with $\phi\in C_b^6(\R^2)$, the following bounds hold simultaneously with probability at least $1-C_an^{-a}$ for any specified family of at most $n^M$ comparison arrays, its one- and two-entry modifications in $[-C_0\sqrt L,C_0\sqrt L]$, and $\lambda\in[\lambda_-,\lambda_+]$. Here $a,M>0$ are fixed, $\lambda^\dagger\in[\lambda_-,\lambda_+]$ on this event, and $v_k$ below denotes the derivative of $w_k$.
\begin{equation}\label{S-NG-U-eq5}
 \norm u+\norm v\le C,\qquad
 \max_i|b_i^\top v|\le C_aL^C,
 \qquad \max_{k\le q}(|w_k|+|v_k|)\le C_aL^Cn^{-1/2}.
\end{equation}
\begin{equation}\label{S-NG-U-eq11}
 \norm{u_S}+\norm{u_{\epsilon,S}}\le C_aL^Cn^{-(r+1)/2},
 \quad
 \norm{\eta_S}_2+\norm{\eta_{\epsilon,S}}_2\le C_aL^Cn^{-r/2}.
\end{equation}
\begin{equation}\label{S-NG-U-eq12}
 |\partial_SG_n|\le C_aL^Cn^{-1-r/2},\qquad
 |\partial_S\lambda^\dagger|\le C_aL^Cn^{-1-r/2}.
\end{equation}
\begin{equation}\label{S-NG-U-eq19}
 |\partial_ST_F^G|+|\partial_S\Phi|
                       \le C_{a,\phi}L^Cn^{-(r+1)/2}.
\end{equation}
The constants are uniform over $\mathcal K$ and the allowed
coordinate distributions. They may depend on $M,a,C_0$, the fixed
derivative orders and the bounds for $\phi$ through order six.
\end{lemma}
\begin{proof}
\OptProofStep{1}{First derivatives for deleted rows and columns}
At $\epsilon=0$, let $H=\nabla_u^2\mathcal F_\epsilon(u)$ and $\mathfrak g=\partial_\epsilon\nabla_u\mathcal F_\epsilon(u)$ with $u$ fixed. Then, from curvature, $\|B^\top g'/n\|\le C$, and \eqref{S-NG-U-eq4}, we see that $v=-H^{-1}\mathfrak g$ is bounded. For modified arrays, we apply Lemma~\ref{S-NG-U-geometry-lemma} after deleting their at most two rows. After restoring rows of norm $O(\sqrt n)$, we get positive semidefinite Hessian terms, and the design bound still holds.

At fixed deterministic $\lambda$ and divisor $n$, deleting row $(b^\top,y)$ leaves remaining design $B_0$ and retained $u_0,H_0,v_0=-H_0^{-1}\mathfrak g_0$ independent of the omitted row, and $\|u_0\|+\|H_0^{-1}\|+\|\mathfrak g_0\|\le C$ holds with failure probability at most $Ce^{-cn}$. From these quantities, we define
\[
 l=b^\top H_0^{-1}b/n,\qquad T+l\ell_R'(y,T)=b^\top u_0,\qquad
 \Delta=-\ell_R'(y,T)H_0^{-1}b/n.
\]
We obtain $\|\Delta\|\le Cn^{-1/2}$, $\|B_0\Delta\|_2\le C$ and $\|B_0\Delta\|_\infty\le C_a\sqrt{L/n}$ from conditional sub-Gaussian projections, with at most $C_0\sqrt L\|H_0^{-1}b_j\|/n$ per fixed entry. At $u_0+\Delta$, we cancel the incoming score by the retained linearized score, and hence we get
\[
 Cn^{-1}\|B_0\|_{\op}\|B_0\Delta\|_\infty\|B_0\Delta\|_2+C\|\Delta\|^2
 \le C_a\sqrt L/n .
\]
By curvature on the ball of radius $R_1+1$, we get
\begin{equation}\label{S-NG-U-eq6}
 u-u_0=\Delta+e,\quad \norm e\le C_a\sqrt L/n,
 \quad\max_{j\ne i}|b_j^\top(u-u_0)|\le C_a\sqrt{L/n}.
\end{equation}
We subtract the response equations to get
\[
 H(v-v_0)=-(\mathfrak g-\mathfrak g_0)-(d_i bb^\top/n+E)v_0,\qquad
 E=H-H_0-d_i bb^\top/n .
\]
We observe that $\|\mathfrak g-\mathfrak g_0\|=O(n^{-1/2})$ by the perturbation Hessian bound, and from~\eqref{S-NG-U-eq6}, $\|B_0\|_{\op}^2/n\le C$, and bounded penalty derivatives we get $\|E\|_{\op}\le C_a\sqrt{L/n}$. Specifically, by the uncentred conditional sub-Gaussian projection bound already proved, we have $|b^\top v_0|\le C_a\sqrt L$, and we invert $H$ to obtain $\|v-v_0\|\le C_a\sqrt{L/n}$ and $|b^\top v|\le C_a\sqrt L$ and apply a row union bound with one column restriction.

Note that at the restricted fit $u^0(\epsilon)$ with $w_k=0$ and $v^0=\partial_\epsilon u^0|_0$, the full-space gradient and its derivative satisfy
\begin{equation}\label{S-NG-U-eq8}
\begin{gathered}
 \nabla\mathcal F_\epsilon(u^0)=s_ke_k,
 \qquad s_k=P_n A_{ik}L_\epsilon'(Y_i,b_i^\top u^0),\\
 H_0v^0+\mathfrak g_0=s'_ke_k,
 \qquad s'_k=P_n A_{ik}\{d_i^0b_i^\top v^0+g_i^{\prime,0}\}.
\end{gathered}
\end{equation}
Here, $H_0,\mathfrak g_0$ are full-space derivatives at $u^0$, $d_i^0=s'(b_i^\top u^0)$, and $g_i^{\prime,0}=\partial_xg(Y_i,b_i^\top u^0)$ for empirical perturbations and zero for population perturbations. Radial and population gradients have zero $k$th coordinate at $w_k=0$. The coefficients in $s_k,s_k'$ are measurable without column $k$ and have norm $O(\sqrt n)$ by the row and design bounds. From conditional concentration, we obtain $|s_k|+|s_k'|\le C_aL^C/\sqrt n$, with $O(L^C/n)$ from fixed entries. By curvature, we obtain $\|u-u^0\|\le C|s_k|$, and
$H_u(v-v^0)=-s'_ke_k-(\mathfrak g_u-\mathfrak g_0)-(H_u-H_0)v^0 .$
The gradient and penalty differences are $O(\|u-u^0\|)$, and the empirical Hessian contribution is at most $C\|Bv^0\|_\infty\|u-u^0\|$ by the design bound. From the row bound and inverse curvature, we obtain $|v_k|\le C_aL^C/\sqrt n$ because $v_k^0=0$. Finally, we apply row and column union bounds to prove \eqref{S-NG-U-eq5}.

\OptProofStep{2}{Entry derivatives of the fit and statistic}
For a direction $a=(i_a,k_a)$, let $\mathcal P(S)$ denote the
partitions of $S$, and let $q_i=L_\epsilon'(Y_i,\eta_i)$.
In this recursion, let $d_i=L_\epsilon''(Y_i,\eta_i)$, which equals
$s'(\eta_i)$ at zero. Then
\begin{equation}\label{S-NG-U-eq13}
 (\eta_S)_i=b_i^\top u_S+
       \sum_{a\in S}\ind\{i=i_a\}(u_{S\setminus a})_{k_a},\qquad
 \partial_Sq_i=\sum_{\pi\in\mathcal P(S)}
 L_\epsilon^{(|\pi|+1)}(Y_i,\eta_i)
                              \prod_{C\in\pi}(\eta_C)_i.
\end{equation}
We separate $Hu_S$ in the differentiated score, where the remaining empirical terms are
\[
 P_nb_i L_\epsilon^{(j+1)}\prod_{l=1}^j\eta_{C_l}
 \quad(j\ge2,\ \textstyle\sum_l|C_l|=r),\qquad
 n^{-1}e_{k_a}\partial_{S\setminus a}q_{i_a},\qquad
 n^{-1}d_{i_a}b_{i_a}(u_{S\setminus a})_{k_a}.
\]
In the case $r=1$, we have $Hu_a=-(e_{k_a}q_{i_a}+d_{i_a}b_{i_a}w_{k_a})/n$, and we combine~\eqref{S-NG-U-eq5} and~\eqref{S-NG-U-eq13} to get $\|u_a\|\le C_aL^C/n$ and bound $\eta_a$. For the induction, from bounded loss derivatives and
\[
 \left\|n^{-1}B^\top\prod_l\eta_{C_l}\right\|
 \le\frac{\|B\|_{\op}}n\|\eta_{C_1}\|_2\prod_{l>1}\|\eta_{C_l}\|_\infty
 \le C_aL^Cn^{-(r+1)/2}
\]
we bound the first empirical term. Next, we combine~\eqref{S-NG-U-eq13} and $\max_i\|b_i\|\le C\sqrt n$ to get the same order for the other empirical terms, where each penalty term has $j\ge2$ nonempty factors and order $O(L^Cn^{-(r+j)/2})$, and we invert $H$ with~\eqref{S-NG-U-eq13} to complete the induction.

We observe that $\partial_\epsilon d_i=L_0'''b_i^\top v+g_i''$ at zero with $g_i''=0$ for population perturbations, and by~\eqref{S-NG-U-eq5} and the tensor bound after~\eqref{S-NG-U-eq4} it holds that $\|H_\epsilon\|_{\op}\le C_aL^C$. Note that in each remainder we differentiate a proper factor or bounded tensor, insert $Bv$, or replace $w_k$ by $v_k$, and hence by the same induction we preserve the power of $n$ and get~\eqref{S-NG-U-eq11}.

We define $f_n(\epsilon)=\mathcal F_\epsilon(u_\epsilon)$ using
the local stationary solution for population perturbations on exceptional
arrays. At zero, stationarity gives
\begin{equation}\label{S-NG-U-eq17}
 \partial_\epsilon f_n=G_n,
 \qquad \partial_{ik}f_n=L_\epsilon'(Y_i,\eta_i)w_k/n.
\end{equation}
We apply $r-1$ entry derivatives and one $\epsilon$ derivative to the second identity, where by~\eqref{S-NG-U-eq5} a residual derivative of order $j$ is $O(L^Cn^{-j/2})$ and the coefficient factor is $O(L^Cn^{-(r-j)/2})$, and hence we divide by $n$ to get~\eqref{S-NG-U-eq12} for $G_n$.

We differentiate the trace equation and observe from Lemma~\ref{S-NG-U-smooth} that $\partial_S\lambda^\dagger$ has coefficient at least $c_\lambda$, with $O(L^Cn^{-1-r/2})$ for the term without a root derivative. Note that every other term contains two proper root derivatives or one root and one empirical derivative and is $O(L^Cn^{-2-r/2})$ by induction. We thus get the root bound and multiply by $\sqrt n$ for $T_F^G$ in~\eqref{S-NG-U-eq19}, with $O(L^Cn^{-(r+j)/2})$ for each $j$-factor term in $\partial_S\Phi$.

\OptProofStep{3}{Uniformity over modifications and Taylor remainders}
We observe from Lemma~\ref{S-NG-U-smooth} that on each array we bound the fit, root, and inverse derivatives by fixed powers of $n$. Next, we combine~\eqref{S-NG-U-eq4} and the recursions with bounded penalty derivatives to get the bound $Cn^{D_0}(1+\|B\|_F)^{D_0}$ through the next order and up to one $\epsilon$ derivative at zero, and hence from bounded entry moments we get $L^2$ norm at most $Cn^{D_1}$ with any required entry power.

For modifications of two entries, we delete their rows and apply Lemma~\ref{S-NG-U-geometry-lemma}. When we restore these rows, curvature and $\|B\|_{\op}\le C\sqrt n$ are preserved with exponentially small failure probability, and also after another row deletion or column restriction. In Step~1, we condition on the incoming projections and bound the fixed entries separately.

In the case $\|B\|_F\le Cn$, we bound the next derivatives by $n^{D_2}$ and apply mesh $n^{-D_2-12}$ in $\lambda$ and the modified entries for $O(n^{-12})$ interpolation error before the partition formula for $\Phi$. Here, we evaluate $\phi$ and its derivatives only through order six, and $D_0,D_1,D_2$ do not depend on the tail exponent.

We choose $M_0$ with $n^{M_0}$ above all arrays, grids, deletions, and index lists up to length six in the $O(n^2)$ outer and $n$ inner replacements and $O(n^3)$ cubic comparisons. By a union bound from $C_an^{-(a+M_0)}$, we get simultaneous failure probability $C_an^{-a}$. Next, by the Cauchy--Schwarz inequality, we bound the exceptional expectation by $Cn^{M_0+D_1-a/2}=o(n^{-7})$ for $a>2D_1+2M_0+16$, and for large $C_0$ we obtain this bound outside $[-C_0\sqrt L,C_0\sqrt L]$ from sub-Gaussian tails and the Taylor derivative bound. We apply these derivative and remainder bounds to the moment comparisons of Lemma~\ref{S-NG-U-risk}. This completes the proof.
\end{proof}

\subsection{Third moments and the realized risk}\label{S-null-third}
Next, we let $\mu_{3k}=\E A_{1k}^3$ and $S(w)=\sum_k\mu_{3k}w_k^3$, and for any fixed $u$ we define $R_F^P(u)=\E_{Z,Y,A}\ell_F(Y,b+t^\top Z+w^\top A)$ over an independent observation from the original distribution.

\begin{lemma}\label{S-NG-U-risk}
Under the conditions of Theorem~\ref{S-NG-U-main}, the bounded Lipschitz distance between the distributions of $(T_B^G,T_R^G)$ under the original and Gaussian null coordinates tends to zero. Under the original distribution, $\sqrt n S(w)\to0$ in $L^2$ and $\sqrt n\{R_F^P(u)-R_F^G(u)\}\pto0$ for $F\in\{B,R\}$. These conclusions hold uniformly over $\mathcal K$ and the allowed coordinate distributions.
\end{lemma}
\begin{proof}
When we reverse column $k$ in~\eqref{S-NG-U-eq1}, only $w_k$ changes sign and the scores, root, and Gaussian-reference risks are unchanged, and thus any $g$ odd under this reversal has conditional mean zero when we replace that column by Gaussians. Next, we expand each entry replacement to degree two, and from moment matching with bounded third moments and Step~3 of the proof of Lemma~\ref{S-NG-U-mixed}, we obtain
\[
 \begin{aligned}
 \max_j|\partial_{jk}^3g|\le CL^Cn^{-\nu}
 \quad\Longrightarrow\quad
 |\E g|&\le\sum_{j=1}^n\{CL^Cn^{-\nu}+o(n^{-5})\}\\
       &=CL^Cn^{1-\nu}+o(n^{-4}).
 \end{aligned}
\]
Note that on this event, we bound the derivatives and fix one extra entry at zero in exceptional expectations.

Next, for $\Phi=\phi(T_B^G,T_R^G)$ with $\phi\in C_b^6(\R^2)$, we expand all $nq$ replacements to degree three and get fourth-order error $O(L^C/\sqrt n)$ by~\eqref{S-NG-U-eq19} at $r=4$. Specifically, each cubic term is $(\E A_{ik}^3)\E g/6$ for $g=\partial_{ik}^3\Phi|_{A_{ik}=0}$ odd in the other column entries, and at $r=6$ with the column bound, it holds that
\[
 \max_{j\ne i}|\partial_{jk}^3 g|\le C_\phi L^Cn^{-7/2},
 \qquad |\E g|\le C_\phi L^Cn^{-5/2}+o(n^{-4}).
\]
We observe from $nq=O(n^2)$ that the cubic sum tends to zero. For unit bounded Lipschitz $f$ and $f_\delta(x)=\E f(x+\delta G_2)$ with $G_2\sim N_2(0,I_2)$, we have $\|f_\delta-f\|_\infty\le\delta\E\|G_2\|$ and $\|\partial^\iota f_\delta\|_\infty\le C_\iota\delta^{-|\iota|}$ through order six, and hence this step follows as $n\to\infty$ and then $\delta\downarrow0$.

When $k\ne l$, $w_k^3w_l^3$ is odd in column $k$, and in its third entry derivatives we get products of six coefficient derivatives of total order three, each of order $L^Cn^{-9/2}$ by~\eqref{S-NG-U-eq5} and~\eqref{S-NG-U-eq11}. With the column and radius bounds, we thus obtain
\[
 |\E w_k^3w_l^3|\le CL^Cn^{-7/2}+o(n^{-4}),\qquad
 \E w_k^6\le CL^Cn^{-3}.
\]
Note that for bounded $\mu_{3k}$, we have
\[
 \begin{aligned}
 \E S(w)^2
 &=\sum_k\mu_{3k}^2\E w_k^6+\sum_{k\ne l}\mu_{3k}\mu_{3l}\E(w_k^3w_l^3)\\
 &\le CL^C(qn^{-3}+q^2n^{-7/2})+o(n^{-2})=o(n^{-1}).
 \end{aligned}
\]

For fixed $u$, conditional on $(Z,Y)$ we replace the independent test coordinates with fourth-order remainder $O(\sum_kw_k^4)$. From bounded $\ell_F^{(6)}$ and $\ell_F^{(5)}$, in the cubic coefficient we expand $\ell_F'''$ to second order up to $O(\sum_{l\ne k}|w_l|^3)$ and restore Gaussian coordinate $k$ up to $O(w_k^2)$, and hence we get
\[
 R_F^P(u)-R_F^G(u)=\tfrac16\E\{\ell_F'''(Y,b+t^\top Z+\|w\|G)\}S(w)
    +O\!\left(\sum_kw_k^4+\Big\{\sum_k|w_k|^3\Big\}^2\right).
\]
On the fixed ball, $\sum_k|w_k|^5\le C\sum_kw_k^4$, and by \eqref{S-NG-U-eq5}, we have
\[
 \sum_kw_k^4+\left(\sum_k|w_k|^3\right)^2
 \le\|w\|_\infty^2(\|w\|^2+\|w\|^4)=O_P(L^C/n).
\]
From the $L^2$ bound for $\sqrt nS(w)$, we get the risk comparison. This completes the proof.
\end{proof}

\subsection{A scale-mixture counterexample}\label{S-NG-mixture}

\begin{proposition}\label{S-NG-prop-mixture}
Let $0<a<1$, let $S\sim\operatorname{Bernoulli}(a)$ and
$G_p\sim N_p(0,I)$ be independent, and define $X=SG_p/\sqrt a$.
Assume that the logistic model is correct with intercept and coefficient
vector zero. The predictors are isotropic and have uniformly sub-Gaussian
projections. If $p/n\to\kappa\in(0,a/2)$, the fit exists and is unique and
bounded with probability tending to one. Then there exist deterministic limits
$b_B(\kappa),b_R(\kappa)$ such that
\begin{equation}\label{S-NG-eq-mixturebias}
 \begin{gathered}
 \widehat B-B_{\rm ext}\pto b_B(\kappa),\qquad
 b_B(\kappa)=-\frac{1-a}{4a}\kappa^2+O(\kappa^3),\\
 \widehat R-R_{\rm ext}\pto b_R(\kappa),\qquad
 b_R(\kappa)=-\frac{3(1-a)}{2a}\kappa^2+O(\kappa^3),
 \end{gathered}
\end{equation}
where the expansions hold as $\kappa\downarrow0$ with $a$ fixed.
\end{proposition}

\begin{proof}
\OptProofStep{1}{The active Gaussian sample}
We observe that $\E XX^\top=I_p$ by independence, and for unit $u$ with $K^2=8/(3a)$, we have
\[
 \E e^{(u^\top X)^2/K^2}
 =1-a+a\{1-2/(aK^2)\}^{-1/2}=1+a\le2.
\]
Let $I_1=\{i:S_i=1\}$, $n_1=|I_1|$, $a_n=n_1/n$, and $v=\beta/\sqrt a$. Note that the probability of $0<n_1<n$ tends to one, and on this event it holds that
\begin{equation}\label{S-NG-eq-mix-activecriterion}
 L_n(b,v)=a_nL_{A,n}(b,v)+(1-a_n)L_{0,n}(b),\qquad
 L_{0,n}(b)=\log(1+e^b)-\overline Y_{0,n}b ,
\end{equation}
where $L_{A,n}$ denotes the average over the $n_1$ active rows and $\overline Y_{0,n}$ the prevalence on inactive rows.

Conditional on $I_1$, the active design is Gaussian with $p/n_1\pto\delta=\kappa/a<1/2$, and null symmetry, convexity, and Theorem~\ref{S-NG-thm-head}(a) give $h_Z=1/2$ and $b_{A,n}\pto0$, with Lemmas~\ref{S-NG-NG-mixed-geometry} and~\ref{S-NG-lem-localization} holding uniformly near $\delta$. From $a_n\pto a>0$, $L_{0,n}\ge0$, comparison at zero, active coercivity, and full rank, the active and mixture minimizers are unique and lie in a fixed ball where the Hessian lower bound $\nabla^2L_n\succeq a_nc_AI$ holds. We integrate from the active minimizer $u_{A,n}$ and use $\overline Y_{0,n}\pto1/2$ to obtain
\begin{equation}\label{S-NG-eq-mix-fitcomparison}
 \begin{aligned}
 \|\widehat u_n-u_{A,n}\|
 &\le\|\nabla L_n(u_{A,n})\|/(a_nc_A)
 =\frac{(1-a_n)|s(b_{A,n})-\overline Y_{0,n}|}{a_nc_A}=o_P(1).
 \end{aligned}
\end{equation}
Here, on the active sample we have empirical mean $O_P(1)\|\widehat u_n-u_{A,n}\|^2$ and conditional external mean $\|\widehat u_n-u_{A,n}\|^2$ for squared score differences. Specifically, we combine \eqref{S-NG-eq-fo-scorelimits} with the independent-row representation to get the active limits, and the inactive score tends to zero.

\OptProofStep{2}{Scalar correction and limits of the loss}
Let $h_0(x)=s(-x)$ and $v=r^2$. For the active signed score we have $M=rG+A h_0(M)$ with $G\sim N(0,1)$ and $A>0$. Because~\eqref{S-state-eq} and~\eqref{S-NG-eq-newstate} coincide at ratio $\delta$, by Proposition~\ref{S-null} and~\eqref{S-nullstate}, we obtain
\begin{equation}\label{S-NG-eq-mix-activeexpansion}
 A=4\delta+8\delta^2+O(\delta^3),\qquad
 v=4\delta+12\delta^2+O(\delta^3).
\end{equation}
Let $c$ be the population trace root for the mixture. Because the trace increases from zero to one, we get $\widehat\lambda\pto c$ by bracketing. By its inactive term, we have $c\le4\kappa/(1-a-\kappa)$ for $\kappa<1-a$. From $\E s'(M)=1/4-v/16+O(\delta^2)$ and $\E s'(M)^2=1/16+O(\delta)$, we obtain
\begin{equation}\label{S-NG-eq-mix-scalarroot}
 \begin{aligned}
 \kappa&=(1-a)c/(4+c)+a\E\{cs'(M)/(1+cs'(M))\}\\
       &=c/4-(acv+c^2)/16+O(\kappa^3),\\
 c&=4\kappa+8\kappa^2+O(\kappa^3).
 \end{aligned}
\end{equation}
Here, in the last line we use $av=4\kappa+O(\kappa^2)$ and the first-order value $c=4\kappa+O(\kappa^2)$.

The adjusted signed-score limits are $K=M-ch_0(M)$ and $-c/2$ on active and inactive rows, with external limits $rG$ and zero. For $g_B=h_0^2$ and $g_R(x)=\log(1+e^{-x})$, from bounded derivatives, second moments and \eqref{S-NG-eq-mix-fitcomparison}, we obtain
\[
 b_F(\kappa)=(1-a)\{g_F(-c/2)-g_F(0)\}+a\E\{g_F(K)-g_F(rG)\},\quad F\in\{B,R\}.
\]

In $K=rG+(A-c)h_0(M)$ use $h_0(x)=1/2-x/4+x^3/48+O(x^5)$ and the moments in Step~2 of the proof of Proposition~\ref{S-null} to obtain
\begin{equation}\label{S-NG-eq-mix-moments}
 \begin{aligned}
 \E K&=(A-c)/2-A(A-c)/8+O(\kappa^3),\\
 \E K^2&=v+(c-A)v/2+(A-c)^2/4+O(\kappa^3),\\
 \E K^3&=3(A-c)v/2+O(\kappa^3),\qquad \E K^4=3v^2+O(\kappa^3).
 \end{aligned}
\end{equation}
With $\delta=\epsilon^2$, by Step~1, we get $A=\epsilon^2L$, $r=\epsilon R$ for smooth even $L,R$ with polynomially bounded proximal derivatives. The trace equation is invariant under $(\epsilon,G)\mapsto(-\epsilon,-G)$ and has $c$ derivative $1/4$ at zero, and hence we get smooth even $c$. By six derivatives under Gaussian expectation, we get $O(\epsilon^6)=O(\kappa^3)$ remainders, and this holds for odd Brier terms.

Apply \eqref{S-NG-eq-mix-moments} to the fourth-order Taylor polynomials of $g_B=h_0^2$ and $g_R$, using $g_R'=-h_0$. Then substitute \eqref{S-NG-eq-mix-activeexpansion} and \eqref{S-NG-eq-mix-scalarroot} to obtain
\[
 \begin{aligned}
 b_B(\kappa)
 &=\frac{c-aA}{8}+\frac{3aA^2-4aAc+c^2}{64}+O(\kappa^3)
 =-\frac{1-a}{4a}\kappa^2+O(\kappa^3),\\
 b_R(\kappa)
 &=2b_B(\kappa)-\frac{a(A-c)v}{16}+O(\kappa^3)
 =-\frac{3(1-a)}{2a}\kappa^2+O(\kappa^3).
 \end{aligned}
\]
This proves \eqref{S-NG-eq-mixturebias}. This completes the proof.
\end{proof}

\section{Inference for penalized log loss}\label{S-moderate}
\subsection{Risk expansion and studentization}
For the observations, fixed feature map, risk $R(u)$, and deterministic penalty $\Lambda_n\succeq0$ in Section~\ref{sec:general-inference}, let $P_nf=n^{-1}\sum_i f(z_i,Y_i)$. The penalized fit is
\begin{equation}\label{S-reader-fit}
 \widehat u=\arg\min_u\{P_n\ell_Y(z^\top u)+u^\top\Lambda_nu/2\}.
\end{equation}

For $u_n^*$ in Theorem~\ref{thm:moderate}, let $G_*=\E\{s'(z^\top u_n^*)zz^\top\}+\Lambda_n$. Define the normalized approximate leave-one-out (ALO) score by
\begin{equation}\label{S-reader-loo}
\begin{aligned}
 F_{-i}(v)=\frac1{n-1}\sum_{j\ne i}\ell_{Y_j}(z_j^\top v)+\frac12v^\top\Lambda_nv,\qquad
 \widetilde\eta_i=z_i^\top\left[\widehat u-
 \{\nabla^2F_{-i}(\widehat u)\}^{-1}\nabla F_{-i}(\widehat u)\right].
\end{aligned}
\end{equation}

With the notation of Algorithm~2, let $H=P_n(w_i z_i z_i^\top)$. The adjusted score, loss, and average are
\begin{equation}\label{eq:normalized-alo}
 \widetilde\eta_i=\eta_i+
 \frac{q_ih_i+c_i-z_i^\top M^{-1}r}{1-w_ih_i},\qquad
 A_i=\ell_{Y_i}(\widetilde\eta_i),\qquad \widehat C_{\rm ALO}=P_nA_i.
\end{equation}

Let $\widehat J_c$ be the covariance of $q_i z_i$, centered with divisor $n$. Define
\begin{equation}\label{S-reader-variance}
\begin{aligned}
 \widehat V_{\rm ALO}&=P_n(A_i-\widehat C_{\rm ALO})^2+\frac2n\operatorname{tr}[\{(H+\Lambda_n)^{-1}\widehat J_c\}^2],\\
 I_{\rm ALO}&=\left[\widehat C_{\rm ALO}\pm z_{1-\alpha/2}\sqrt{\widehat V_{\rm ALO}/n}\right]\cap[0,\infty).
\end{aligned}
\end{equation}

We center Takeuchi information criterion (TIC) intervals at $\widehat C_{\rm TIC}$ with standard error $\{\widehat V_{\rm TIC}/n\}^{1/2}$ and define $\widehat V_{\rm TIC}$ by divisor-$n$ sample variance ${\rm Var}_n\{\ell_{Y_i}(\eta_i)\}$ around the raw-loss sample mean plus the \eqref{S-reader-variance} trace term. The estimate \citep{commenges2015universal} is
\begin{equation}\label{S-reader-tic}
\widehat C_{\rm TIC}=P_n\ell_Y(z^\top\widehat u)+\allowbreak\operatorname{tr}\{(H+\Lambda_n)^{-1}\widehat J_c\}/n.
\end{equation}

We define $G=H+\Lambda_n$, $\psi_i=q_iz_i$, and $\bar\psi=P_n\psi_i$.

\begin{proof}[Proof of Theorem~\ref{thm:moderate}]
Let $Pf=\E f(z,Y)$ and $u_*=u_n^*$. For the score and its covariance, define
\[
 \psi_*=\{s(z^\top u_*)-Y\}z,\quad m=P\psi_*=-\Lambda_nu_*,
 \quad \xi=\psi_*-m,\quad g=P_n\xi,\quad J=P\xi\xi^\top.
\]
Let $H_*=P\{s'(z^\top u_*)zz^\top\}$, $H_n^*=P_n\{s'(z^\top u_*)zz^\top\}$ and $A=(H_*+\Lambda_n)^{-1}$, where constants depend only on $K,D,c,\sigma_0$. By directional second moments, we get $\|m\|\le K\sqrt{\log2}$ and $\E\|g\|^2\le Cd/n$. For unit $a$, from integration of $\Prob(|z^\top a|>u)\le2e^{-u^2/K^2}$, we obtain $\E|z^\top a|^{2k}\le2k!K^{2k}$. After weighting by $s'$ or squared residuals and centering, the moments are bounded by $k!C^k$, and the exponential series is at most $e^{C\lambda^2}$ for $|\lambda|\le C^{-1}$. By exponential Markov on a $1/4$-net of size at most $9^d$, we get
\[
\|H_n^*-H_*\|_{\op}+\|P_n\xi\xi^\top-J\|_{\op}
 =O_P(\sqrt{d/n}),\qquad \|P_nzz^\top\|_{\op}=O_P(1).
\]

Let $Z_{(k)}$ be the $k$th largest $|z_i^\top a|$. By a union over $k$-subsets,
$\Prob\{Z_{(k)}>u\}\le\exp\{k\log(2en/k)-ku^2/K^2\}.$
Let $u_k^2=K^2\{\log(2en/k)+(t+\log n)/k\}$. The bounds $\sum_k\log(2en/k)^{3/2}\le Cn$ and $\sum_kk^{-3/2}\le C$ give a cube sum at most $CK^3\{n+(t+\log n)^{3/2}\}$. On a $1/2$-net of size at most $5^d$, we replace $t$ by $t+d\log5$. By the seminorm property of $a\mapsto\{P_n|z^\top a|^3\}^{1/3}$, we obtain, with probability at least $1-e^{-t}$,
\begin{equation}
\label{S-moderate-cubic}
 M_{3,n}:=\sup_{\|a\|=1}P_n|z^\top a|^3
 \le CK^3\{1+(d+t+\log n)^{3/2}/n\}=O_P(1).
\end{equation}
The same calculation bounds the population third moment.

For $F_n(u)=P_n\ell_Y(z^\top u)+u^\top\Lambda_nu/2$, by the bounded derivatives and H\"older's inequality, we have
$\|\nabla^2F_n(u_*+v)-(H_n^*+\Lambda_n)\|_{\op}\le CM_{3,n}\|v\|.$
On a small fixed ball, curvature is at least $c/2$ with probability $1-o(1)$. At radius $\epsilon$, we have $F_n(u_*+v)-F_n(u_*)\ge c\epsilon^2/4-\|g\|\epsilon>0$, and hence the minimum is interior and global by convexity. By the fixed Hessian kernel $\ker(P_nzz^\top)\cap\ker\Lambda_n$, we get uniqueness. With $\delta=\widehat u-u_*$, by integrated score equations, we get
\begin{equation}
\label{S-moderate-fit}
 0=g+(H_n^*+\Lambda_n)\delta+O_P(\|\delta\|^2),\quad
 \|\delta\|=O_P(\sqrt{d/n}),\quad \delta=-Ag+O_P(d/n).
\end{equation}

We use $(P_n-P)\psi_*=g$ for linear term $g^\top\delta=-g^\top Ag+O_P((d/n)^{3/2})$ and \eqref{S-moderate-cubic} to bound remaining terms by $\|H_n^*-H_*\|_{\op}\|\delta\|^2+M_{3,n}\|\delta\|^3=O_P((d/n)^{3/2})$. Note that for $\ell_u=\ell_Y(z^\top u)$, we have
\begin{equation}
\label{S-moderate-risk}
 P_n\ell_{\widehat u}-R(\widehat u)
 =(P_n-P)\ell_{u_*}-g^\top Ag+O_P\{(d/n)^{3/2}\}.
\end{equation}
We combine H\"older's inequality, \eqref{S-moderate-cubic}, and bounded derivatives of $s'$ and $(s-Y)^2$ to bound Hessian and score-covariance changes by $C\|\delta\|$ and mean-score changes by $\|P_nzz^\top\|_{\op}\|\delta\|/4$. With centering, concentration, and $G-G_*=H-H_*$, we get
\[
\|H-H_*\|_{\op}+\|\widehat J_c-J\|_{\op}=O_P(\sqrt{d/n}),\qquad
 \frac{\operatorname{tr}(G^{-1}\widehat J_c)}n
 =\frac{\operatorname{tr}(AJ)}n+O_P\{(d/n)^{3/2}\}.
\]
By independence and $P\xi=0$,
\[
\E(g^\top Ag)=\frac{\operatorname{tr}(AJ)}n,\qquad
 \Var(g^\top Ag)=\frac{2(n-1)}{n^3}\operatorname{tr}(AJAJ)
                  +\frac{\Var(\xi^\top A\xi)}{n^3}.
\]
From directional moments, we get $\operatorname{tr}(AJAJ)\le Cd$ and $\Var(\xi^\top A\xi)\le Cd^2$. By the variance identity, we have $\|g^\top Ag-\operatorname{tr}(AJ)/n\|_{L^2}\le C(\sqrt d/n+d/n^{3/2})$. From this bound and \eqref{S-moderate-risk}, we obtain
\begin{equation}
\label{S-moderate-expansion}
 \widehat C_{\rm TIC}-R(\widehat u)
 =(P_n-P)\ell_{u_*}
 +O_P\{\sqrt d/n+d/n^{3/2}+(d/n)^{3/2}\}.
\end{equation}
Because $\ell_{u_*}\le\log2+|z^\top u_*|$, the fourth moments are bounded. From Lyapunov's condition, $\sigma_n^2\ge\sigma_0^2$, and $d^{3/2}/n\to0$, we get the normal limit.

At the exact fit, the Newton identity in Section~\ref{sec:general-inference} with $r=0$ gives \eqref{eq:normalized-alo}. From $M=G-\Lambda_n/n$, $0\preceq\Lambda_n\preceq G$, and $\Lambda_n\widehat u=-\bar\psi$, we obtain
\[
 \begin{gathered}
 (1-1/n)G\preceq M\preceq G,\qquad
 0\preceq M^{-1}-G^{-1}\preceq G^{-1}/(n-1),\\
 \|\Lambda_n\widehat u\|=\|\bar\psi\|\le\|P_nzz^\top\|_{\op}^{1/2}=O_P(1).
 \end{gathered}
\]
Directional tails and fourth moments yield
\[
 \begin{gathered}
 \max_i\|z_i\|^2=O_P(d+\log n),\qquad P_n\|z\|^4=O_P(d^2),\\
 \max_iw_ih_i=o_P(1),\qquad P_nh_i^2=O_P(d^2/n^2),\qquad P_nc_i^2=O_P(n^{-2}).
 \end{gathered}
\]
Taylor expansion and $\bar\psi=-\Lambda_n\widehat u$ give
\[
\widehat C_{\rm ALO}-P_n\ell_{\widehat u}
 =P_n(q_i^2h_i+q_ic_i)+O_P(d^2/n^2)
 =\frac{\operatorname{tr}(M^{-1}\widehat J_c)}n+O_P(d^2/n^2).
\]
We observe from inverse bounds that $0\le n^{-1}\operatorname{tr}\{(M^{-1}-G^{-1})\widehat J_c\}\le\operatorname{tr}(G^{-1}\widehat J_c)/\{n(n-1)\}=O_P(d/n^2)$, and hence we get $\widehat C_{\rm ALO}-\widehat C_{\rm TIC}=O_P(d^2/n^2)=o_P(n^{-1/2})$.

Next, we apply Lipschitz continuity for studentization to get
\[
P_n|\ell_{\widehat u}-\ell_{u_*}|^2
 \le\delta^\top(P_nzz^\top)\delta=O_P(d/n),\qquad
 P_n|A_i-\ell_{Y_i}(\eta_i)|^2
 \le P_n|\widetilde\eta_i-\eta_i|^2=O_P(d^2/n^2).
\]
Specifically, by the Cauchy--Schwarz inequality, it holds that
\[
 |P_nA_i^2-P_n\ell_{u_*}^2|
 \le\{P_n(A_i-\ell_{u_*})^2P_n(A_i+\ell_{u_*})^2\}^{1/2}=o_P(1).
\]
By the preceding $L_2$ approximation for both raw fitted and adjusted losses, the oracle sample variance law of large numbers, and the $o_P(1)$ trace term, both variance estimators are consistent and \eqref{eq:moderate-clt} follows from Slutsky's theorem.

When an inexact fit has gradient $r=o_P(n^{-1/2})$, we bound optimizer error by $O_P(\|r\|)$ via local curvature and convexity. In \eqref{eq:normalized-alo}, we thus get linear and quadratic errors $-(1-1/n)\bar\psi^\top M^{-1}r=o_P(n^{-1/2})$ and $O_P(\|r\|^2)$ in the residual correction, and the failure probability tends to zero. This completes the proof.
\end{proof}

\subsection{Uniform coverage for bounded features}\label{S-bounded-ridge}
Because the clinical dimension is fixed, the coordinatewise bounded clinical transformation \eqref{eq:bounded-features} satisfies the bounded full Euclidean norm assumption of Corollary~\ref{cor:bounded-ridge}.

For a fixed $m$, let $\widehat{u}_j$ denote the coefficient of the $j$th candidate fit and let $I_j$ denote its $(1-\alpha/m)$ confidence interval for $j \in \{1,\dots,m\}$. If each candidate satisfies Theorem~\ref{thm:moderate} or Corollary~\ref{cor:bounded-ridge}, then for any selected index $\widehat{j} \in \{1,\dots,m\}$, by the union bound we obtain
\[
\Prob(R(\widehat{u}_{\widehat{j}}) \notin I_{\widehat{j}}) \le \sum_{j=1}^m \Prob(R(\widehat{u}_j) \notin I_j) = \alpha + o(1).
\]

\begin{proof}[Proof of Corollary~\ref{cor:bounded-ridge}]
Let $\bar Y=P_nY$, $u_*=u_n^*=(a_*,\beta_*^\top)^\top$, and $b_0=(2\log2/\rho_0)^{1/2}$. On the penalized $\log2$-sublevel set, we have $\|\beta\|\le b_0$ and $\pi_0|a|\le R(a,\beta)+L\|\beta\|\le\log2+Lb_0$, and hence we get a unique minimum by compactness and a positive definite penalized Hessian. When we replace $\pi_0$ by $\bar Y\wedge(1-\bar Y)>0$, we get the empirical conclusion.

From the intercept score equation, we have $\E s(a_*+\phi(X)^\top\beta_*)=\pi$, and because $|\phi(X)^\top\beta_*|\le Lb_0$, we obtain
$|a_*|\le a_0:=\log\{(1-\pi_0)/\pi_0\}+Lb_0,\qquad |z^\top u_*|\le M_0:=a_0+Lb_0 .$
With $w_0=s'(M_0)>0$,
\[
\begin{aligned}
(a,v)^\top G_*(a,v)&\ge w_0\E(a+\phi(X)^\top v)^2+\rho_0\|v\|^2,\\
a^2+\|v\|^2&\le2\E(a+\phi(X)^\top v)^2+(2L^2+1)\|v\|^2
\end{aligned}
\]
giving $G_*\succeq\min\{w_0/2,\rho_0/(2L^2+1)\}I$.

For $Q=\ell_Y(z^\top u_*)$, $D_Y=1-2Y$, and $f(t)=1-e^{-t}$ with $t\ge0$, the identity $s(z^\top u_*)-Y=D_Yf(Q)$ and the intercept equation give $\E D_Yf(Q)=0$. The 1-Lipschitz property and $f(Q)\ge s(-M_0)$ give
\[
|1-2\pi|s(-M_0)\le|\Cov\{D_Y,f(Q)\}|
 \le\sqrt{\Var\{f(Q)\}}\le\sqrt{\Var(Q)},
\]
giving $\Var(Q)\ge4\pi_1^2s(-M_0)^2$.

Let $Pf=\E f(z,Y)$, $\ell_u=\ell_Y(z^\top u)$, $L_0=(1+L^2)^{1/2}$, and $\mathcal A_n=\{\pi_0/2\le\bar Y\le1-\pi_0/2\}$, with $\Prob(\mathcal A_n^c)\le2e^{-n\pi_0^2/2}$ by Hoeffding's inequality. On $\mathcal A_n$, the coefficient and curvature bounds hold empirically with $\pi_0/2$, and hence we get the Hessian lower bound $c_1I$ between $u_*$ and $\widehat u$ independently of $d_n$. For $g=(P_n-P)\{(s(z^\top u_*)-Y)z\}$ and $\Delta=\widehat u-u_*$, we have
\[
 \begin{gathered}
  \E\|g\|^2\le L_0^2/n,\quad
  \|\Delta\|\le c_1^{-1}\|g\|=O_P(n^{-1/2}),\quad
  (P_n-P)\ell_{\widehat u}=(P_n-P)\ell_{u_*}+O_P(n^{-1}),
 \end{gathered}
\]
uniformly over the distribution class. By a Taylor expansion with linear term $g^\top\Delta$, we obtain the loss expansion because $\|z\|\le L_0$ implies a uniform bound on the loss Hessians.

For ALO, from $M\succeq(1-1/n)(H+\Lambda_n)$ and $\|\Lambda_n\widehat u\|=\|\bar\psi\|\le L_0$, we obtain
\[
\max_i h_i+\max_i|c_i|\le C/n,\qquad
 \widehat C_{\rm ALO}-P_n\ell_{\widehat u}=O_P(n^{-1}).
\]
Because $\operatorname{tr}\widehat J_c\le L_0^2$,
\[
\frac2n\operatorname{tr}\{((H+\Lambda_n)^{-1}\widehat J_c)^2\}
 \le \frac Cn,\qquad
 \widehat V_{\rm ALO}-\operatorname{Var}(Q)\pto0
\]
from $P_n|A_i-\ell_{Y_i}(z_i^\top u_*)|^2=O_P(n^{-1})$ and the uniformly bounded oracle losses. Combining the loss expansions gives $\widehat C_{\rm ALO}-R(\widehat u)
  =(P_n-P)\ell_{u_*}+O_P(n^{-1})$. Combining the uniform bound on the oracle losses with the positive variance lower bound, we establish the uniform central limit theorem and studentization for any $d_n$. One class is absent with probability at most $2(1-\pi_0)^n$. 

Let $\widehat u^{(-i)}$ minimize $F_{-i}$ in \eqref{S-reader-loo}, and for the single Newton update and its loss, we define
\[
 \widetilde u_i=\widehat u-\{\nabla^2 F_{-i}(\widehat u)\}^{-1}\nabla F_{-i}(\widehat u),
 \qquad A_i=\ell_{Y_i}(z_i^\top\widetilde u_i).
\]
We use the gradient identity in Section~\ref{sec:general-inference} at $r=0$. By $\|\psi_i\|\le L_0$ and $\|\Lambda_n\widehat u\|=\|P_n\psi_i\|\le L_0$, we get
\[
\nabla F_{-i}(\widehat u)=-\frac{\psi_i+\Lambda_n\widehat u}{n-1},\qquad \max_{1\le i\le n}\|\nabla F_{-i}(\widehat u)\|\le \frac{2L_0}{n-1}.
\]
On $\mathcal A_n$ for large $n$, all deletion prevalences lie in $[\pi_0/4,1-\pi_0/4]$, and we have $\nabla^2F_{-i}\succeq c_2I$ between $\widehat u$ and $\widehat u^{(-i)}$ by the coefficient bounds. We use $\|z_j\|\le L_0$, $|s''|\le1/4$, and quadratic penalty to obtain $\|\nabla^2F_{-i}(u)-\nabla^2F_{-i}(v)\|_{\op}\le CL_0^3\|u-v\|$, and combine integrated curvature with Taylor expansion to get
\[
\max_{1\le i\le n}\|\widehat u^{(-i)}-\widehat u\|\le \frac{2L_0}{c_2(n-1)}\le \frac{C}{n},\qquad
\max_{1\le i\le n}\|\widetilde u_i-\widehat u^{(-i)}\|\le \frac{CL_0^3}{2c_2}\max_{1\le i\le n}\|\widehat u^{(-i)}-\widehat u\|^2\le \frac{C}{n^2}.
\]

Define
\[
L_i^-=\ell_{Y_i}(z_i^\top\widehat u^{(-i)}),\qquad C_{\mathrm{LOO}}=P_n L_i^-,\qquad V_{\mathrm{LOO}}=P_n(L_i^--C_{\mathrm{LOO}})^2.
\]
On $\mathcal A_n$, by the $L_0$-Lipschitz property of the losses and risk and the previous coefficient bounds, we obtain
$\max_i|A_i-L_i^-|\le C/n^2,\qquad \max_i|R(\widehat u^{(-i)})-R(\widehat u)|\le C/n.$
By uniform boundedness of $A_i,L_i^-$ and $P_n(A_i-\widehat C_{\rm ALO})^2=P_nA_i^2-(P_nA_i)^2$, we get $|P_n(A_i-\widehat C_{\rm ALO})^2-V_{\mathrm{LOO}}|\le C\max_i|A_i-L_i^-|\le C/n^2$. From these bounds, \eqref{S-reader-variance}, and the $C/n$ trace bound, we obtain
\[
\begin{gathered}
|\widehat C_{\rm ALO}-C_{\mathrm{LOO}}|=O_P(n^{-2}), \quad
|P_n R(\widehat u^{(-i)})-R(\widehat u)|=O_P(n^{-1}), \quad
\widehat V_{\rm ALO}-V_{\mathrm{LOO}}=O_P(n^{-1}),
\end{gathered}
\]
uniformly. Note that the all-pairs estimator of \citet[Theorem~5]{bayle2020confidence} equals $V_{\mathrm{LOO}}$. From the positive lower bound on $\Var(Q)$, the target and confidence-limit differences are $o_P(n^{-1/2})$, and hence we obtain first-order equivalence of the intervals by the uniform central limit theorem.

This completes the proof.
\end{proof}

\section{Additional simulation and clinical results}\label{S-bio-additional}
\subsection{Joint regions and Brier errors}\label{S-joint-evaluation}

By Theorem~\ref{S-xa-clt} and Proposition~\ref{S-xa-variance-consistency},
\[
 e_n=\sqrt n\begin{pmatrix}\widehat B-B_{\rm ext}\\\widehat R-R_{\rm ext}\end{pmatrix},\qquad
 e_n^\top\widehat\Sigma^{-1}e_n\dto\chi^2_2.
\]
From the $0.95$ quantile of $\chi^2_2$, we construct the joint regions in Table~\ref{tab:joint-covariance} under the design of Section~\ref{subsec:interval-comparisons}. In Table~\ref{S-tab-sim-brier}, we report 200-replication mean errors for the ten settings of Table~\ref{tab:sim-logistic}. We subtract the realized Brier loss from each estimate.

\begin{table}[htbp]
{\centering
\setlength{\arrayrulewidth}{0.3pt}
\setlength{\tabcolsep}{4pt}
\caption{Mean errors in estimating Brier loss.}\label{S-tab-sim-brier}
\begin{tabular}{@{}rrrrrrrr@{}}
\toprule
$\theta$ & $\gamma$ & $\kappa$ & $n$ & Harrell & .632 & 10-fold CV & One-fit\\
\midrule
$0$ & $1$ & $0.05$ & $1000$ & $-0.0006$ & $-0.0008$ & $+0.0018$ & $-0.0003$ \\
$0$ & $1$ & $0.1$ & $1000$ & $-0.0026$ & $-0.0012$ & $+0.0028$ & $-0.0005$ \\
$0$ & $1$ & $0.2$ & $1000$ & $-0.0080$ & $+0.0031$ & $+0.0101$ & $+0.0007$ \\
$0$ & $0.3$ & $0.05$ & $1000$ & $-0.0007$ & $-0.0008$ & $+0.0018$ & $-0.0002$ \\
$0$ & $0.3$ & $0.1$ & $1000$ & $-0.0037$ & $-0.0020$ & $+0.0034$ & $-0.0006$ \\
$0$ & $0.3$ & $0.2$ & $1000$ & $-0.0141$ & $-0.0043$ & $+0.0064$ & $-0.0021$ \\
$-2$ & $1$ & $0.05$ & $1000$ & $+0.0007$ & $+0.0005$ & $+0.0020$ & $+0.0003$ \\
$-2$ & $1$ & $0.1$ & $1000$ & $+0.0014$ & $+0.0035$ & $+0.0044$ & $+0.0010$ \\
$0$ & $1$ & $0.1$ & $2000$ & $-0.0024$ & $-0.0011$ & $+0.0032$ & $-0.0002$ \\
$0$ & $0.3$ & $0.1$ & $2000$ & $-0.0034$ & $-0.0018$ & $+0.0034$ & $-0.0001$ \\
\bottomrule
\end{tabular}\par}
\begin{tabnote}
CV, cross-validation. Equal rounded values are from separate 200-replication runs per row.
\end{tabnote}
\end{table}

\subsection{Gaussian predictors}\label{S-bio-gaussian}

In Section~\ref{sec:numerical}, mean errors of Harrell's AUC estimator are within 1.6 Monte Carlo standard errors of $\Delta_A$, and for balanced settings the bootstrap slope factor exceeds the limiting external slope by $0.03$--$0.31$. Over 500 replications per coverage setting, root mean squared error ratios of $\widehat F$ to ALO are $0.994$--$1.013$ for AUC, Brier loss, and log loss. Note that at $(\theta,\gamma,\kappa)=(-2,1,0.2)$ outside the bootstrap existence region, $99.0\%$ of bootstrap fits fail and 145 of 200 samples have no successful fit. On the other hand, we compute $\widehat F$ for each sample and obtain mean AUC error $-0.001$.

In Table~\ref{tab:cov-onefit} from 500 replications per setting, we get Bonferroni joint coverage of $95.4$--$98.0\%$ with the covariance correction and $79.0$--$96.0\%$ without it. Next, in the design of Table~\ref{tab:noma-intervals} at $(n,\gamma)=(600,1)$ over 100 replications, we observe $16\%$ Brier coverage for the two-stage interval of \citet[Algorithm~2]{noma2021confidence} with 200 outer and 100 inner resamples.

\subsection{Non-Gaussian predictors}\label{subsec:sim-robust}
We consider $n=1000$ and $(\theta,\gamma,\kappa)\in\{(0,1,0.1),(-2,1,0.05),(0,0.3,0.1)\}$, where two equal nonzero coefficients satisfy $\beta_n^\top\Sigma_n\beta_n=\gamma^2$. Specifically, predictors are Gaussian with first-order autoregressive (AR) correlation $0.5$, independent standardized Bernoulli($0.3$), alternating Bernoulli and Gaussian, or independent standardized $t_5$. Over 1,000 replications, mean AUC error is $0.014$--$0.047$ for Harrell's estimator and $-0.0025$ to $0.0002$ for $\widehat A$, and $\widehat A$ has lower root mean squared error than Harrell's, .632, and ten-fold cross-validation in each setting. In Table~\ref{tab:robust-intervals}, we report coverage over 500 replications.

\begin{table}[htbp]
{\centering
\setlength{\arrayrulewidth}{0.3pt}
\caption{Coverage ranges (\%) for Algorithm~1 across predictor distributions.}\label{tab:robust-intervals}
\begin{tabular}{@{}lrrrr@{}}
\toprule
 & \multicolumn{2}{c}{Brier} & \multicolumn{2}{c}{Log loss}\\
\cmidrule(lr){2-3}\cmidrule(lr){4-5}
Distribution & One-fit & Independent & One-fit & Independent\\
\midrule
Gaussian AR & 95.0--95.8 & 85.4--94.2 & 94.4--96.0 & 85.6--92.6\\
Bernoulli & 94.0--96.0 & 84.2--94.6 & 94.0--96.0 & 84.0--93.0\\
Mixed & 93.6--97.0 & 84.2--96.4 & 93.6--96.4 & 84.2--95.2\\
$t_5$ & 93.6--95.2 & 84.2--93.6 & 93.4--95.2 & 84.0--92.6\\
\bottomrule
\end{tabular}\par}
\begin{tabnote}
AR, first-order autoregressive.
\end{tabnote}
\end{table}

\rightskip=0pt plus .35em We consider four additional designs under Theorem~\ref{thm:optboot-nongaussian} with $X_1\sim\operatorname{Bernoulli}(0.2)$ and $X_2=(1+0.5X_1)\min(E,5)$ for standard exponential $E$ independent of $X_1$, where $X_2$ is skewed and dependent on $X_1$. Specifically, we let the other $p-2$ predictors be independent of $(X_1,X_2)$, Gaussian with autoregressive correlation $0.5$ in the Gaussian design, or independent Rademacher, standardized Bernoulli($0.2$), or standardized $\min(E,5)$ in the other three. For standardized $\widetilde X_1,\widetilde X_2$ and $c^2=\Var(\widetilde X_1+\widetilde X_2)$, we define log odds $\theta+\gamma\{(\widetilde X_1+\widetilde X_2)/c+(X_3+X_4)/\sqrt3\}/\sqrt2$ in the Gaussian design and $\theta+\gamma(\widetilde X_1+\widetilde X_2)/c$ in the other three, with other coefficients zero. We use $\theta=-1$, $\gamma=1$, $n\in\{1000,2000\}$, $\kappa\in\{0.1,0.2\}$, and 1,000 replications per setting. Note that $h_Z$ in \eqref{S-NG-eq-headboundary} rounds to $0.39$ (Gaussian) and $0.38$ (other three) and exceeds both $\kappa$ values. We get finite fits and average conditional expected loss given predictors on an independent test sample of size $2\times10^5$ ($10^6$ in the Gaussian design) to evaluate realized Brier and log losses. Table~\ref{tab:nongaussian-components} reports nominal $95\%$ coverage ranges over the two $\kappa$ values for One-fit intervals with $\widehat\Sigma$ and Independent intervals without its rank-one term. Here we observe One-fit marginal coverage $93.3$--$96.5\%$ across the 16 settings ($94.3$--$95.8\%$ at $n=2000$) and Independent coverage $83.7$--$93.7\%$. Next, we obtain $92.2$--$95.2\%$ coverage for $95\%$ Wald ellipses with $\widehat\Sigma$ and $84.7$--$91.9\%$ without its rank-one term.

\setlength{\textfloatsep}{10pt plus 2pt minus 2pt}
\begin{table}[htbp]
\centering
\setlength{\arrayrulewidth}{0.3pt}
\caption{Coverage ranges (\%) for Algorithm~1 under predictors with non-Gaussian components.}\label{tab:nongaussian-components}
\begin{tabular}{@{}lrrrrr@{}}
\toprule
 & & \multicolumn{2}{c}{Brier} & \multicolumn{2}{c}{Log loss}\\
\cmidrule(lr){3-4}\cmidrule(lr){5-6}
Remaining predictors & $n$ & One-fit & Independent & One-fit & Independent\\
\midrule
Gaussian & 1000 & 94.4--94.6 & 87.4--91.5 & 94.3--95.4 & 84.6--90.5\\
Gaussian & 2000 & 95.4--95.8 & 87.8--91.1 & 94.8--94.9 & 85.1--90.7\\
Rademacher & 1000 & 94.9--95.1 & 89.3--92.5 & 94.7--95.4 & 86.0--91.5\\
Rademacher & 2000 & 94.3--95.5 & 87.0--91.9 & 95.0--95.7 & 84.6--91.8\\
Bernoulli & 1000 & 93.6--94.9 & 87.6--90.5 & 93.3--95.0 & 83.7--89.4\\
Bernoulli & 2000 & 95.6--95.7 & 87.4--92.6 & 95.3--95.4 & 85.7--91.1\\
Truncated exponential & 1000 & 94.8--96.3 & 88.3--93.7 & 93.6--96.5 & 85.7--92.2\\
Truncated exponential & 2000 & 95.0--95.7 & 90.1--92.5 & 95.5--95.8 & 85.8--92.1\\
\bottomrule
\end{tabular}
\end{table}

\subsection{Ridge regression}\label{subsec:ridge-numerics}
For the clinical fits, we use the fixed centers $c_j$ and scales $s_j$ from Section~\ref{subsec:clinical-design} in
\begin{equation}\label{eq:bounded-features}
 \phi_j(x)=3\tanh\!\left[\frac13\operatorname{asinh}\{(x_j-c_j)/s_j\}\right],
 \qquad s_j>0.
\end{equation}

We provide the centers and scales for 21 numerical predictors in \texttt{fixed\_preprocessing.csv}. In SUPPORT, we use those centers for missing values after treating albumin above 15 as missing.

In Algorithm~2, we transform two independent standard Cauchy covariates by~\eqref{eq:bounded-features} with zero centers and unit scales under log odds $a+b\{\phi_1(X)+\phi_2(X)\}/\sqrt2+c\phi_1(X)\phi_2(X)$ for $(b,c)\in\{(0,0),(1,0),(1,0.3)\}$, where we choose $a$ for prevalence $0.26$. In 1,000 replications at $n=100,400$, we fit only main effects on $\phi_1(X),\phi_2(X)$ at $\rho=0.05$, and all fits succeed with $94.3$--$95.7\%$ coverage for $I_{\rm ALO}$ from Corollary~\ref{cor:bounded-ridge}. Next, at $p=\lfloor\sqrt n\rfloor$ with independent standardized Bernoulli($0.3$) predictors and log odds $-1+(X_1+X_2)/\sqrt2+aX_1X_2$ for $a\in\{0,0.5\}$, we form TIC intervals from Theorem~\ref{thm:moderate} with $\widehat C_{\rm TIC}$ and unadjusted losses in~\eqref{S-reader-variance}, and in 500 replications of unpenalized main-effects fits we get $91.4$--$92.0\%$ coverage at $n=200$ and $94.4$--$95.6\%$ at $n=800,3200$.

\subsection{Comparison with cross-validation}\label{S-practical-cv}
In Section~\ref{subsec:practical-comparators}, for the 500 Pima samples at each of $n=100,200$, we form naive ten-fold cross-validation (CV) intervals from mean held-out log loss $\pm1.96$ sample standard deviations of the $n$ individual held-out losses (divisor $n-1$) divided by $\sqrt n$, and clip the lower endpoint at zero. Specifically, for CV and $I_{\rm ALO}$, we get median widths $0.173,0.115$ and $0.191,0.122$ and median CPU times $0.00247$--$0.00372$ and $0.000469$--$0.000660$ seconds.

\subsection{Penalty selection}\label{app:penalty-selection}

For the full fits, we use $\rho=0.01$ as selected in Section~\ref{subsec:practical-comparators}. The log-loss point estimates and Bonferroni $95\%$ intervals equal $0.4746$ $[0.4615,0.4877]$ for SUPPORT and $0.5266$ $[0.4766,0.5767]$ for Pima.

Coverage by sample size is in \texttt{checks/*SELECTION\_SUMMARY.json} within the practical archive.

\clearpage
\phantomsection
\addcontentsline{toc}{part}{References}
\bibliographystyle{abbrvnat}
\bibliography{references}
\end{document}